\documentclass[twocolumn]{aastex631}

\usepackage{amsmath}
\usepackage{commath}
\usepackage{xcolor}
\usepackage{hyperref}

\usepackage[english]{babel}

\newcommand{\vect}[1]{\boldsymbol{#1}}

\def\siiv     {\ensuremath{\text{Si\,\textsc{iv}}}}

\def\oiv    {\ensuremath{\text{O\,\textsc{iv]}}}}

\def\civ     {\ensuremath{\text{C\,\textsc{iv}}}}

\def\mgii     {\ensuremath{\text{Mg\,\textsc{ii}}}}

\def\nv     {\ensuremath{\text{N\,\textsc{v}}}}

\def\feii     {\ensuremath{\text{Fe\,\textsc{ii}}}}

\def\feiii    {\ensuremath{\text{Fe\,\textsc{iii}}}}

\def\ciii     {\ensuremath{\text{C\,\textsc{iii]}}}}

\def\halpha {\ensuremath{\text{H}\alpha}}

\def\ncand {1032}  

\def\nqsobinned{121}

\def\ncandgood{640}

\def\nqsolit{41}
\def\nfollowed{274}
\def\noqsophot{79}

\def\nfollowedspec{195}
\def\nqsospec{84}
\def\nqso{125}
\def\nnotfollowed{325}

\shorttitle{The Pan-STARRS\,1 $z\approx6$ QLF}
\shortauthors{Schindler et al.}
\graphicspath{{./}{figures/}}

\begin{document}

\title{The Pan-STARRS1 $\mathbf{z>5.6}$ Quasar Survey: III. The $\mathbf{z\approx6}$ Quasar Luminosity Function}

\correspondingauthor{Jan-Torge Schindler}
\email{schindler@strw.leidenuniv.nl}
\author[0000-0002-4544-8242]{Jan-Torge Schindler}
\affiliation{Max Planck Institut f\"ur Astronomie, K\"onigstuhl 17, D-69117, Heidelberg, Germany}
\affiliation{Leiden Observatory, Leiden University, PO Box 9513, 2300 RA Leiden, The Netherlands}

\author[0000-0002-2931-7824]{Eduardo Ba{\~n}ados}
\affiliation{Max Planck Institut f\"ur Astronomie, K\"onigstuhl 17, D-69117, Heidelberg, Germany}

\author[0000-0002-7898-7664]{Thomas Connor}
\affiliation{Jet Propulsion Laboratory, California Institute of Technology, 4800 Oak Grove Drive, Pasadena, CA 91109, USA}

\author[0000-0002-2662-8803]{Roberto Decarli}
\affiliation{INAF --- Osservatorio di Astrofisica e Scienza dello Spazio, via Gobetti 93/3, I-40129, Bologna, Italy}

\author[0000-0003-3310-0131]{Xiaohui Fan}
\affiliation{Steward Observatory, University of Arizona, 933 N Cherry Ave, Tucson, AZ 85721, USA}

\author[0000-0002-6822-2254]{Emanuele Paolo Farina}
\affiliation{Gemini Observatory, NSF’s NOIRLab, 670 N A’ohoku Place, Hilo, Hawai'i 96720, USA}

\author[0000-0002-5941-5214]{Chiara Mazzucchelli}
\affiliation{Núcleo de Astronomía, Facultad de Ingeniería y Ciencias, Universidad Diego Portales, Av. Ejército 441, Santiago, 8320000, Chile}

\author[0000-0002-2579-4789]{Riccardo Nanni}
\affiliation{Leiden Observatory, Leiden University, PO Box 9513, 2300 RA Leiden, The Netherlands}

\author[0000-0003-4996-9069]{Hans-Walter Rix}
\affiliation{Max Planck Institut f\"ur Astronomie, K\"onigstuhl 17, D-69117, Heidelberg, Germany}

\author[0000-0003-2686-9241]{Daniel Stern}
\affiliation{Jet Propulsion Laboratory, California Institute of Technology, 4800 Oak Grove Drive, Pasadena, CA 91109, USA}

\author[0000-0001-9024-8322]{Bram P. Venemans}
\affiliation{Leiden Observatory, Leiden University, PO Box 9513, 2300 RA Leiden, The Netherlands}

\author[0000-0003-4793-7880]{Fabian Walter}
\affiliation{Max Planck Institut f\"ur Astronomie, K\"onigstuhl 17, D-69117, Heidelberg, Germany}



\begin{abstract}
We present the $z\!\approx\!6$ type-1 quasar luminosity function (QLF) based on the Pan-STARRS1 (PS1) quasar survey. The PS1 sample includes \nqso{} quasars at $z\approx5.7-6.2$ with $-28\lesssim M_{1450}\lesssim-25$. Complemented by 48 fainter quasars from the SHELLQs survey, we evaluate the $z\approx6$ QLF over $-28\lesssim M_{1450}\lesssim-22$. Adopting a double power law with an exponential evolution of the quasar density ($\Phi(z)\propto10^{k(z-6)}$; $k=-0.7$), we use a maximum likelihood method to model our data. We find a break magnitude of $M^*=-26.38_{-0.60}^{+0.79}\,\text{mag}$, a faint end slope of $\alpha=-1.70_{-0.19}^{+0.29}$, and a steep bright end slope with $\beta=-3.84_{-1.21}^{+0.63}$.
Based on our new QLF model we determine the quasar comoving spatial density at $z\!\approx\!6$ to be $n( M_{1450}<-26)=1.16_{-0.12}^{+0.13}\,\text{cGpc}^{-3}$. In comparison with the literature, we find the quasar density to evolve with a constant value of $k\approx-0.7$ from $z\approx7$ to $z\approx4$.
Additionally, we derive an ionizing emissivity of $\epsilon_{912}(z=6) =7.23_{-1.02}^{+1.65}\times 10^{22}\,\text{erg}\,\text{s}^{-1}\text{Hz}^{-1}\text{cMpc}^{-3}$ based on the QLF measurement. Given standard assumptions and the recent measurement of the mean free path of \citet{Becker2021} at $z\approx6$ we calculate an HI photoionizing rate of $\Gamma_{\text{HI}}(z{=}6) \approx 6 \times10^{-16}\,\text{s}^{-1}$, strongly disfavoring a dominant role of quasars in hydrogen reionization.
\end{abstract}



\section{Introduction} \label{sec:intro}
Quasars are rapidly accreting supermassive black holes (SMBHs) at galaxy centers, which shine as the most luminous non-transient light sources in the Universe.
At low redshift tight correlations between the SMBH mass and its host galaxy's central properties raised attention on the role of AGNs in galaxy evolution \citep[see][ for a review]{Kormendy2013}.
Specifically, feedback during bright quasar phases has been identified as a prominent avenue to establish this relationship \citep[e.g.,][]{DiMatteo2005}.
In this context, understanding the evolution of quasars has received growing attention in recent years, especially their evolution in the early Universe.
Following the discovery of the first $z\gtrsim6$ quasars \citep{Fan2001c} it was quickly realized that SMBHs with masses of $M_{\text{BH}}\approx 10^9\,\text{M}_\odot$ already exist less than $1\,\text{Gyr}$ after the Big Bang, placing constraints on their formation. These have been significantly tightened by the discovery of the most distant quasars at $z\approx7.5$ known today \citep{Banados2018, YangJinyi2020, WangFeige2021b}.
While single sources highlight the open questions with regard to SMBH formation \citep[see][ for a review]{Inayoshi2020}, understanding the full demographics at high redshifts will play a key role in addressing them with the quasar luminosity function (QLF), the main observational statistic to characterize their population.
%

At lower redshifts \citep[e.g.,][]{Boyle1988, Boyle2000, Pei1995}, the QLF is most effectively described by a broken double power law, which has been widely adopted at higher redshifts as a good representation of most quasar samples \citep[e.g.,][]{Richards2006, Shen2012, Ross2013}.
In this context the QLF is described by a break magnitude, a normalization, and two power--law slopes.
%
At $z\approx6$ the most comprehensive measurement of the QLF was presented in \citet{Matsuoka2018c} using a combined sample built from the Subaru High-$z$ Exploration of Low-Luminosity Quasars \citep[SHELLQs;][]{Matsuoka2016} project and previous QLF analyses \citep{Jiang2008, Jiang2016, Willott2010a} with a total of 112 sources, covering a redshift range of $5.7 < z < 6.5$ and luminosities of $-30 \lesssim  M_{1450}/\text{mag} \lesssim -22$.
Placed in context with the lower redshift QLF literature \citep[e.g.,][]{Richards2006, Croom2009, Glikman2011, Shen2012, Ross2013, Akiyama2018, Schindler2018, Schindler2019a, Boutsia2021, PanZhiwei2022} the recent results underline the (exponential) increase \citep{Schmidt1995,Fan2001c} in quasar activity from $z\approx6$ to its peak at $z=2-3$ \citep{Richards2006, Kulkarni2019, ShenXuejian2020}.
%
%
The highest redshift constraint on the QLF at $z\approx6.7$ \citep{WangFeige2019a} indicates an even more rapid decline of quasar activity at $z>6.5$ with consequences for upcoming quasars surveys at $z>8$ (e.g., based on the Euclid mission wide survey; \citet{EuclidPrepV2019, EuclidPrepI2021}).

Quasars, or more generally active galactic nuclei (AGN), and star formation are the major sources of ultra-violet (UV) radiation that drive the reionization of intergalactic hydrogen.
The role of quasars in this process, as inferred from QLF number counts, has been a matter of debate in the literature.
Type-1 UV QLFs at $z\approx5$ \citep[e.g.,][]{McGreer2013, McGreer2018, Yang2016, KimYongjung2020, ShinSuhyun2020}, at $z\approx6$ \citep[e.g.,][]{Jiang2008, KimYongjung2015, Jiang2016, Matsuoka2018c}, at $z\approx6.5$ \citep{WangFeige2019a}, and from the redshift compilation of \citet{Kulkarni2019} provide substantial evidence for a subdominant contribution of quasars compared to star formation to reionization at $z\gtrsim5$. These results are supported by the analysis of the bolometric QLF based on multi-wavelength data sets \citep{ShenXuejian2020}.
On the other hand, \citet{Giallongo2015}, \citet{Giallongo2019}, and \citet{Grazian2020} find high quasar number densities for lower luminosity sources, $-22.5 \leq  M_{1450} \leq -18.5$. These studies are based on multi-wavelength selected sources from the Cosmic Assembly Near-IR Deep Extragalactic Legacy Survey (CANDELS) GOODS-South, GOODS-North, and EGS fields.
The analysis in \citet{Giallongo2015} and \citet{Giallongo2019} is largely based on photometric candidates, whereas the study of \citet{Grazian2020} has spectroscopy for their two sources.
Based on these number densities at the faint-end the authors argue that quasars could be the dominant source of ionizing photons at $z\approx4-6$, which is supported by the analysis of \citet{Grazian2022} based on the QUBRICS quasar survey \citep{Calderone2019, Boutsia2020}.
However, a range of independent deep X-ray studies report significantly lower number densities for the faint quasar population \citep[e.g.,][]{Weigel2015, Cappelluti2016, Vito2016, RicciF2017, Parsa2018}, challenging the results of \citet{Giallongo2015} and \citet{Giallongo2019}.

In this work we present a new measurement of the type-1 UV QLF at $z\approx6$ based on the selection strategy and discoveries from the Pan-STARRS distant quasar survey \citep{Banados2014, Banados2016, Banados2022}. Our quasar sample includes \nqso{} sources at $z\approx5.7-6.2$ within a luminosity range of $-28\lesssim M_{1450}\lesssim-25$, more than doubling the number counts of previous samples in this range \citep[][SDSS]{Jiang2016}. Combining the new sample with lower luminosity sources from SHELLQs \citep{Matsuoka2018c}, we present the most precise measurement of the type-1 UV QLF at these redshifts to date.

In Section\,\ref{sec:data} we review the quasar selection of the PS1 distant quasar survey and present the new quasar sample used in this work. Section\,\ref{sec:analysis} discusses the resulting quasar selection function and completeness. We present the QLF in Section\,\ref{sec:results} and discuss the implications with regard to quasar evolution and reionization in Section\,\ref{sec:discussion}. Finally, we summarize this work in Section\,\ref{sec:conclusions}.
Interested readers can find the mathematical framework for our QLF analysis described in detail in Appendix\,\ref{app:qlfanalysis},  whereas Appendix\,\ref{app:qsomodel} expands on the discussion of our quasar model used for the completeness calculation.
In this work we adopt a $\Lambda$CDM cosmology with $\text{H}_0=70\,\text{km}\,\text{s}^{-1}\,\text{Mpc}^{-1}$, $\Omega_{\text M}=0.3$, and \mbox{$\Omega_\Lambda=0.7$}.
All magnitudes are reported in the AB photometric system.

\section{Data}\label{sec:data}

The foundation of the quasar searches this work builds upon is the $3\pi$ Steradian Survey \citep{Chambers2016} that was carried out by the Panoramic Survey Telescope and Rapid Response System Telescope \verb|#|1 \citep[PS1,][]{Kaiser2002, Kaiser2010}.
From 2009 to 2015 the PS1 $3\pi$ survey imaged the sky above a declination of $-30^{\circ}$ in the five filter bands $g_{\text{P1}}, r_{\text{P1}}, i_{\text{P1}}, z_{\text{P1}}, y_{\text{P1}}$. The full data releases of the PS1 $3\pi$ survey are hosted by the Barbara A. Mikulski Archive for Space Telescopes (MAST) at the Space Telescope Science Institute\footnote{\url{https://panstarrs.stsci.edu}}.

\subsection[The PS1 3 PI Steradian Survey PV2 catalog]{The PS1 3$\pi$ Steradian Survey PV2 catalog}
The PS1 $3\pi$ distant quasar survey \citep{Banados2014, Banados2016} based their selection on the internal pre-releases of the stacked PS1 $3\pi$ photometry, in particular the second internal data release PV2. According to \citet[][Section 2.1]{Banados2016} the PV2 $5\sigma$ median limiting magnitudes are ($g_{\text{P1}}, r_{\text{P1}}, i_{\text{P1}}, z_{\text{P1}}, y_{\text{P1}}$) $= (23.2, 23.0, 22.7, 22.1, 21.1)$.
The quasar selection uses the stacked PSF magnitude signal to noise ratio (SNR$_{x}$) and the ($3\sigma$) limiting magnitude ($m_{\text{lim, x}}$) of a given band $x$ as selection criteria. These properties are derived from quantities in the PV2 catalog. We briefly describe them here to provide context for Section\,\ref{sec:modelps1phot}, where we discuss the modeling of simulated quasar photometry to assess the selection function completeness.
The SNR$_{x}$ is calculated from the stacked PSF magnitude $1\sigma$ uncertainty ($\sigma_{m,x}$) of the same filter band $x$:
\begin{gather}
     \text{SNR}_{x} = \frac{2.5}{\ln{10} \times \sigma_{m,x}} \ . \label{eq:snr}
\end{gather}

The $3\sigma$ limiting magnitude $m_{\text{lim, x}}$ is derived from the band zero point (\texttt{zp}$_x$) and the $1\sigma$ uncertainty on the stacked PSF fit instrumental flux (\texttt{PSF\_INST\_FLUX\_SIG}):
\begin{gather}
    m_{\text{lim},x} = -2.5\times \log_{10}(3 \times \text{\texttt{PSF\_INST\_FLUX\_SIG}}_{x}) + \text{\texttt{zp}}_{x} \ . \label{eq:mlim}
\end{gather}
The \texttt{PSF\_INST\_FLUX\_SIG} is a PV2 catalog property from the internal data release, but not provided in the public data release on MAST.
The band $x$ zero point, \texttt{zp}$_x$, only depends on the stacked exposure time (\texttt{EXPTIME}$_x$) in that filter band:
\begin{gather}
    \text{\texttt{zp}}_x = 25 + 2.5 \times \log_{10}(\text{\texttt{EXPTIME}}_x) \ . \label{eq:zp}
\end{gather}

All stacked magnitudes are corrected for Galactic extinction using the \citet{Schlegel1998} dust map with the corrections of \citet{Schlafly2011}. The quasar photometric selection is conducted on the \textit{dereddened} stacked magnitudes.

\subsection{Quasar candidate selection}\label{sec:qso_selection}
\begin{figure*}[ht]
    \centering
    \includegraphics[width=0.9\textwidth]{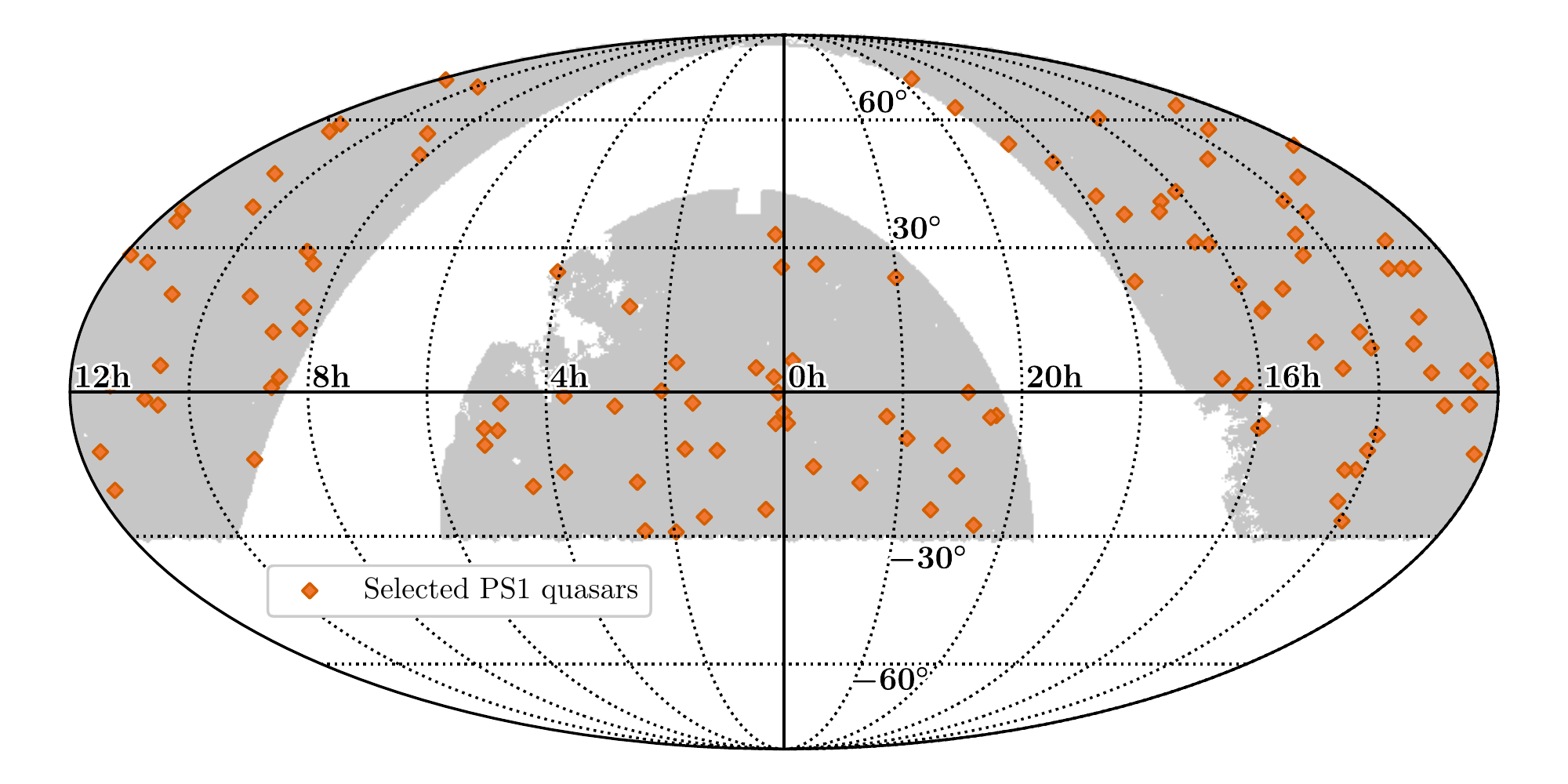}
    \caption{Mollweide projection of the PS1 quasar survey area ($20803\,\text{deg}^2$) considered for the QLF analysis. Right ascencsion and Declination are noted in hours and degrees. Regions included in the quasar selection based on the PS1 PV2 catalog are shaded grey using a HEALPix tesselation with Lvl 10 (12,582,912 HEALPix cells over the entire sky, see Table\,\ref{tab:surveyarea}). The M31 mask is included (RA 0h\,42m\,44s, Dec +41$^{\circ}$\,16' 9$\arcsec$). High redshift quasars (Section\,\ref{sec:qso_selection}) are shown as filled orange diamonds.
    }
    \label{fig:coverage}
\end{figure*}
We follow the quasar selection criteria of \citet[][ Section 2.1.1]{Banados2016}, focusing on the search of quasars at $5.7 \lesssim z \lesssim 6.2$. The selection criteria are applied to the PS1 PV2 catalog generated from PS1 image stacks.

For completeness, we briefly describe the selection here.
First, we exclude sources which have been flagged as suspicious \citep[see][Table\,6]{Banados2014} by the Imaging Processing Pipeline \citep{Magnier2020DataProcessing, Magnier2020PixelAnalysis}. Additionally, we require $85\%$ of  the normalized point-spread-function (PSF) flux in the $i_{\text{P1}}$, $z_{\text{P1}}$, and $y_{\text{P1}}$ bands to be located in unmasked pixels (\texttt{PSF\_QF} $>0.85$).
This quality cut leans towards a more complete selection by including some lower-quality measurements \citep{Magnier2020PixelAnalysis}. We will refer to these requirements on the PV2 catalog as our ``photometric quality selection'' hereafter.

The Milky Way plane has traditionally been avoided by quasar surveys using criteria based on Galactic latitudes \citep[e.g.,][]{Fan1999a, Jiang2016}. The main reason is that the high source density and stronger Galactic extinction leads to unreliable photometry for extragalactic background sources.
Following \citet{Banados2014} we impose a Galactic latitude limit of $|b| > 20^\circ$ and additionally select only sources with modest degrees of Galactic reddening as determined from the \citet{Schlegel1998} dust map cross-matched to the PS1 PV2 source catalog, our ``extinction selection'' criterion:
\begin{gather}
    E(B-V) < 0.3 \label{eq:ext_criterion} \ .
\end{gather}
Additionally, we exclude all sources around M31 ($7^\circ < \text{R.A.} < 14^\circ; 37^\circ < \text{Decl.} < 43^\circ$) as their inclusion results in a large number of candidates that are most likely stars associated with M31.
Figure\,\ref{fig:coverage} shows a Mollweide projection of the sky with the PS1 quasar survey coverage shown in gray. Confirmed high redshift quasars selected as described in this section are highlighted as orange diamonds.
A description of the resulting survey area is provided in Section\,\ref{sec:selectionfunction}, where we discuss the different contributions of the selection criteria to the survey selection function.

Apart from M--, L--, and T--dwarf stars (or brown dwarfs), the main contaminants for high-redshift quasars are low redshift galaxies, which mostly appear extended under the PS1 observing conditions. The PS1 $3\pi$ website\footnote{\href{https://panstarrs.stsci.edu/}{https://panstarrs.stsci.edu/}} lists median seeing conditions of $1.11\arcsec, 1.07\arcsec,\ \text{and}\ 1.02\arcsec$ for the $i_{\text{P1}}$, $z_{\text{P1}}$, and $y_{\text{P1}}$ filter bands, respectively. To reject extended sources we adopt the ``morphology criterion'' as discussed in \citet[][their Section\,2.1]{Banados2016}. We keep sources whose absolute difference between the aperture and PSF magnitudes, $|f_{\text{ext}}|$, is below a value of $0.3$ in either the PS1 \textit{z}-- or \textit{y}--band:
\begin{gather}
    |f_{\text{ext},z}| < 0.3\ \text{OR}\ |f_{\text{ext},y}| < 0.3 \label{eq:morph_criterion} \ .
\end{gather}
\citet{Banados2016} tested this criterion against known spectroscopic stars and galaxies in SDSS \citep[DR12;][]{Alam2015} and quasars \citep[SDSS DR10 quasar catalog][]{Paris2014}. This criterion removes the majority ($92\%$) of galaxies, while retaining $92\%$ of stars and $97\%$ of quasars.

The focus of this quasar luminosity function analysis is the redshift range $5.7 \lesssim z \lesssim 6.2$. At these redshifts quasars can be efficiently differentiated from brown dwarfs by applying color criteria to the PS1 \textit{i}--, \textit{z}--, and \textit{y}--band stacked magnitudes \citep{Banados2014}.
We summarize the ``photometric selection'' criteria discussed in \citet{Banados2014} for the $5.7 \lesssim z \lesssim 6.2$ range below.

\begin{gather}
    \text{SNR}_{z_{\text{P1}}} > 10 \label{eq:phot_criterion1}\\
    \text{SNR}_{y_{\text{P1}}} > 5 \\
    \label{eq:phot_criterion2}\\
    (z_{\text{P1}}-y_{\text{P1}}) < 0.5 \label{eq:phot_criterion3} \\
    ((\text{SNR}_{i_{\text{P1}}} \ge 3\ \text{AND}\ ({i_{\text{P1}}}-{z_{\text{P1}}}) > 2.0)\  \text{OR}\ \nonumber \\  (\text{SNR}_{i_{\text{P1}}} < 3\ \text{AND}\ (i_{\text{P1,lim}}-z_{\text{P1}}) > 2.0)) \label{eq:phot_criterion4} \\
    (\text{SNR}_{r_{\text{P1}}} < 3\ \text{OR}\   (r_{\text{P1}}-z_{\text{P1}}) > 2.2) \\
    \text{SNR}_{g_{\text{P1}}} < 3 \ ,\label{eq:phot_criterion5}
\end{gather}
These criteria are applied to the stacked, \textit{dereddened} PS1 magnitudes.

After all the above selection criteria are applied to the PS1 PV2 photometry, we perform forced photometry on the PS1 stacked and single-epoch images \citep[see Section 2.2 and 2.3 in][]{Banados2014}. We remove all sources where the forced photometry is inconsistent with the reported values in the PS1 PV2 catalog.
Effectively this removes $80\%$ of the candidates mainly due to discrepancies in the $i_{\text{P1}}$-band \citep[][their Section\,2.2]{Banados2014}.


After the steps above, which are all automatic,  this yields a total of \ncand{} candidates. The PS1 photometry of the \textit{z}--band and \textit{y}--band has been taken close in time to each other. Therefore, we visually inspect their stacked and single--epoch images to exclude bright spurious  sources (e.g., moving objects) that appear in only one single-epoch \textit{z}--band \textit{y}--band pair. %
EB and JTS visually inspected all \ncand{} candidates independently and assigned a rank from 1 (good photometry) to 4 (inconsistent/erroneous photometry).
Sources that are clearly detected in the $z$-band, have consistent single epoch measurements, and a $i-z$ color in agreement with the catalog are given rank 1.
Sources with rank 2 usually have a fainter detection in the $z$-band with their stacked $y$-band measurement close to the the limit of our SNR requirement, such that the single epoch images are harder to assess.
Sources with rank 3 usually have some issues with the data. For example, a clear $z$-band detection is only visible in 1 or 2 epochs. Sources that are given rank 4 show clear data artifacts.
The final candidates catalog includes sources for which the summed rank is 2, 3, or 4. This results in a total of \ncandgood{} quasar candidates for follow-up observations, 73 sources with summed rank 2, 117 sources with summed rank 3 and 450 sources with summed rank 4.
All sources in the final candidate catalog received an individual rank of either 1 or 2 from either EB or JTS.
We exclude a total of 202 sources with summed visual rank 5 and 190 sources with an even higher summed visual rank ($>5$) from further follow-up.

%

The full quasar selection procedure can be summarized in seven individual steps:
\begin{enumerate}
    \item Source detection in PS1 PV2 catalog
    \item Photometric quality selection
    \item Extinction selection and area exclusion
    \item Morphological selection
    \item Photometric selection
    \item Photometry consistency check
    \item Visual inspection
\end{enumerate}
Each of these selection steps has an impact on the selection function of the survey, which we discuss in Section\,\ref{sec:selectionfunction}.


\subsection{The PS1 high redshift quasar sample}\label{sec:sample}
\begin{figure}
    \centering
    \includegraphics[width=0.5\textwidth]{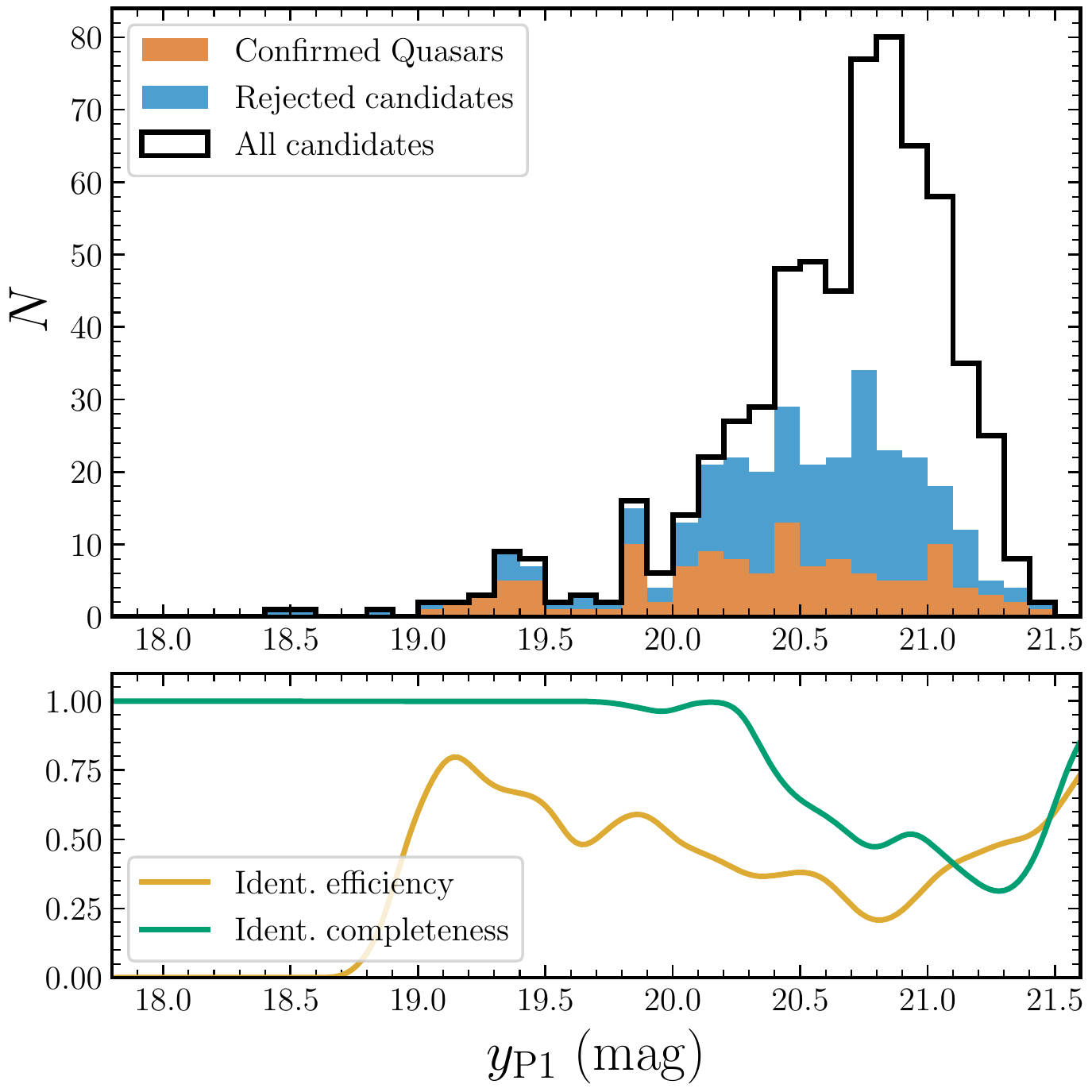}
    \caption{\textit{Upper panel:}  Histogram of all quasar candidates selected for follow-up observations (black solid line) as a function of dereddened \textit{y}--band magnitude. Quasars confirmed from the literature or through our observations are shown in orange.  Candidates rejected by observations are shown in light blue stacked on the confirmed quasars.
    \textit{Lower panel:} Kernel density estimates of the selection efficiency (yellow) and the adjusted identification completeness (green, Equation\,\ref{eq:completeness2}) as a function of dereddened \textit{y}--band magnitude.
    The observed completeness and efficiency increase towards the faint end. The KDE extrapolates this beyond the observed magnitude range.}
    \label{fig:spec_completeness}
\end{figure}

%
The detailed follow--up strategy along with descriptions of the photometric and spectroscopic data is presented in \citet{Banados2014, Banados2016, Banados2022}.
Of the \ncandgood{} good quasar candidates, \nqsolit{} are published quasars in the literature and \nfollowed{} sources have been photometrically or spectroscopically followed up.
The photometric follow-up rejected \noqsophot{} candidates.
These sources were ruled out as good candidates due to their red $y_{\text{P1}}-J$ color ($y_{\text{P1}}-J>1$) or if the follow-up photometry did not meet the main selection criteria (Equations\,\ref{eq:phot_criterion1}--\ref{eq:phot_criterion5}).
Our confirmation spectroscopy identified \nqsospec{} quasars among the \nfollowedspec{} observed sources. Due to limited telescope time \nnotfollowed{} quasar candidates have not been followed up, yet.
We display the full identification statistics in Figure\,\ref{fig:spec_completeness}. The upper panel shows histograms of the confirmed quasars ($N_{\text{CQ}}(y_{\text{P1}})$), the rejected candidates ($N_{\text{RC}}(y_{\text{P1}})$; photometry and spectroscopy) and the total number of candidates.
%
%
We estimate the identification efficiency, $\text{Eff}_{\text{ID}}(y_{\text{P1}})$, as the ratio between confirmed candidates to the total number of identified candidates,
\begin{gather}
    \text{Eff}_{\text{ID}} = \frac{N_{\text{CQ}}}{N_{\text{CQ}} + N_{\text{RC}}} \label{eq:efficiency}
\end{gather}
We express the identification completeness, $\text{S}_{\text{ID}}(y_{\text{P1}})$, as the number of observed quasars divided by the number of expected quasars given our efficiency:
\begin{gather}
    \text{S}_{\text{ID}} = \frac{N_{\text{CQ}}}{N_{\text{CQ}} + \rm{Eff}_{\rm{ID}}\times N_{\text{NoID}}}  \ .
\end{gather}
To calculate a continuous distribution of the completeness and efficiency as a function of $y_{\text{P}1}$, we use a Gaussian kernel density estimate. Kernel density estimation (KDE) provides a non-parametric representation of the probability density function of a random variable. In comparison to binning it is smooth and independent of the end points of the bins. For the full sample we determined the best bandwidth via cross-validation to be around 0.1. We will use this value for all KDE estimates related with the identification completeness.
In the follow-up campaigns sources were prioritized by visual rank. This leads to a bias in the completeness between samples of different rank (see Figure\,\ref{fig:spec_completeness_ranks} in the Appendix).
To mitigate the bias we calculate the expected number of quasars for each rank individually, accounting for the different efficiencies of each sample.
We sum the number of expected quasars for each of the three samples (ranks 2,3, and 4) and then calculate and adjusted identification completeness:
\begin{gather}
    \text{S}_{\text{ID}} = \frac{N_{\text{CQ}}}{N_{\text{CQ}} + \sum_{r=2}^{4} \rm{Eff}_{\rm{ID,r}}\times N_{\text{NoID},r}} \ , \label{eq:completeness2}
\end{gather}
We show the efficiency and the adjusted identification completeness in the bottom panel of Figure\,\ref{fig:spec_completeness}.
%
The selection efficiency of the full sample is highest at the bright end of the candidate distribution and then declines towards fainter magnitudes.
The low number statistics at the faint end (faintest bin the top panel) result in an upturn of the efficiency at the faint end.
%
%
Our sample has been followed up with very high identification completeness as the bright end ($y_{\text{P}1}<20.0$).
Towards fainter magnitudes the completeness declines with two visible minima at $y_{\text{P}1}=20.75$ and $y_{\text{P}1}=21.3$, where it reaches only 50\% and $30\%$, respectively.
At the faint end, $y_{\text{P}1}>21.3$, our follow-up becomes more complete again and the identification completeness increases.
%

The full quasar sample used in our QLF analysis consists of the \nqso{} confirmed PS1-selected quasars. Their sky distribution is shown in Figure\,\ref{fig:coverage} and we provide a complete list in the Appendix (Table\,\ref{tab:qsos}).
The quasars have redshifts in the range of $z=5.54$ to $6.31$ with a median of $5.88$. Their dereddened \textit{z}--band (\textit{y}--band) magnitudes are within $18.68$ to $21.53$ ($19.03$ to $21.47$) with a median of $20.36$ ($20.37$).

\section{Analysis}\label{sec:analysis}

The analysis of the QLF requires us to quantify our quasar selection function  (see Section\,\ref{sec:qso_selection}), including a correction for the incomplete spectroscopic follow-up (Section\,\ref{sec:sample}).
In order to realistically evaluate the photometric selection for type-1 quasars we simulate quasar photometry taking into account the properties of the PS1 PV2 catalog (e.g., inhomogeneous depth). In particular, to apply the photometric selection criteria from Equations\,\ref{eq:phot_criterion1}--\ref{eq:phot_criterion5}, we need to produce observed (error-prone) magnitudes, the signal--to--noise ratio and the limiting magnitudes.
We describe the photometric modeling in Section\,\ref{sec:photomodeling} and continue to derive K--correction terms based on these models to estimate absolute magnitudes with the QLF quasar sample in Section\,\ref{sec:kcorrection}. The selection function is then evaluated in Section\,\ref{sec:selectionfunction}.


\subsection{Modeling PS1 quasar photometry} \label{sec:photomodeling}

\begin{figure*}
    \centering
    \includegraphics[width=\textwidth]{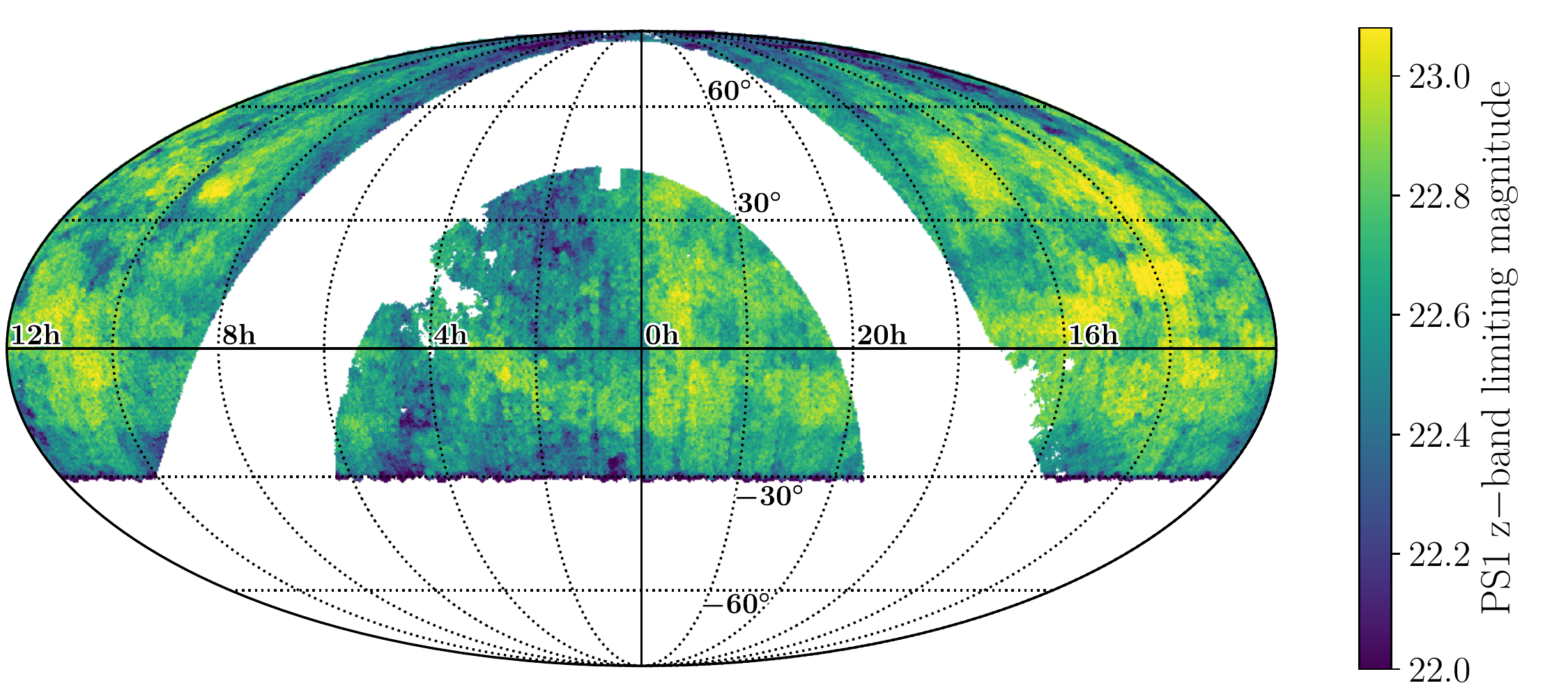}
    \caption{Mollweide projection map of the \textit{z}--band $3\sigma$ limiting magnitude per pixel of the PS1 PV2 survey area (median value per pixel). The map is created from all PV2 sources that obey the quality flags and morphology selection criteria using a HEALPix tesselation with 786,432 pixels (Lvl=8) over the entire sky. The color range extends from the 1st percentile to the 99th percentile value of all HEALPix pixels, highlighting the strong inhomogeneities in the \textit{z}--band across the survey footprint.}
    \label{fig:zbanddepth}
\end{figure*}


\subsubsection{Modeling quasar photometry with \texttt{simqso}} \label{sec:simqso}
To build a sample of simulated quasar photometry we are using a forked version of the python package \texttt{simqso}\footnote{\url{https://github.com/jtschindler/simqso}} \citep{McGreer2021}, originally presented  in \citet{McGreer2013, McGreer2018}, which contains updates to the emission line defaults and includes photometric error models for additional surveys. The code constructs artificial quasar spectra from a parametric model of different type-1 quasar spectral components. The parameters are informed from the general knowledge of quasar spectra at all redshifts and follow the assumption that quasar spectral energy distributions do not evolve with redshift \citep[e.g.,][]{Jiang2006, Shen2019a, YangJinyi2021}.
The quasar model is built from a power-law continuum, quasar emission lines, an iron pseudo-continuum component, a dust component, and a component modeling absorption of neutral hydrogen in the intergalactic medium.
Using \texttt{simqso} we have chosen the parameters of the different components to produce a reliable high-redshift quasar model specifically developed for this project.

We construct the continuum from a set of broken power laws ($f_{\nu} = A \times \nu_{\alpha}$)
that are designed to reproduce the continuum emission of the \citet{Selsing2016} quasar template and the broad band photometry of known SDSS quasars \citep{Schneider2010, Paris2017}.
Break points and mean power law slopes are listed in Table\,\ref{tab:breakpoints_slopes}.
For each spectrum the mean power law slopes are sampled by drawing values from a Gaussian distribution around the mean slope ($\alpha$) with a dispersion of $\sigma = 0.3$.
\begin{table}[ht]
\footnotesize
\centering
\caption{Wavelength break points and slopes for the simulated quasar spectrum}
\label{tab:breakpoints_slopes}
 \begin{tabular}{cc}
  \tableline
  Break Points in \AA & Slope before Break Point ($\alpha_{\nu}$) \\
  \tableline
  \tableline
  1200 & -1.50 \\
  2850 & -0.50 \\
  3645 & -0.60 \\
  6800 & 0.00 \\
  9000 & 0.30 \\
  30,000 & 0.30 \\
  $>30,000$ & 2.00 \\
 \end{tabular}
\end{table}
We are modeling a total of 65 individual lines (e.g., \nv{} ), line complexes (e.g., \siiv{}+\oiv{}), and individual iron (\feii{} and \feiii{}) multiplets. The majority of these lines are modeled with a single Gaussian and only five emission lines (Ly$\alpha$, \halpha{}, \civ{}, \ciii{}, \mgii{}) are constructed from two Gaussians, a narrow and a broad component.
%
%
Correlations between the equivalent width of some emission lines and the quasar luminosity, i.e., the Baldwin effect \citep{Baldwin1977}, are taken into account.
To approximate the emission from the large range of iron transitions seen in quasar spectra we included a composite of iron templates. This composite was last updated in \texttt{simqso} by \citet{Yang2016} and uses the \citet{Vestergaard2001} iron template at $1250\,\text{\AA--}2200\,\text{\AA}$, the \citet{Tsuzuki2006} iron template at $2200\,\text{\AA--}3500\,\text{\AA}$, and the \citet{Boroson1992a} iron template at $3500\,\text{\AA--}7500\,\text{\AA}$.
Following \citet{LyuJianwei2017b}, we add three blackbody components  with temperatures of $1800\,\text{K},\ 880\,\text{K}$, and $285\,\text{K}$ to model dust emission.
We have adjusted the equivalent width scaling of some emission lines, the amplitude scaling of some regions of the iron template, and the amplitudes of the blackbody dust emission components in order to fully reproduce the \citet{Selsing2016} quasar template and the optical to infrared mean quasar colors from the combined sample of the SDSS DR7 \citep{Schneider2010} and DR12 \citep{Paris2017} quasar catalogs.
%
To model absorption due to neutral hydrogen in the intergalactic medium, we use the Lyman-$\alpha$ forest model of \citet{McGreer2013}. We use this model to construct 10,000 different quasar sight lines up to $z=7$ that are randomly paired with simulated quasars to produce the absorption signatures blueward of the quasar's Lyman--$\alpha$ line.

The simulation that we use as a basis for the K-correction to absolute magnitudes and the selection function analysis consists of a uniform grid in absolute magnitude at 1450\,\AA, $-30 \le M_{1450} \le -23$, and redshift, $5.0 \le z \le 7.0$, with 56 intervals along the axis of absolute magnitude and 40 intervals along the redshift axis. Each cell is uniformly populated with 88 quasars for a total of 197,120 quasars.
The resulting quasar spectra are then multiplied with the PS1 filter bands to produce synthetic magnitudes.
\begin{figure*}[ht!]
    \centering
    \includegraphics[width=0.48\textwidth]{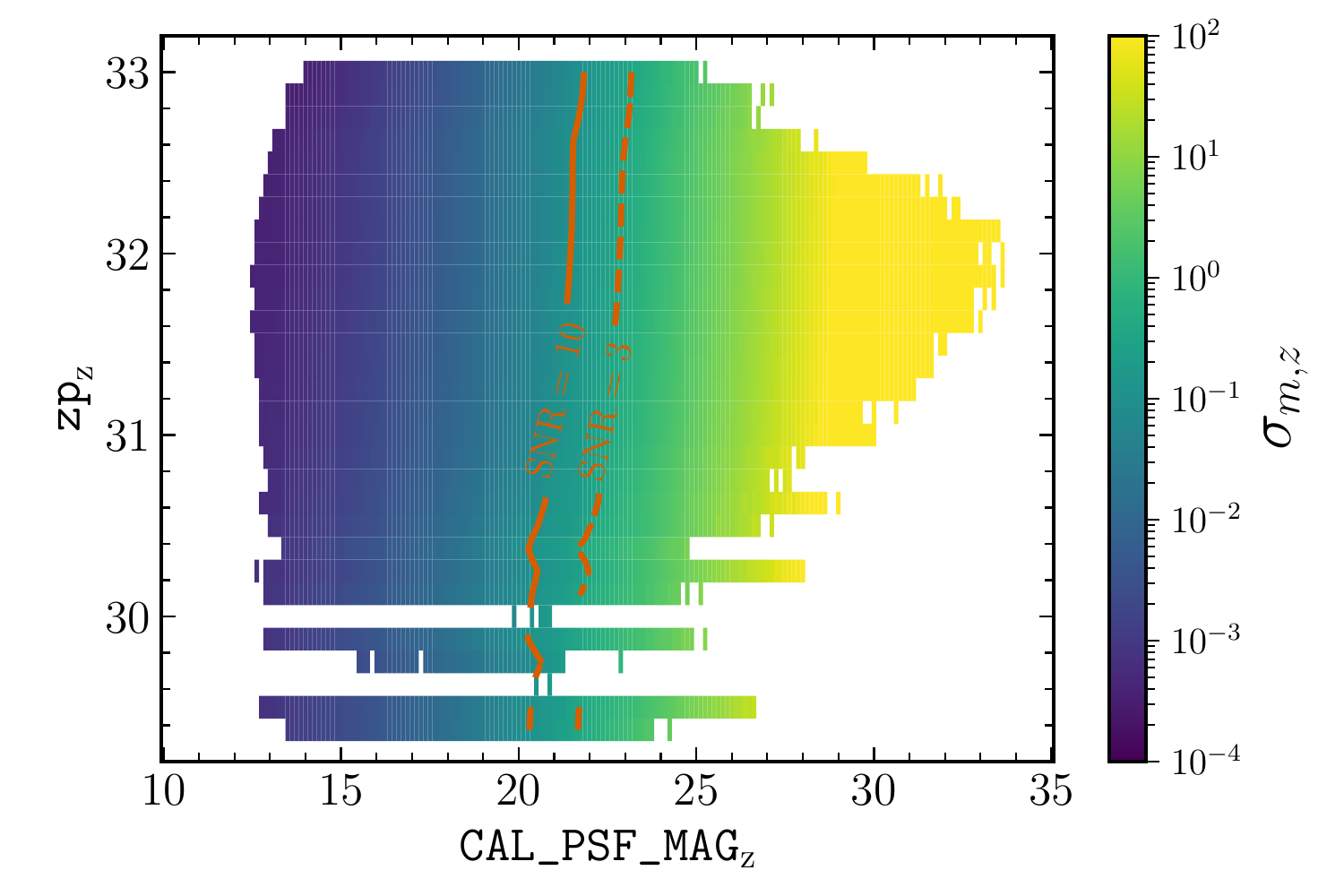}
    \includegraphics[width=0.48\textwidth]{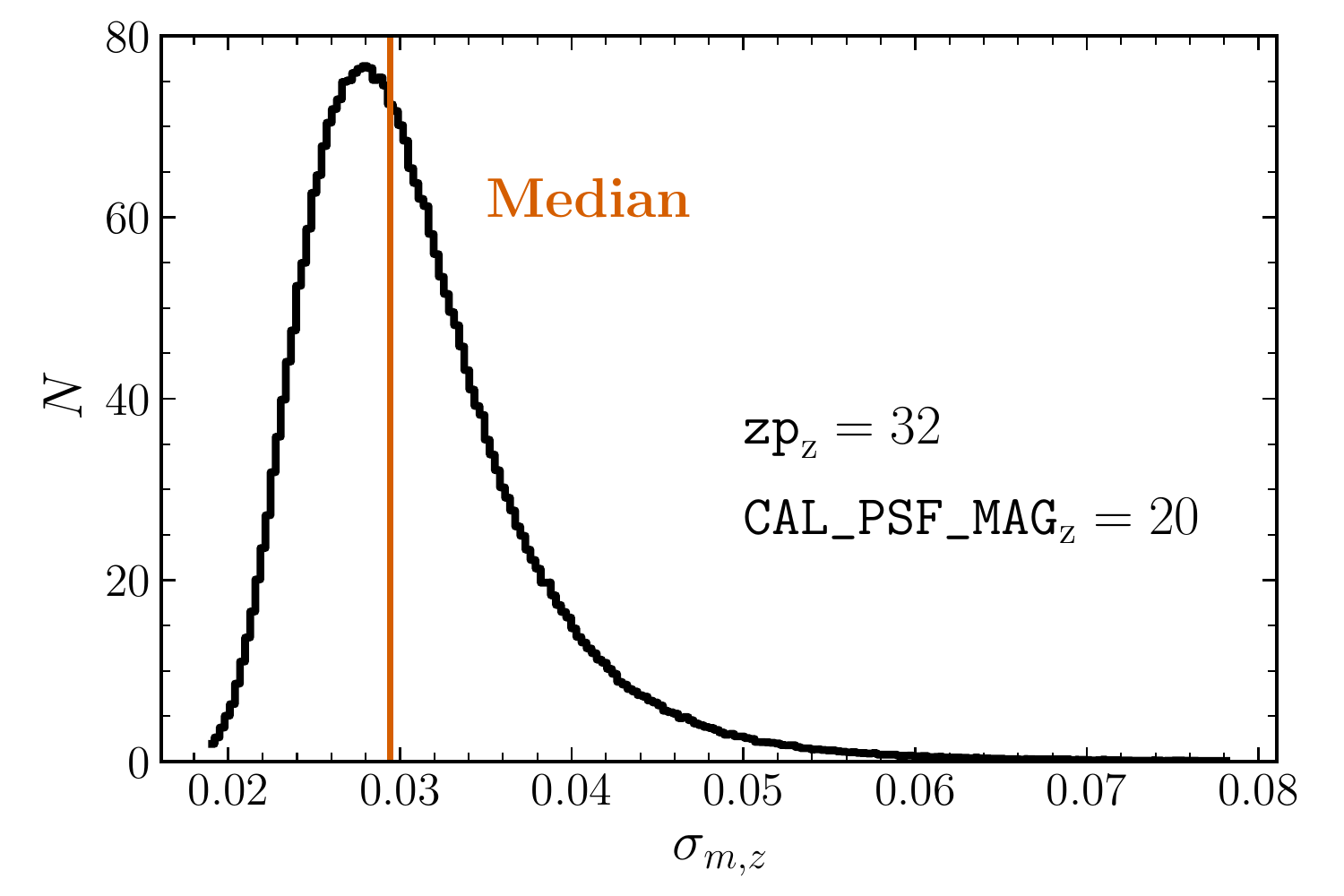}
    \caption{\textit{Left:} 2D histogram showing the median stacked PSF magnitude error in the \textit{z}--band $\sigma_{m,z}$ per bin as a function of \textit{z}--band zero point \texttt{zp}$_{z}$ and stacked PSF magnitude \texttt{CAL\_PSF\_MAG}$_{z}$. Lines showing the constant PSF magnitude signal--to--noise ratio at $\text{SNR}=10$ and $\text{SNR}=3$ highlight that  $\sigma_{m,z}$ is dependent on both \texttt{CAL\_PSF\_MAG}$_{z}$ and \texttt{zp}$_{z}$.
    \textit{Right:} Histogram of the PSF magnitude error in the \textit{z}--band $\sigma_{m,z}$ within the 2D bin centered on \texttt{zp}$_{z}=32$ and \texttt{CAL\_PSF\_MAG}$_{z}=20$. Even though the 2D histogram (left) shows a clear relation of the median $\sigma_{m,z}$ with both \texttt{zp}$_{z}$ and \texttt{CAL\_PSF\_MAG}$_{z}$, the individual $\sigma_{m,z}$ values within a bin show a significant dispersion around the median.}
    \label{fig:magerrmap}
\end{figure*}
%


%
Recent studies \citep{YangJinyi2021, Bischetti2022} report an increased BAL quasar fraction at $z \gtrsim 6$ compared to lower redshifts.
Similar to previous completeness calculations for $z>6$ QLF measurements \citep{Jiang2016, WangFeige2019a, Matsuoka2018c}, our quasar model does not take BAL quasars into account.
If our quasar selection would be biased against BAL quasars, it would be necessary to account for their different spectral properties in the completeness calculation.
Around $40\%$ (6) of the 14 BAL quasars in the \citet{Bischetti2022} sample have weak absorption (balnicity index $\rm{BI}<1000\,\rm{km}\rm{s}^{-1}$), which does not affect the quasar colors beyond the scatter included in our quasar model.
On the other hand, strong broad
\civ{} and \siiv{} absorption troughs ($\rm{BI}>5000\,\rm{km}\rm{s}^{-1}$) will reduce the flux between the \civ{} and the Lyman-$\alpha$ line significantly. At $z>5.8$ the \siiv{} line will be in the $y_{\rm{PS1}}$-band and the \civ{} line will lie redward of that. Hence, the strongest BAL features will diminish the $y_{\rm{PS1}}$-band flux, resulting in a bluer $z_{\rm{PS1}}-y_{\rm{PS1}}$ color. As a consequence, our color selection requiring $i_{\rm{PS1}}-z_{\rm{PS1}}>2.0$ and $z_{\rm{PS1}}-y_{\rm{PS1}}<0.5$ also includes strong BAL quasars naturally without the need to add them to our quasar model.
This inherently high completeness to BAL quasars in our color selection may play a role in the higher BAL fractions discovered at $z>6$ \citep{YangJinyi2021, Bischetti2022}.

There is evidence that quasars with weak emission lines (or weak-line quasars; WLQ) could make up $\sim10\%$ of the population at $z\gtrsim 6$ \citep[e.g., ][]{Banados2016,Shen2019a}, which is higher than what is observed at lower redshifts \citep{DiamondStanic2009}. The weaker Ly$\alpha$ and \nv{} emission lines can result in bluer $i_{\rm{PS1}}-z_{\rm{PS1}}$ color and a redder $z_{\rm{PS1}}-y_{\rm{PS1}}$ color for $z>5.6$ quasars. In the recent literature on $z>6$ QLFs \citep[e.g.,][]{Jiang2016, WangFeige2019a, Matsuoka2018c} a larger fraction of WLQs has not been taken into account, except by \citet{Matsuoka2018c} who down-scaled the Ly$\alpha$ line of their model spectra to be more inclusive of quasars with weaker emission lines.
%
%

The spread in quasar spectral properties covered within the random sampling of \texttt{simqso} already covers a significant fraction of emission line variation. Nonetheless, quasars with very weak lines \citep[e.g., the blue line in Fig.\,10 of][]{Banados2016} are  underrepresented in our quasar model (and in all models used for previous QLFs) and may be missed at $z>6.0$ due to their redder $z_{\rm{PS1}}-y_{\rm{PS1}}$ (see Figure\,\ref{fig:qsogen_simqso} left). This effect introduces a small systematic uncertainty of $<10\%$ to our QLF analysis, implying that we might be missing $<10\%$ of quasars at $6.0\lesssim z\lesssim6.2$ due to their weak emission lines.

One of the main assumptions for our determination of the selection function is the \texttt{simqso} quasar model itself. To test the robustness of this model we compare it to the quasar model presented by \citet[][hereafter \texttt{qsogen}]{Temple2021}. \cite{EuclidPrepV2019}  used \texttt{qsogen} to predict the quasar yields in the Euclid survey.  Both \texttt{simqso} and \texttt{qsogen} are designed to reproduce the median colors of quasars at $z<5$.
The $z_{\rm{P1}}-y_{\rm{P1}}$ quasar color of both our \texttt{simqso} model and \texttt{qsogen} agrees well with each other. The random sampling of continuum and emission line properties in the \texttt{simqso} model fully encompasses the variation in $z_{\rm{P1}}-y_{\rm{P1}}$ color introduced by the spread in emission line properties parametrized by the \texttt{qsogen} \texttt{emline\_type} parameter.
The main difference is the treatment of absorption due to the neutral intergalactic medium, which leads to small differences in the $z_{\rm{P1}}-y_{\rm{P1}}$ color at $z>5.8$ and larger differences in the $i_{\rm{P1}}-z_{\rm{P1}}$ color over $z=5.2-6.6$. In comparison to PS1 quasar colors the \texttt{simqso} model colors provide a more faithful representation of the empirical data points, whereas the default \texttt{qsogen} model results in a bluer $i_{\rm{P1}}-z_{\rm{P1}}$ color.
We refer interested readers to a more detailed discussion in Appendix\,\ref{app:qsomodel}.

\subsubsection{Simulating PS1 photometric uncertainties}\label{sec:modelps1phot}

The \texttt{simqso} package allowed us to generate synthetic quasar photometry. However, to fully assess the selection function, we need to take the photometric errors into account. For surveys with an approximately homogeneous depth one can derive simple magnitude error relations to sample the photometric error from.
However, as shown in Figure\,\ref{fig:zbanddepth} the PV2 photometric z--band 3$\sigma$ limiting magnitude, (Equation\,\ref{eq:mlim}), has strong variations depending on the sky position. This is also true for all other PS1 bands (e.g., see Figure\,\ref{fig:ybanddepth} in Appendix\,\ref{app:suppfigures}) and the depth variations do not necessarily correlate with each other across the different filter bands.
As a consequence we cannot define a single magnitude error relation to apply to all simulated quasar photometry.
Instead we adopt a sampling approach where we randomly associate observed catalog properties with simulated quasars that allow us to calculate the error properties necessary to evaluate the photometric selection criteria on our simulated sample.
Our method is similar in spirit to the approach discussed in Section\,3.1 of \citet{Yang2016}, which was also adopted for PS1 in \citet{WangFeige2019}.

The methodology allows us to calculate the stacked PSF magnitude uncertainty ($\sigma_{m}$), which also provides us with the signal-to-noise ratio (see Equation\,\ref{eq:snr}) and the $3\sigma$ limiting magnitude ($m_{\text{lim}}$) for each simulated quasar. Using $\sigma_{m}$ we can then construct PSF magnitudes from the synthetic photometry and evaluate the photometric selection function from the simulated quasar sample, including photometric uncertainties.

We begin with the PV2 catalog after we applied the quality flags, the extinction selection and the morphology selection. This guarantees that we are using properties of point sources most similar to our quasar candidates that also have good quality photometry within our chosen footprint.
In order to sample these properties we use the Hierarchical Equal Area isoLatitute Pixelation \citep[\texttt{HEALPix}\footnote{\url{http://HEALPix.sf.net/}};][]{Gorski2005} as implemented in the \texttt{healpy} \citep{Zonca2019} python package to associate each source in our clean catalog with a small area on the sky, i.e., a HEALPix cell.
%
For this purpose we subdivided the sky into a map of 196,608 (Lvl=7) cells, of which 100,766 are filled with at least one source. We then sample source positions as uniformly as possible across the survey footprint and retrieve their filter band zero points and their Galactic reddening value, $E(B-V)$, from the \citet{Schlegel1998} dust map for each of the 197,120 simulated quasars.
Each of the simulated quasars is then randomly associated with the noise properties of a real source in the catalog.

We have used the full cleaned PV2 catalog to investigate the relations between $\sigma_{m}$, the \texttt{zp}, and the stacked PSF magnitude (\texttt{CAL\_PSF\_MAG}).
Figure\,\ref{fig:magerrmap} (left) shows a 2D histogram of the median $\sigma_{m,z}$ as a function of \texttt{zp}$_{z}$ and \texttt{CAL\_PSF\_MAG}$_{z}$.
The figure shows the expected dependence of $\sigma_{m,z}$ on \texttt{CAL\_PSF\_MAG}$_{z}$ and \texttt{zp}$_{z}$, which is further highlighted by the slightly diagonal lines of constant SNR.
We note that within each of the 2D bins a range of $\sigma_{m,z}$ values exists as depicted in Figure\,\ref{fig:magerrmap} (right).
In order to approximate the stacked PSF magnitude error for a simulated quasar we first use the sampled $E(B-V)$ to redden the synthetic magnitude. The reddened magnitude and the sampled \texttt{zp} then determine the bin in the $\sigma_{m}$(\texttt{zp}, \texttt{CAL\_PSF\_MAG}) 2D histogram (e.g., Figure\,\ref{fig:magerrmap}, left) from which we randomly draw a $\sigma_{m}$ value given the 1D $\sigma_{m}$ distribution in that particular bin (e.g., Figure\,\ref{fig:magerrmap}, right). This is done for all PS1 filter bands. In rare instances where the combination of the sampled \texttt{zp} and simulated \texttt{CAL\_PSF\_MAG} does not exist in the cleaned PV2 catalog, we adopt the maximum value of the median $\sigma_{m}$ in the 2D histogram. This can be the case for synthetic magnitudes in dropout bands (e.g. the PS1 g-band) that exceed the observed range of values. From $\sigma_{m}$ we calculate the SNR for each given band using Equation\,\ref{eq:snr}.
Once we have associated a $\sigma_{m}$ for all synthetic quasars and all bands, we perturb the reddened synthetic magnitude by drawing from a Gaussian distribution with the magnitude as the mean and the associated $\sigma_{m}$ as the $1\sigma$ uncertainty. In a last step we subtract the reddening from the perturbed magnitudes to retrieve the appropriate dereddened magnitudes used for the photometric selection function evaluation.

\begin{figure}
    \centering
    \includegraphics[width=0.48\textwidth]{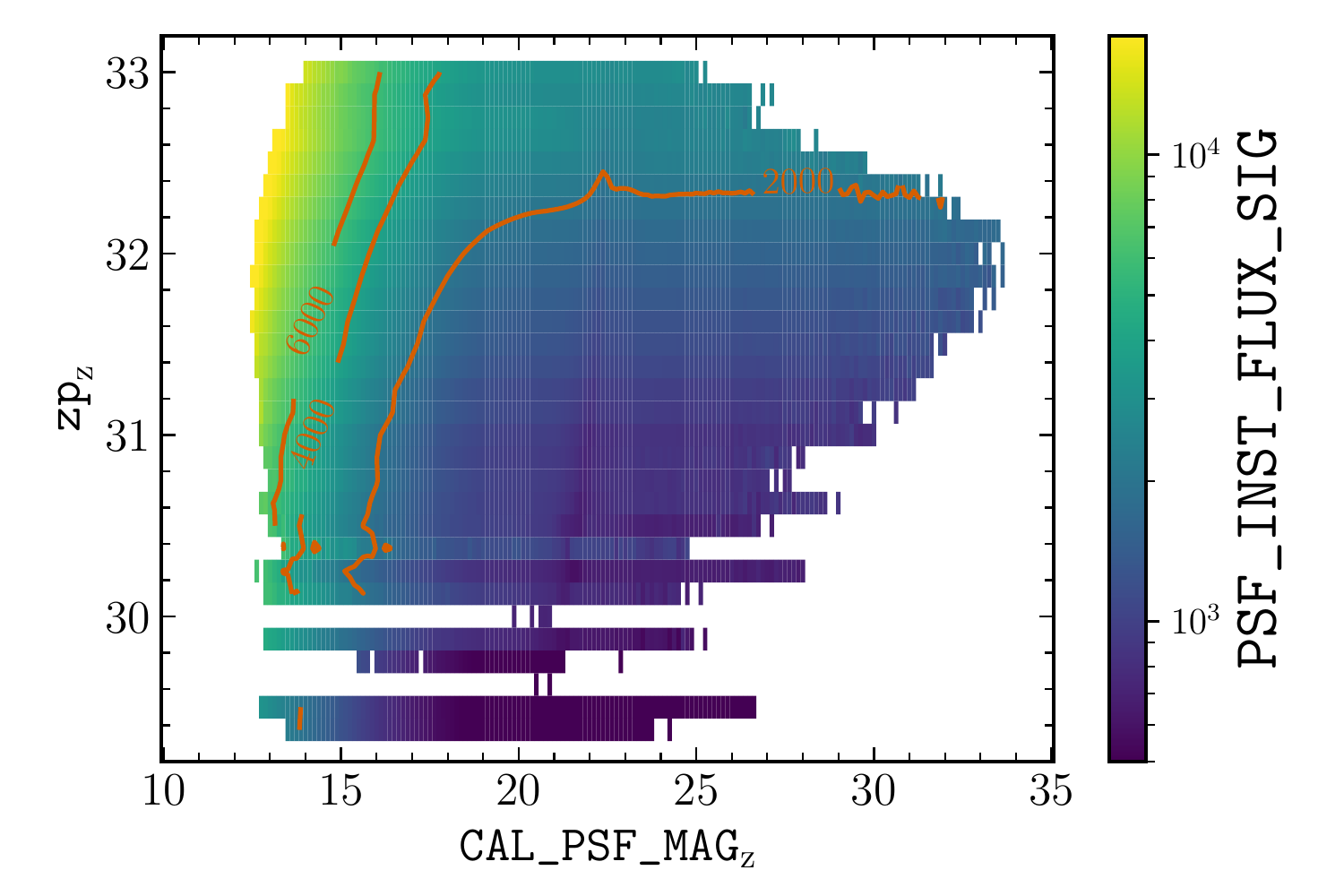}
    \caption{2D histogram showing the stacked PSF fit instrumental flux uncertainty \texttt{PSF\_INST\_FLUX\_SIG}$_{z}$ per bin as a function of \textit{z}--band zero point \texttt{zp}$_{z}$ and stacked PSF magnitude \texttt{CAL\_PSF\_MAG}$_{z}$.
    Dark red lines show values of constant \texttt{PSF\_INST\_FLUX\_SIG}$_{z}$ (2000, 4000, 6000).
    Similar to the magnitude error in the \textit{z}--band $\sigma_{m,z}$ (Figure\,\ref{fig:magerrmap}), \texttt{PSF\_INST\_FLUX\_SIG}$_{z}$ is dependent on both \texttt{zp}$_{z}$ and \texttt{CAL\_PSF\_MAG}$_{z}$.}
    \label{fig:psfinstfluxsigmap}
\end{figure}

In order to assess the $3\sigma$ limiting magnitude $m_{\text{lim}}$ (see Equation\,\ref{eq:mlim}) we not only require knowledge of \texttt{zp}, but also of the stacked PSF fit instrumental flux uncertainty, \texttt{PSF\_INST\_FLUX\_SIG}, at the source position.
Similar to $\sigma_{m}$, \texttt{PSF\_INST\_FLUX\_SIG} can also be mapped as a function of \texttt{zp} and \texttt{CAL\_PSF\_MAG} as shown in Figure\,\ref{fig:psfinstfluxsigmap}. The figure shows the median z-band \texttt{PSF\_INST\_FLUX\_SIG} per bin and depicts that it is also dependent on both \texttt{zp} and \texttt{CAL\_PSF\_MAG}.
In order to associate a \texttt{PSF\_INST\_FLUX\_SIG} value with a simulated quasar we proceed just as we have done for the magnitude error above. Then we calculate the limiting magnitudes for all simulated quasars according to Equation\,\ref{eq:mlim}.


\subsection{K-correction and absolute magnitudes}\label{sec:kcorrection}

\begin{figure}
    \centering
    \includegraphics[width=0.48\textwidth]{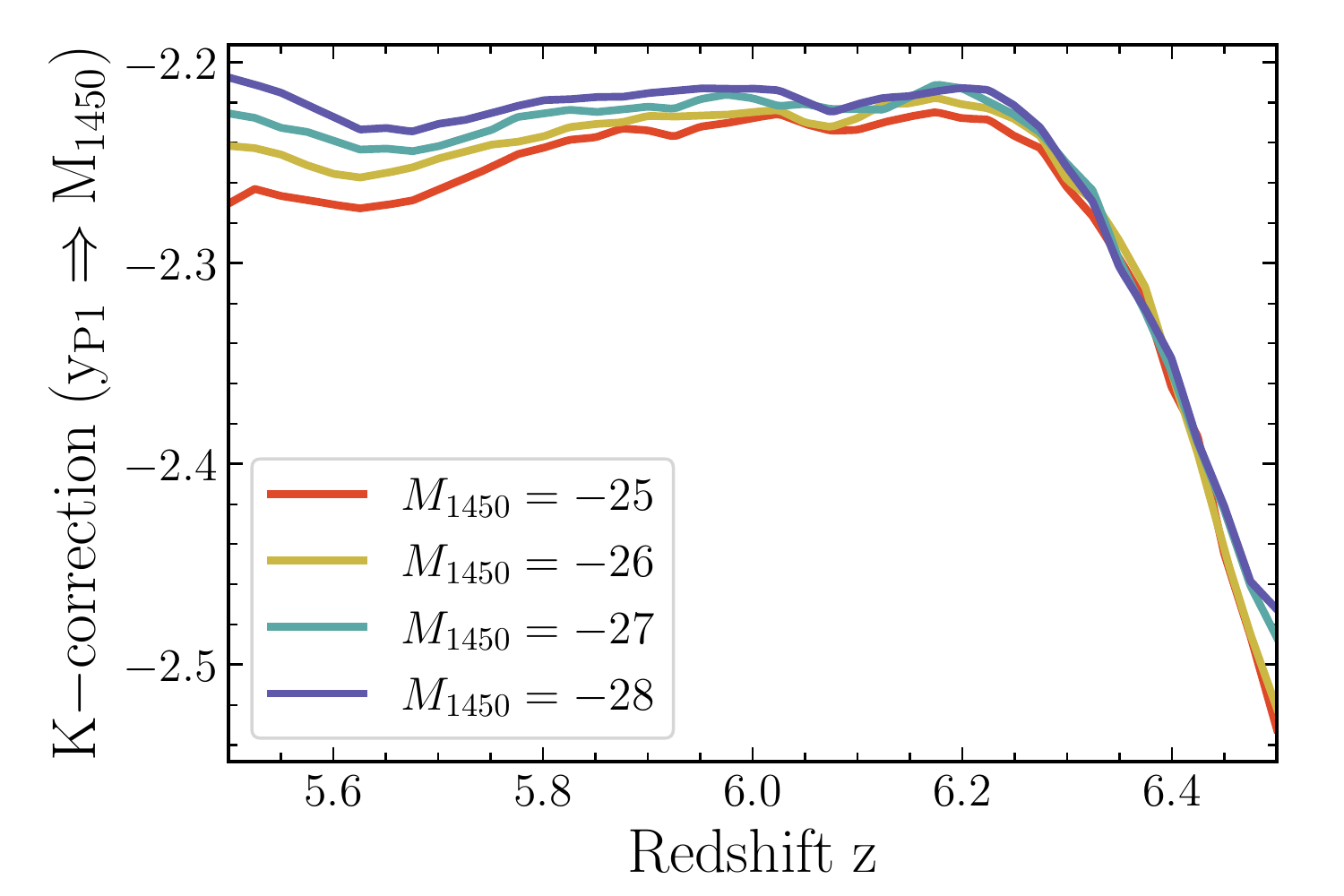}
    \caption{Luminosity dependent K-correction factor derived from the simulated quasar sample to correct the observed dereddened PS1 stacked PSF \textit{y}--band magnitude, $y_{\text{P1}}$, to obtain the absolute monochromatic magnitude measured at rest-frame $1450\,\text{\AA}$, $M_{1450}$. The colored lines show the K--correction factor for different luminosities as a function of redshift. The luminosity dependence is most pronounced at $z\approx5.5-5.8$, where the \civ{} emission line falls into the PS1 \textit{y}-band.}
    \label{fig:kcorr}
\end{figure}

In order to evaluate the QLF as a function of absolute monochromatic magnitude measured at rest-frame $1450\,\text{\AA}$ ($M_{1450}$), we need to be able to calculate $M_{1450}$ from the observed dereddened PS1 stacked PSF magnitudes for our quasar sample.
The conversion to rest-frame $M_{1450}$ requires a term that accounts for the changing filter response as the quasar spectrum is being redshifted, the K-correction.
We compute a K-correction based on our simulated quasar photometry as a function of $M_{1450}$ and quasar redshift. As the PS1 \textit{z}--band cuts out at around $9300\,\text{\AA}$, the rest--frame wavelength of $1450\,\text{\AA}$ falls into the PS1 \textit{y}--band above $z\approx5.4$. Therefore, we derive relations between $M_{1450}$, $z$, and the dereddened \textit{y}--band magnitude $y_{\text{P1}}$ from our grid of simulated quasars. We then retrieve the $y_{\text{P1}}$ K--correction factor by interpolation from this relation.
Figure\,\ref{fig:kcorr} shows the luminosity dependent K-correction factor as a function of redshift for different quasar luminosities. The median absolute deviation (MAD) of the K--correction factor over the entire grid, a measure of its dispersion due to variety in different quasar spectra simulated by the code, is $K_{\text{MAD}}=0.07\,\text{mag}$.
We use this K--correction to calculate $M_{1450}$ from $y_{\text{P1}}$ for all quasars in our QLF sample. The $M_{1450}$ values for each individual quasar are listed in Table\,\ref{tab:qsos}.
%

%
Ideally, one would like to measure the flux at $1450\,\text{\AA}$ directly from the quasar spectra to derive $M_{1450}$. Unfortunately, not all quasars in our sample have the necessary spectral coverage and therefore we conclude that our photometry--based $M_{1450}$ determination provides a more general approach and allows us to treat all quasars in our sample uniformly.
%
%


\subsection{The selection function}\label{sec:selectionfunction}
In order to understand the completeness of our quasar selection as a function of absolute magnitude $M_{1450}$ and redshift $z$ we need to quantify the selection function following the strategy laid out in Section\,\ref{sec:qso_selection}.
The full survey selection function $S(\vect{q})$ can be written as the product of the selection functions for all independent selection steps. The parameter vector $\vect{q}$ contains all the catalog properties, which are used in the different selections.
\begin{equation}
\begin{split}
S(\vect{q}) = & S_{\Omega}(\vect{q}) \times  S_{\text{morph}}(\vect{q}) \times S_{\text{phot}}(\vect{q}) \times  \\ & S_{\text{qf}}(\vect{q}) \times S_{\text{ID}}(\vect{q})
\end{split} \label{eq:surveyselectionfunction}
\end{equation}
We have separated
the selection function for the extinction cut and area exclusion $S_{\Omega}$,
the selection function based on morphology $S_{\text{morph}}$ (Equation\,\ref{eq:morph_criterion}),
the photometric selection $S_{\text{phot}}$ (Equations\,\ref{eq:phot_criterion1}--\ref{eq:phot_criterion5}),
the quality flag criteria $S_{\text{qf}}$,
and the selection introduced by our incomplete follow--up $S_{\text{ID}}$. Before we present the full survey selection function, we will discuss their individual contributions.


\subsubsection{The survey area estimate and Galactic extinction}
The Galactic extinction $E(B-V)(\alpha, \delta)$ is a function of right ascension $\alpha$ and declination $\delta$ of the sources. We apply a selection criterion, Equation\,\ref{eq:ext_criterion}, to construct our sample, which effectively excludes sky regions with high Galactic extinction. Thus, the extinction selection function only reduces the available survey area irrespective of redshift and luminosity of the sources:
\begin{gather}
    S_{\Omega,\text{ext}}(E(B-V), \alpha, \delta) = \begin{cases}
      1 & \forall\ E(B-V)(\alpha, \delta)< 0.3\\
      0 & \forall\ E(B-V)(\alpha, \delta) \ge 0.3 \
    \end{cases}
\end{gather}
We already noted above (Section\,\ref{sec:qso_selection}) that in addition to the extinction cut, regions close to the Galactic plane ($|b|<20\,\text{deg}$),
\begin{gather}
    S_{\Omega,\text{gal}}(b) = \begin{cases}
      1 & |b|<20\,\text{deg} \\
           0 & \text{otherwise} \ ,
    \end{cases}
\end{gather}
as well as the area around M31,
\begin{gather}
    S_{\Omega,\text{M}31}(\alpha, \delta) = \begin{cases}
     0 & \forall\ 7^\circ < \alpha < 14^\circ\ \text{AND}\ 37^\circ < \delta < 43^\circ \\
      1 & \text{otherwise} \ ,
    \end{cases}
\end{gather}
are excluded for the final quasar selection as well.
The complete area selection function is then a product of the three criteria above:
\begin{equation}
    S_{\Omega} = S_{\Omega,\text{ext}} \times S_{\Omega,\text{M}31} \times S_{\Omega,\text{gal}} \ .
\end{equation}

To estimate the area included by our selection criteria we again utilize a \texttt{HEALPix} tesselation of the sky.
%
Our area estimate is based on 328,479,702 sources of the PS1 PV2 catalog, which passed the photometric quality criteria, the morphology selection and were included in the area defined by the extinction selection and the Galactic plane and M31 exclusion regions.
Using \texttt{HEALPix} we divide the sky into a grid of curvilinear, equal--sized quadrilaterals.
We calculate the number of HEALPix cells which are populated by at least one source and sum up the area of all of those cells for a total area estimate. We vary the HEALPix cell sizes to understand the resolution effects on the total estimated area.
At the lowest resolution, the sky is represented by 12 HEALpix cells. For the next resolution level these cells are each divided into four sub-cells. Hence, the total number of HEALpix cells $N_{\text{pix}}$ depends on the resolution level $\text{lvl}$ following $N_{\text{pix}}= 12\times4^{\text{lvl}}$.
At too low resolution levels the survey area is overestimated as the coarse cells cannot capture the fine structure of the Galactic extinction map. As we proceed to finer resolutions the number of HEALpix cells approaches the number of sources for the area estimate. At even higher resolution the size of the HEALpix cells gets smaller than the area between adjacent sources in the survey and we are effectively undersampling the area. As a result the survey area estimate decreases rapidly.
Table\,\ref{tab:surveyarea} shows the results from the HEALpix analysis. The columns are the HEALpix level $\text{Lvl}$, the total number of pixels $N_{\text{pix}}$ per level, the number of filled pixels $N_{\text{pix, filled}}$ per level, the area per pixel $\Omega_{\text{pix}}$ and the total filled area per level $\Omega_{\text{total}}$, the effective survey area.  The total area estimate drops significantly above $\text{lvl}=11$, indicating that the HEALPix cell density is approaching the source density and we are beginning to overresolve the area.
Thus, we adopt $\text{lvl}=10$ for our fiducial area estimate and use the differences to the adjacent levels to reflect the uncertainties on our estimate. This results in a survey area of $20803.38^{+61.75}_{-54.45}\,\text{deg}^2$.
This calculated survey area now implicitly takes the extinction selection function into account. We consider the relative uncertainty on the survey area of $\approx0.3\%$ to be negligible and it is not propagated further.
\begin{deluxetable}{crrcc}
\tabletypesize{\footnotesize} 
\tablecaption{Estimates of the total survey area $\Omega_{
m{total}}$\label{tab:surveyarea}}
\tablehead{\colhead{Lvl} &\colhead{$N_{\rm{pix}}$} &\colhead{$N_{\rm{pix,\ filled}}$} &\colhead{$\Omega_{\rm{pix}}$} &\colhead{$\Omega_{\rm{total}}$} \\ 
\nocolhead{} &\nocolhead{} &\nocolhead{} &\colhead{($\rm{deg}^2$)} &\colhead{($\rm{deg}^2$)} 
} 
\startdata 
$6$ & $49,152$ & $25,519$ & $0.83929$ & $21417.93$  \\ 
 $7$ & $196,608$ & $100,766$ & $0.20982$ & $21143.07$  \\ 
 $8$ & $786,432$ & $399,766$ & $0.05246$ & $20970.07$  \\ 
 $9$ & $3,145,728$ & $1,591,062$ & $0.01311$ & $20865.13$  \\ 
 $10$ & $12,582,912$ & $6,345,413$ & $0.00328$ & $20803.38$  \\ 
 $11$ & $50,331,648$ & $25,317,657$ & $0.00082$ & $20750.93$  \\ 
 $12$ & $201,326,592$ & $92,989,674$ & $0.00020$ & $19054.11$  \\ 
 $13$ & $805,306,368$ & $212,082,285$ & $0.00005$ & $10864.22$  \\ 
 $14$ & $3,221,225,472$ & $294,119,359$ & $0.00001$ & $3766.67$  \\ 
 \enddata 
\end{deluxetable} 

\subsubsection{Morphology selection function}
Submillimeter observations indicate that the host galaxies $z\gtrsim5.7$ quasars are often compact with effective (half-light) radii of $R_{e}\approx1.11\,\text{kpc}$ \citep{Neeleman2021}. They are effectively unresolved  by the PS1 photometry.
The selection criterion in Equation\,\ref{eq:morph_criterion} is designed to reject extended contaminants from the selection. The quasar completeness was quantified in \citet[][their Section 2.1 and their Figure 3]{Banados2016} to be 97\%.
%
With the significant detections we require in the PS1 \textit{z}-- and \textit{y}--band (Equations\,\ref{eq:phot_criterion1} and \ref{eq:phot_criterion2}), it is reasonable to assume that this value is independent of apparent magnitude and quasar redshift. The host galaxies of quasars become only prevalent at much fainter magnitudes \citep[][their Figure\,6]{Matsuoka2016} beyond the PS1 detection limit.
Therefore, we adopt the value of 97\% for the quasar selection completeness:
\begin{equation}
     S_{\text{morph}}(f_{\text{ext},z}, f_{\text{ext},y}) = 0.97\ .
\end{equation}

Excluding extended sources introduces a bias to our sample against strongly lensed quasars. Multiple images or the foreground lens galaxy could make the lensed quasar appear extended in the imaging.

\subsubsection{Photometric selection function}
\begin{figure}
    \centering
    \includegraphics[width=0.48\textwidth]{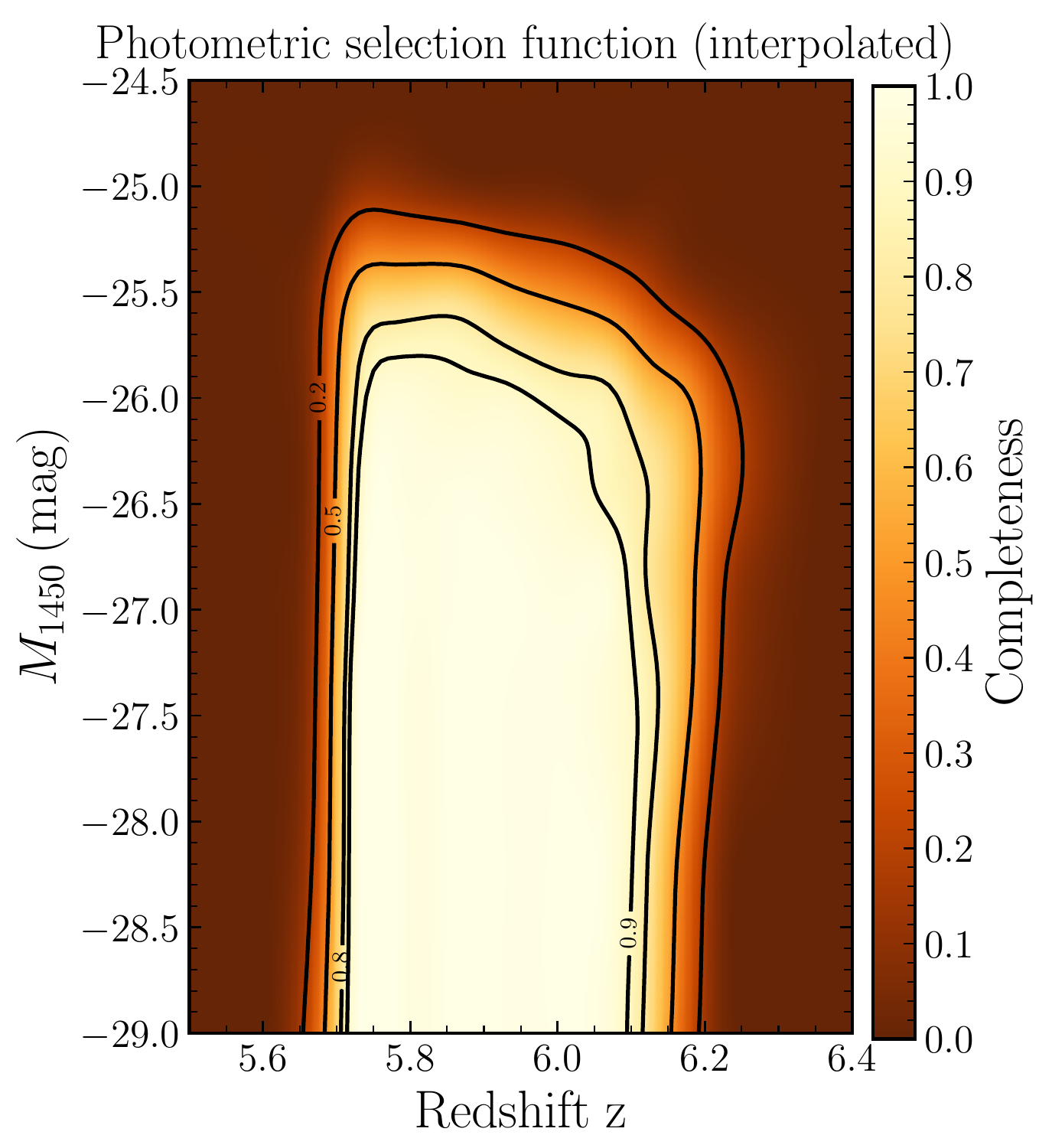}
    \caption{Interpolated photometric selection function evaluated on a redshift ($z$) and absolute magnitude ($M_{1450}$) grid of simulated quasars. Black contours are drawn at a completeness of 20\%, 50\%, 80\% and 90\%. The high completeness at $5.7 \lesssim z \lesssim 6.2$ directly reflects the color selection criteria listed in Section\,\ref{sec:qso_selection}. The faint end is limited by the depth of the PS1 survey and the signal--to--noise ratio criteria in different filter bands.
    }
    \label{fig:photo_selfun}
\end{figure}

The criteria described in Equations\,\ref{eq:phot_criterion1}--\ref{eq:phot_criterion5} form the core of the PS1 photometry $5.7 \lesssim z \lesssim 6.2$ quasar selection.
The photometric selection function is based on the magnitudes and SNRs in different PS1 filter bands and the limiting \textit{i}--band magnitude,
\begin{equation}
\begin{aligned}
S_{\text{phot}}(\vect{q}) = S_{\text{phot}}( & r, i, z, \text{SNR}_g, \text{SNR}_r, \\ & \text{SNR}_i, \text{SNR}_z, \text{SNR}_y,  i_{\text{lim, dr}}) \ .
\end{aligned}
\end{equation}

We apply the selection criteria to our grid of simulated synthetic quasar photometry (Sections\,\ref{sec:simqso} and \ref{sec:modelps1phot}) to evaluate its impact as a function of redshift and absolute magnitude.
We present the resulting photometric selection function in Figure\,\ref{fig:photo_selfun}.
High redshift quasars are commonly selected by the strong flux break at the Lyman$\alpha$ line, where blueward emission is absorbed by neutral gas in the IGM.
The ${i_{\text{P1}}}--{z_{\text{P1}}}$ color criterion (Equation\,\ref{eq:phot_criterion3})  selects quasars with a Lyman$\alpha$ break above redshifts of $z\approx5.6$ and is responsible for the rise of the selection function at this redshift.
The ${z_{\text{P1}}}-{y_{\text{P1}}}$ color criterion (Equation\,\ref{eq:phot_criterion3}) imposes a certain level of continuity between the two filter bands and thus actively deselects quasars beyond $z\approx6.3$.
The width of the redshift transition regions depends on the diversity in quasar spectral properties and the variations of absorbing neutral hydrogen along the line of sight.
The signal-to-noise ratio criteria (Equations\,\ref{eq:phot_criterion1} and \ref{eq:phot_criterion2}) limit the selection in apparent magnitude and results in a redshift dependent absolute magnitude limit. The inhomogeneity of the PS1 depth and the intrinsic scatter of the flux measurements result in a slow decrease with increasing absolute magnitude rather than a sharp break.

\begin{figure}[!ht]
    \centering
    \includegraphics[width=0.5\textwidth]{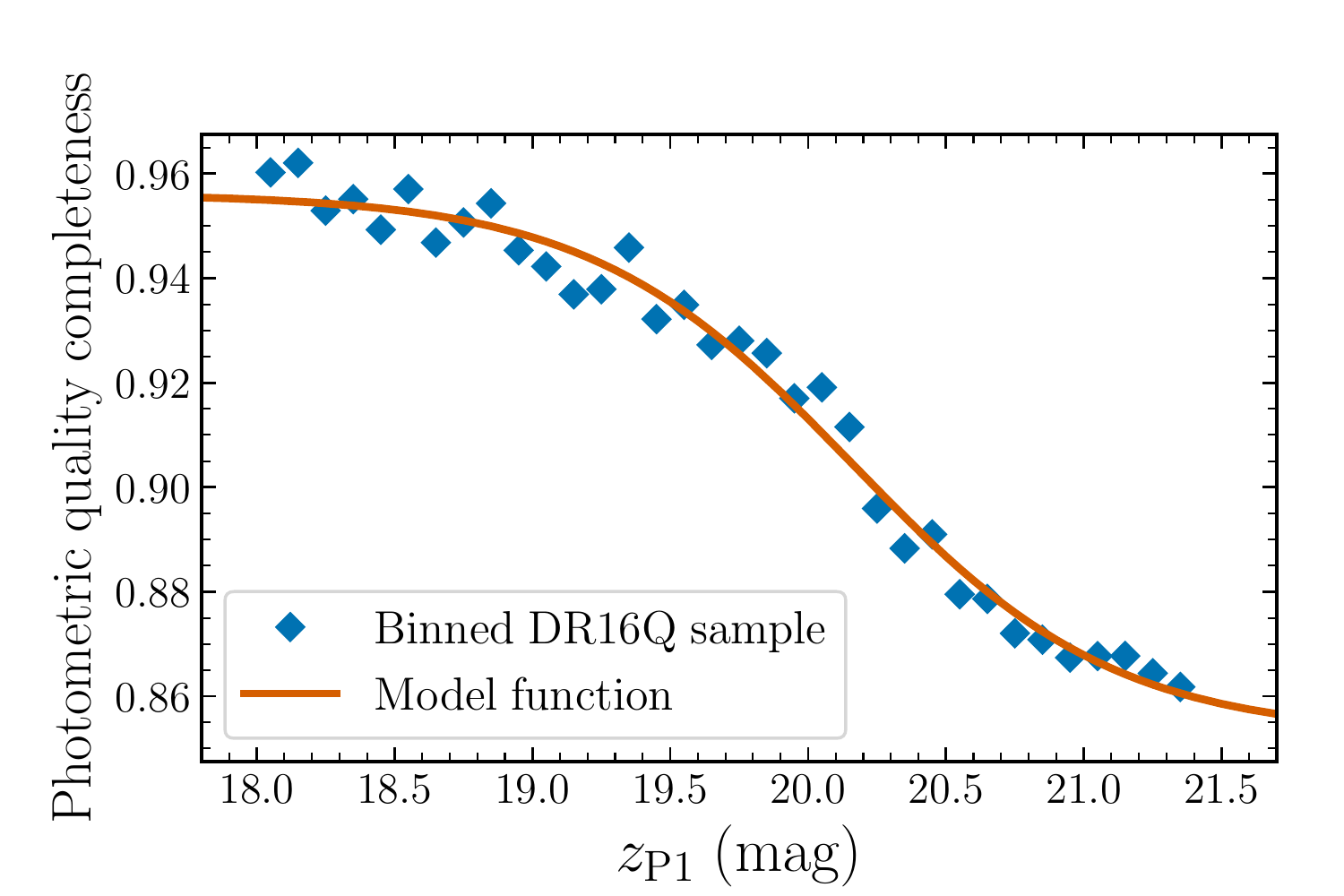}
    \caption{Completeness estimate of our photometric quality selection as a function of the dereddened PS1 \textit{z}--band stacked magnitude $z_{\text{P}1}$ (blue diamonds). The completeness is calculated in 34 magnitude bins ($z_{\text{P}1}{=}18{-}21.5$, $\Delta z_{\text{P}1}{=}0.1$) from a sample of 179,945 $z{=}2.0{-}5.0$ quasars selected from the SDSS DR16 quasar catalog. We approximate the completeness with an hyperbolic tangent function (solid orange line, Equation\,\ref{eq:quality_selfun}) to extrapolate the completeness beyond the measurement limits.
    }
    \label{fig:qf_completeness}
\end{figure}

\begin{figure*}[!ht]
    \centering
    \includegraphics[width=0.48\textwidth]{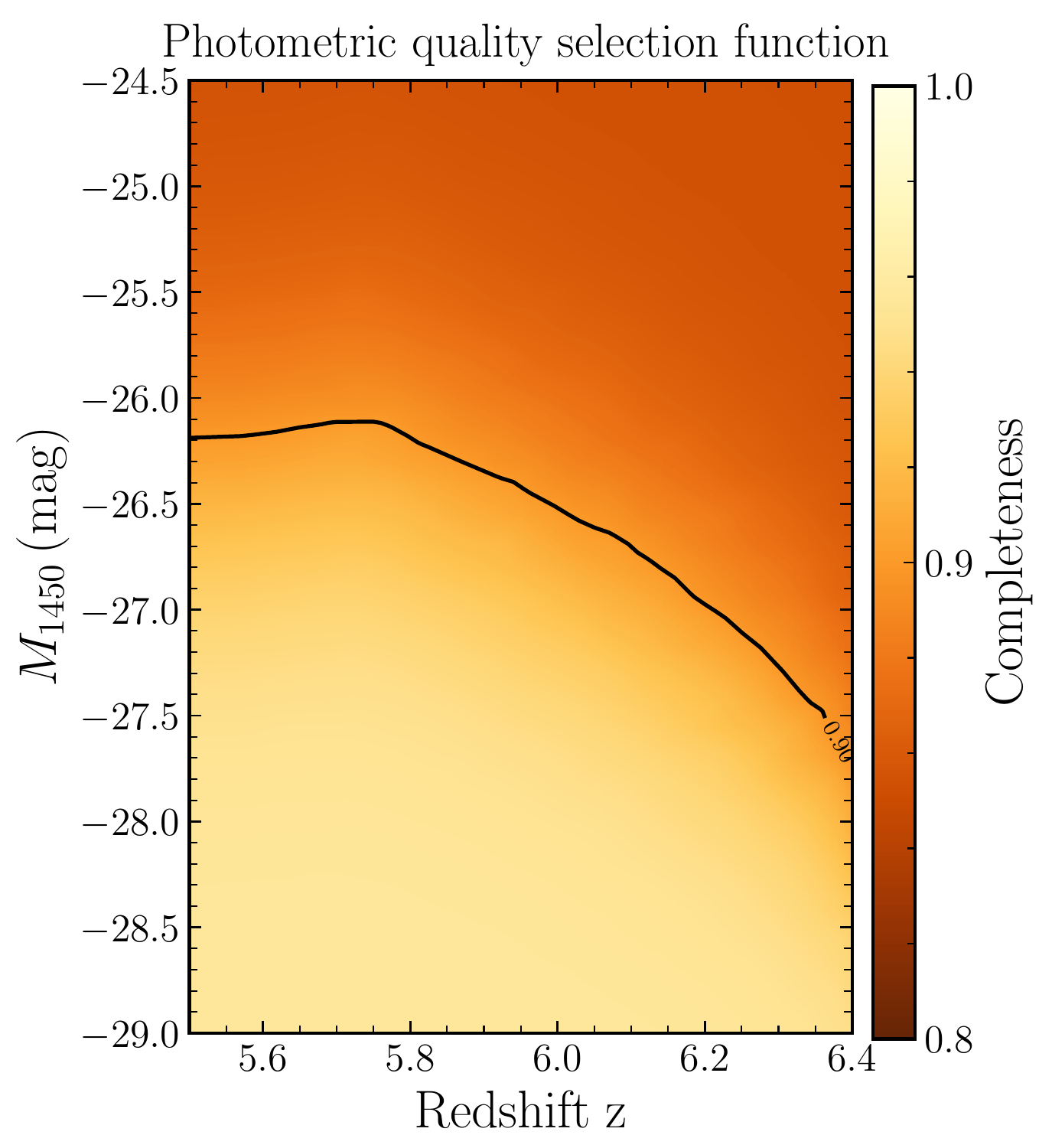}
    \includegraphics[width=0.48\textwidth]{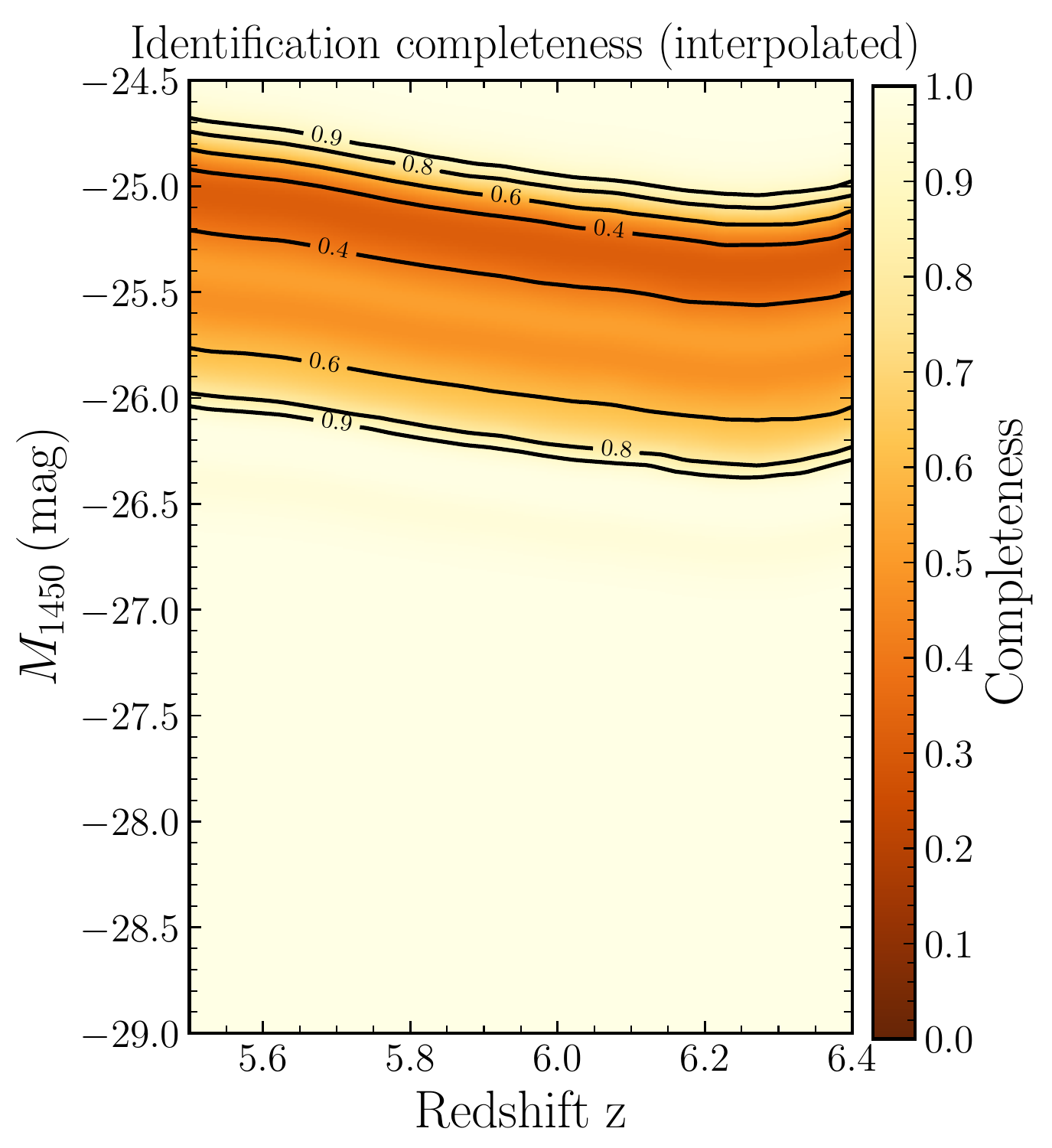}
    \caption{
    \textit{Left:} The photometric quality selection function $S(z_{\text{P1}}(M_{1450}, z)$ (Equation\,\ref{eq:quality_selfun}) as a function of redshift ($z$) and absolute magnitude ($M_{1450}$). We retain a relatively high completeness of $>85\%$ as already indicated by the 1D evaluation in Figure\,\ref{fig:qf_completeness}. In order to emphasize the gradients we have limited the color scaling of the completeness to the interval $[80\%, 100\%]$.
    \textit{Right:} The identification completeness as a function of redshift and absolute magnitude. The low completeness ($<50\%$) at $y_{\text{P}1}=20.5-21.5$ (see Figure\,\ref{fig:spec_completeness}) affects the absolute magnitude range between $M_{1450}\approx-24.9$ to $M_{1450}\approx-26.2$, depending on the redshift. Our identification campaigns are largely complete at the faint end (Figure\,\ref{fig:spec_completeness}), resulting in a rise of the completeness towards the faintest magnitudes. This rise is excluded in our ML fit, for which we use the luminosity range $M_{1450}=-25$ to $-29$.
    }
    \label{fig:quality_spec_selfun}
\end{figure*}

\subsubsection{Photometric quality selection}
We assess the selection function for these photometric quality criteria  (see Section\,\ref{sec:qso_selection}) using two empirical samples of quasars matched to the PS1 PV2 catalog.

As we want to address the completeness of our method, we first build a sample of $z\approx5.7$, which were discovered from other surveys but lie within the PS1 footprint.
After applying the signal-to-noise ratio requirements on the photometry ($\text{SNR}_{z_{\text{P1}}} > 10, \text{SNR}_{y_{\text{P1}}} > 5$) to ensure an appropriate comparison, we retained 76 sources. We applied the quality criteria and find 72 sources to be included in the photometric quality selection, resulting in a completeness of $\sim95\%$.

To verify the completeness with a larger sample we perform an additional test on quasars selected from the SDSS DR16 quasar catalog \citep{Lyke2020}. We require the SDSS quasars to have low extinction ($E(B-V)<0.3$) and follow our candidate signal-to-noise ratio requirements ($\text{SNR}_{z_{\text{P1}}} > 10, \text{SNR}_{y_{\text{P1}}} > 5$). Additionally, we only select sources within the redshift range of $2 < z < 5$. We exclude quasars at $z>5$ from the DR16 quasar catalog as the majority of these sources are objects misclassified by the automatic pipeline.
Catalog quasars at $z<2$ have been preemptively excluded to avoid complications that arise when the host galaxy starts to be resolved. At that point the values of \texttt{PSF\_QF} and the quality flags potentially differ significantly from pure point sources.
The full sample contains 179,945 SDSS quasars, of which 162,770 are retained when applying the photometric quality criteria, resulting in an average completeness of $\sim90\%$.

We further investigate whether the photometric quality selection function is dependent on the dereddened PS1 \textit{z}--band magnitude. We thus calculate the completeness in 34 magnitude bins between $z_{\text{P}1}=18$ to $21.5$ with a bin width of $\Delta z_{\text{P}1}=0.1$. The result shows a  dependence of the completeness with $z_{\text{P}1}$ (see Figure\,\ref{fig:qf_completeness}).
The quality selection function dependency on the \textit{y}--band magnitude reflects the correlation between some quality flags (e.g., \texttt{POORFIT, MOMENT\_SN}, see \citet{Banados2014}, Table\,6) and the lower SNR of fainter sources.
We model the binned measurements with a hyperbolic tangent:
\begin{equation}
    S(z_{\text{P}1}) = a \times \tanh{(\eta \times (z_{\text{P}1} + \phi))} + b \label{eq:quality_selfun}
\end{equation}
and use the \texttt{LMFIT} python package \citep{lmfit2014} to retrieve the best--fit parameters via the Levenberg--Marquardt algorithm. The values are $a =0.052\pm0.003$, $b=0.904\pm0.002$, $\eta=-1.049\pm0.127$, and $\phi=-20.161\pm0.063$. The best--fit model function is shown in Figure\,\ref{fig:qf_completeness} as the solid orange line.



We adopt this parameterization of the photometric quality completeness for the calculation of the full selection function.
For each point in the space of absolute magnitude $M_{1450}$ and redshift $z$, we evaluate the photometric quality selection $S(z_{\text{P1}})$ by calculating the apparent magnitude $z_{P1}(M_{1450}, z)$ at the point using the adopted cosmology and a $z_{P1}$ K-correction factor determined analogously to the $y_{P1}$ K-correction (Section\,\ref{sec:kcorrection}).
The photometric quality selection in the space of redshift and absolute magnitude is shown in the left panel of Figure\,\ref{fig:quality_spec_selfun}. The overall completeness is high, $>85\%$.



\subsubsection{Identification completeness} \label{sec:id_completeness}
We further have to take into account that not all of our \ncandgood{} quasar candidates have been identified through follow-up observations or the literature.
We note that spectroscopic follow-up observations were prioritized by the candidate's visual rank. However, no other selection relevant information e.g., the coordinate range, played a factor in this process.
To account for this we use the weighted identification completeness as a function of $y_{\text{P}1}$ as shown in the bottom panel of Figure\,\ref{fig:spec_completeness}.
%
%
%
Similar to the photometric quality completeness we evaluate the kernel density estimate of the identification completeness $\text{Comp}_{\text{ID}}(y_{\text{P1}})$ in the space of redshift ($z$) and absolute magnitude ($M_{1450}$) by mapping these properties to the dereddened PS1 \textit{y}--band magnitude $y_{\text{P1}}(M_{1450}, z)$ using the adopted cosmology and the $y_{P1}$ K--correction factor.
The right panel of Figure\,\ref{fig:quality_spec_selfun} shows the map of the spectroscopic completeness as a function of redshift and absolute magnitude.
The low identification completeness ($<80\%$)  at apparent magnitudes $y_{\text{P}1}=20.5-21.5$ of our follow-up campaigns (Figure\,\ref{fig:spec_completeness}) is reflected here at absolute magnitudes $M_{1450}\approx-24.9$ to $M_{1450}\approx-26.2$.
The completeness rises towards the faint end, equivalent to the behavior in Figure\,\ref{fig:spec_completeness}, as we have identified many of the faintest candidates.
%

\subsubsection{The full survey selection function}\label{sec:fullselectionfunction}
\begin{figure}
    \centering
    \includegraphics[width=0.5\textwidth]{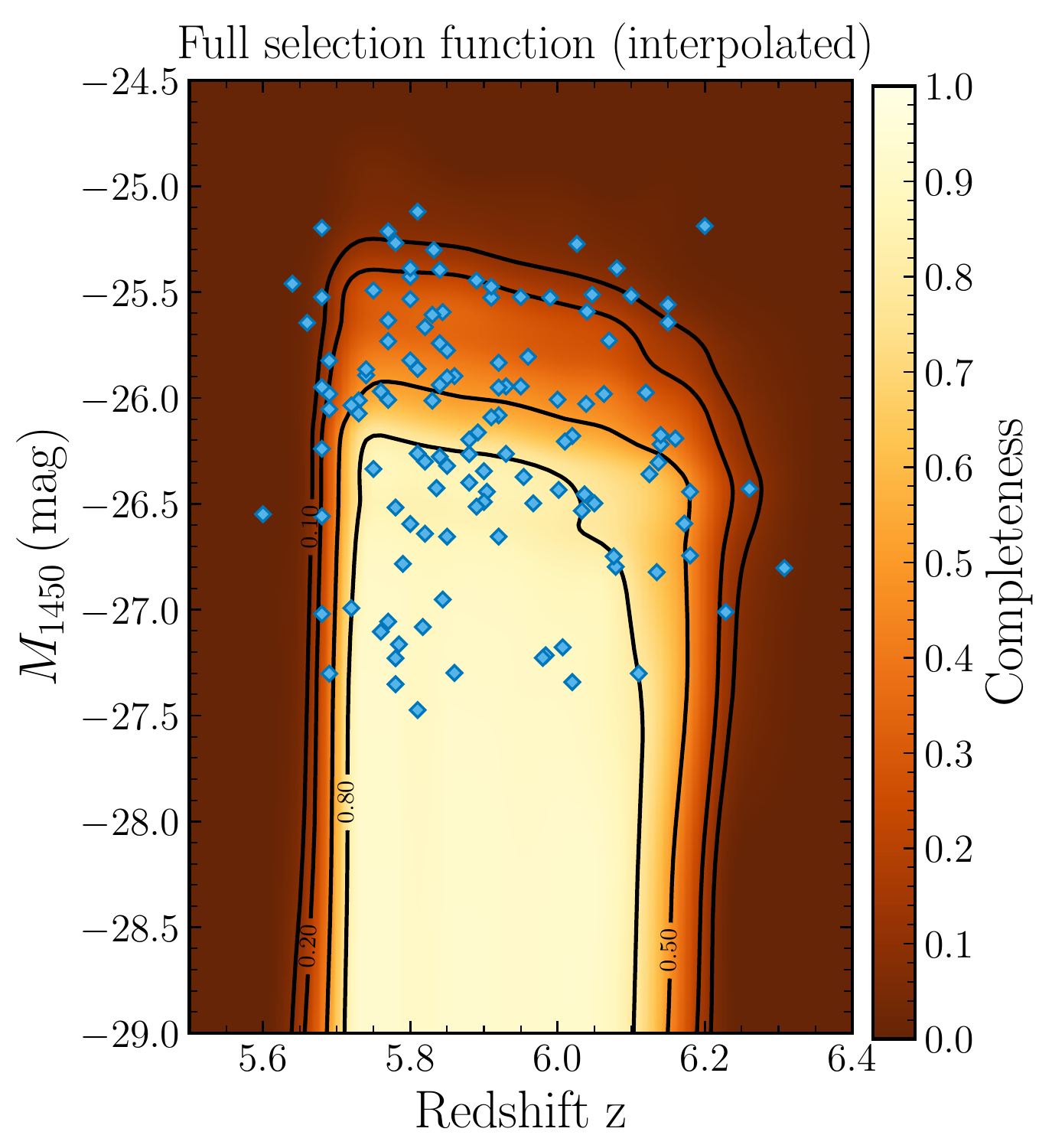}
    \caption{The full PS1 distant quasar survey selection function for the quasar selection focused at $5.7\lesssim z \lesssim 6.2$. We show the \nqso{} confirmed quasars that make up the QLF sample as blue diamonds.
    Black contours are drawn at a completeness of 10\%, 20\%, 50\% and 80\%.
    \label{fig:selfun}
    }
\end{figure}

In order to obtain the full survey selection function, we combine the individual selection functions multiplicatively according to Equation\,\ref{eq:surveyselectionfunction} and present the result in Figure\,\ref{fig:selfun}. The shape of the selection function is dominated by the photometric selection (see Figure\,\ref{fig:photo_selfun}). Our incomplete follow-up identification decreases the completeness at the faint end ($M_{1450}=-26.5$ to $-25$) compared to the pure photometric selection. The impact of the photometric quality selection and morphology selection is more subtle as it decreases the completeness over the full absolute magnitude and redshift range.
We use this selection function to correct for our completeness in the measurements of the quasar luminosity function.

Figure\,\ref{fig:selfun} highlights a few quasars that lie in regions of very low completeness and it is worthwhile to briefly understand why they passed our selection strategy.
The lowest redshift quasar of our sample is PSO224.65067+10.21379 at a spectroscopic redshift of $z\approx5.4$. This source shows strong broad absorption in \nv{} that effectively removes all Ly$\alpha$ flux, mimicking a Ly$\alpha$ break at $z\ge5.7$.
At the high redshift end our sample includes SDSS~J1030+0524 at $z=6.308$. This source as well as some fainter sources at $z\approx6.2$ (PSO184.33893+01.52846 and PSO334.01815-05.00488) have especially strong Ly$\alpha$ flux. Due to the resulting  blue $z_{\text{P}1}-y_{\text{P}1}$ color they are included in our selection (Equation\,\ref{eq:phot_criterion3}).
Finally, our sample includes a number of faint quasars ($ M_{1450}\gtrsim-25.5$) with selection probabilities below $20\%$. This indeed indicates that we expect more than 5 times as many quasars in this parameter region in full agreement with previous QLF results \citep{Willott2010a, Matsuoka2018c}.






\section{Results }\label{sec:results}
In this section we present our QLF measurement at $z\approx6$. As the PS1 distant quasar sample only constrains the bright end of the QLF $( M_{1450}<-25)$, we combine our sources with 48 quasars from the SHELLQs quasar sample presented in \citet{Matsuoka2018c}. We adopt the quasar properties based on their Table\,1 and use their selection function (see their Figure\,9; the electronic data provided by Yoshiki Matsuoka) to correct the SHELLQs sample for incompleteness.
We discuss the PS1 QLF binned as a function of absolute magnitude in  Section\,\ref{sec:binnedqlf}. Then, in Section\,\ref{sec:mlqlf}  we fit a double power law to the combined sample (PS1 + SHELLQs) using a maximum likelihood approach.

\subsection{The binned quasar luminosity function}\label{sec:binnedqlf}
We determine the binned quasar luminosity function over the redshift interval $5.65\lesssim z \lesssim 6.25$ in bins of absolute magnitude at $1450$\,\AA. This redshift range includes a total of \nqsobinned{} of the total \nqso{} quasars in our sample.
For the calculation of the binned QLF we implemented the $1/V_{\text{a}}$ method \citep{Schmidt1968, Avni1980} with the modifications outlined in \citet{Page2000}. We correct the quasar number counts using the completeness from the full quasar selection function described above (Section\,\ref{sec:fullselectionfunction}).
Table\,\ref{tab:binnedqlf} summarizes the results. The columns describe the absolute magnitude bin, the median absolute magnitude $\text{med}( M_{1450})$, the median redshift $\text{med}(z)$, the number of quasars per bin, and the uncorrected and corrected binned QLF values ($\Phi$) with uncertainties reflecting the confidence interval for a Poisson distribution that corresponds to $1\sigma$ in Gaussian statistics.\footnote{We approximately calculate this confidence interval equivalent to Equations\,4 and 5 in \citet{Gehrels1986}.}

\begin{figure*}
    \centering
    \includegraphics[width=0.85\textwidth]{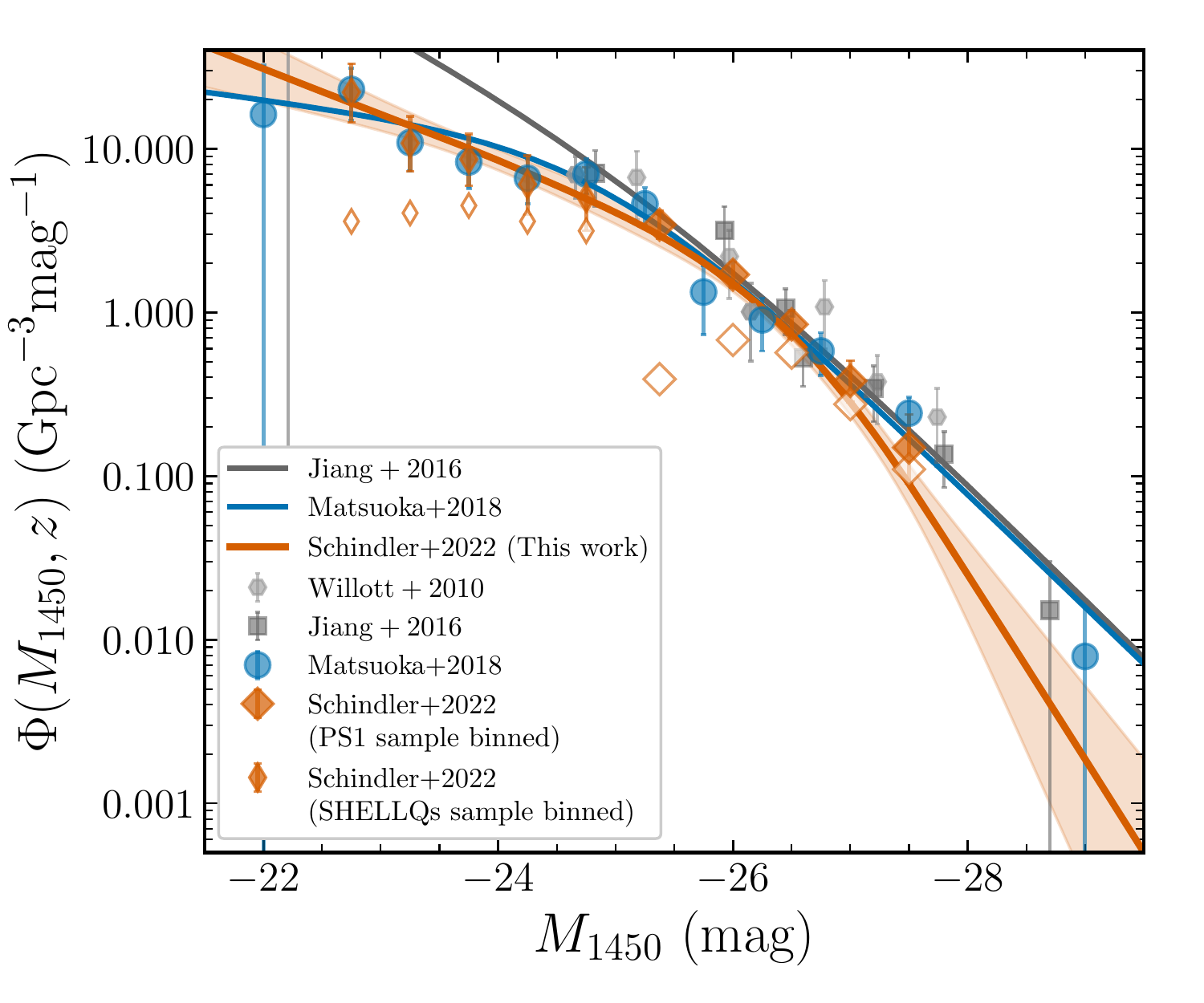}
    \caption{The $z\sim6$ QLF from the combined SHELLQs and PS1 distant quasar survey samples (orange). We show the binned QLF measurements of the SHELLQs sample and the PS1 distant quasar survey as thin and broad orange diamonds, respectively. Open symbols depict the binned QLF not taking the selection function into account. Light grey hexagons, grey squares and blue circles denote the binned measurements from the studies of \citet{Willott2010a}, \citet{Jiang2016} and \citet{Matsuoka2018c}.
    The solid orange line is the median value of the full posterior from our maximum likelihood MCMC double power law (DPL) fit with fixed density evolution ($k=-0.7$), where the shaded region highlights the 16 to 84 percentile uncertainty. For comparison we also show the best-fit DPL models for the \citet{Jiang2016} QLF and the \citet{Matsuoka2018c} QLF as blue and grey solid lines. All parametric QLF fits have been evaluated at $z=6.0$.}
    \label{fig:qlf_comparison}
\end{figure*}

The results of our binned PS1 (and SHELLQs) QLF is depicted as (thin) filled orange diamonds in Figure\,\ref{fig:qlf_comparison} compared to other binned QLF measurements at $z\sim6$ from the literature. We compare our results to the binned QLFs determined by \citet{Willott2010a} (41 quasars, $5.74 < z < 6.42$, light grey hexagons), \citet{Jiang2016} (47 quasars, $5.7 < z \le 6.4$, dark grey squares), and \citet{Matsuoka2018c} (112 quasars, $5.7 \le z \le 6.5$, blue circles).
With a total of \nqsobinned{} quasars, the binned PS1 QLF agrees well with the literature data at $z\approx6$.
We note that our sample covers a narrower and slightly lower redshift range than the previous work in the literature. As a consequence we miss the bright quasar J0100+2802 at $z=6.30$ \citep{Wu2015} and our brightest bin only extends to $ M_{1450}=-27.75$.
The agreement between our binned values of the SHELLQs sample QLF (thin diamonds) and the values from \citet[][blue circles]{Matsuoka2018c} in the magnitude bins centered at $ M_{1450}=-23.25$ and $-23.75$ is excellent. This demonstrates the consistency of our methods and their implementation, but inadvertently results in the blue circles and orange diamonds to completely overlap in Figure\,\ref{fig:qlf_comparison}.
While we use only the SHELLQs data, \citet{Matsuoka2018c} derive the binned QLF from a combination of samples, which explains the differences seen between the thin diamonds and the blue circles.

\subsection{Maximum Likelihood Estimation of the quasar luminosity function}\label{sec:mlqlf}

The measurement of the binned QLF, while agnostic to its underlying shape, is dependent on the choice of binning, both in luminosity and redshift. With our choice of only one redshift bin, the analysis in the previous section could not account for any redshift evolution of the QLF.
Alternatively, we can assume a parametric model for the QLF, including redshift evolution, and constrain the model QLF $\Phi(M, z|\vect{\Theta}_{\text{QLF}})$ and its parameters $\vect{\Theta}_{\text{QLF}}$ by Markov Chain Monte Carlo (MCMC) sampling from the probability of the model QLF given the observed quasar sample $N(M, z)$, $P_{\text{QLF}}\equiv P(N(M, z)|\Phi(M, z| \vect{\Theta}_{\text{QLF}}))$.
We follow \citet[][]{Marshall1983} to derive the logarithmic probability $\ln(P_{\text{QLF}})$. We present the full derivation of the logarithmic probability for a luminosity function model $\Phi(M, z|\vect{\Theta}_{\text{QLF}})$ with a selection function $S(\vect{q})$ in Appendix\,\ref{app:qlfanalysis}.
The logarithmic probability $\ln(P_{\text{QLF}})$ can then be approximated as (see also Equation\,\ref{eq:qlflogprob})
\begin{gather}
\begin{split}
  \ln(P_{\text{QLF}}) \propto & \sum_{j=1}^{N(M,z)} \ln \left[\Phi(M_j, z_j| \vect{\Theta}_{\text{QLF}}) S_j(\vect{q}(M_j, z_j))\right] \\ & - \Lambda(M,z) \ ,
\end{split}
\end{gather}
where $\Lambda(M,z)$ is the quasar incidence rate as given by Equation\,\ref{eq:incidence_3}.
Given a large enough dynamic range in luminosity, the QLF at low-redshift is well approximated by a broken double power law (DPL) \citep[e.g.,][]{Boyle2000},
\begin{equation}
 \Phi(M, z) = \frac{\Phi^{\star}(z)}{10^{0.4(\alpha+1)(M-M^{\star})}+10^{0.4(\beta+1)(M-M^{\star})}} \ ,
\end{equation}
defined by the normalization $\Phi^{\star}$, the break magnitude $M^{\star}$ and the two power law slopes $\alpha$ and $\beta$. By convention, $\alpha$ is most commonly chosen as the faint-end slope, with $\beta$ then describing the bright-end slope.
Generally, all four parameters could evolve with redshift \citep[see e.g.,][]{Kulkarni2019}. Our sample only spans a narrow redshift interval, therefore we only adopt a redshift evolution for the normalization in the form of
\begin{equation}
    \log(\Phi^{\star}(z)) =  \log(\Phi^{\star}(z=6)) + k \times (z-6) \ .
\end{equation}
The parameter $k$  describes the exponential evolution of the quasar density with redshift.
We implement the luminosity function model and the calculation of the logarithmic probability and then use the python package \texttt{emcee} \citep{emcee} for efficient MCMC sampling of the parameter space.
For each model fit we run \texttt{emcee} with 10,000 steps and 50 walkers for a total of 500,000 samples. After discarding the first 1,000 steps for each walker, we retrieve the full flat chain of 450,000 realizations.
For our fit to the PS1 sample we use a luminosity range of $M_{1450}=-29$ to $-25$ and a redshift range of $z=5.65$ to $6.25$, which includes \nqsobinned{} quasars.
The SHELLQs sample covers data with lower luminosities ,$M_{1450}=-27.8$ to $-22$, and a larger redshift range, $z=5.705$ to $6.495$.

Our fiducial model fits the main four DPL parameters based on the combined PS1 and SHELLQs quasar samples, while assuming a fixed value for the normalization evolution $k=-0.7$. Our choice is motivated by previous QLF measurements at $z\approx5-7$ \citep{McGreer2013, Jiang2016, WangFeige2019, PanZhiwei2022} and makes the results easily comparable to \citet{Matsuoka2018c}, where the same value for $k$ was used. The covariance matrix of the fit parameters is shown Figure\,\ref{fig:mlfit_covariance}.
We note that the SHELLQs quasar sample covers a larger redshift range than ours. This is taken into account in our maximum likelihood formulation given our assumption on the redshift evolution.
Figure\,\ref{fig:qlf_comparison} shows our fiducial QLF model compared to the binned QLF values and literature data at $z=6$ \citep{Willott2010a, Jiang2016,  Matsuoka2018c}. The shaded regions include the 16 to 84 percentile range of all 450,000 realizations (our posterior).
The fit results are provided in Table\,\ref{tab:mlfit_results} along with their uncertainties (16 to 84 percentile range). The first two columns of the table specify the model and data used in the fit.
Table\,\ref{tab:litcomp} summarizes selected QLF studies from the literature for comparison.
The figure highlights two characteristics of our new QLF DPL fit. The bright-end slope is significantly steeper than previous measurements and the overall number densities are lower with the orange curve lying beneath the blue and grey curves for the majority of the magnitude range.
\begin{figure}
    \centering
    \includegraphics[width=0.5\textwidth]{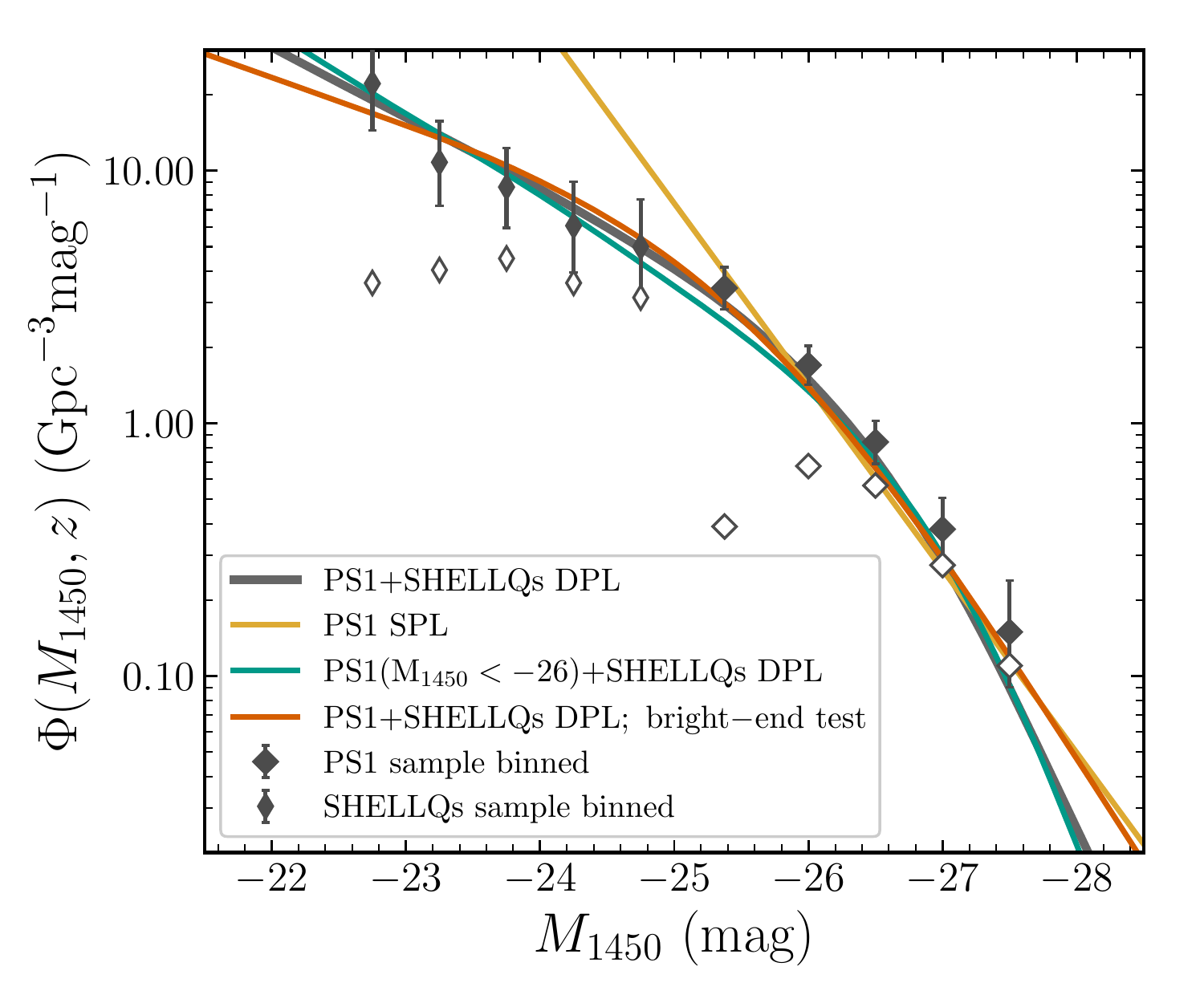}
    \caption{We compare our fiducial QLF double power law model fit (dark grey) to a single power law fit to the PS1 sample (yellow) and a double power law fit to the combined PS1 + SHELLQs sample with the PS1 sample restricted to $ M_{1450}\le-26$ (teal), excluding the range of low spectroscopic completeness. The binned QLF is shown in dark grey with open symbols denoting values not corrected for completeness. The shading indicates the 16 to 84 percentile region of the posteriors.}
    \label{fig:qlf_fit_comp}
\end{figure}

To test the robustness of our fiducial QLF fit results we explore four variations on the model.
The first variation mimics our fiducial model but allows $k$ to vary during the fit. The results are in good agreement with our fiducial model. However, the best fit value for $k$ is $-0.20\pm0.2$ in tension with our assumption of $k=-0.7$. We will continue to discuss this inconsistency below.
In a second variation we fit a single power law (SPL) with $k=-0.7$ to the PS1 quasar sample without the additional SHELLQs data. The PS1 data alone can be well described by this single power law with a slope of $\alpha=-2.81$ and a normalization of $\log\Phi^*(z=6)=-8.85$. Figure\,\ref{fig:qlf_fit_comp} shows the SPL fit (yellow) against our DPL fiducial model (grey).
The third variation uses our fiducial model, but only includes PS1 quasars with $M_{1450}\le-26$ in addition to the SHELLQs sample. It is designed to exclude the region of lower spectroscopic completeness (see right panel in Figure\,\ref{fig:quality_spec_selfun}) to understand its influence on the fit results. All fit parameters change slightly within the uncertainties of the fiducial model. Comparing this variation (teal) to the fiducial model (grey) in Figure\,\ref{fig:qlf_fit_comp} underlines that the differences over the constrained magnitude range are negligible. We conclude that the magnitude range of low spectroscopic completeness has a minor influence on our fiducial results.
Due to the restricted redshift range of the PS1 quasar sample we miss two bright quasars included in \citet{Jiang2016} and \citet{Matsuoka2018c}.
These are J1148+5251 at $z=6.42$ with $M_{1450}=-27.8$ and J0100+2802 at $6.30$ with $M_{1450}=-29.1$. As a test we artificially add these two sources with a nominal redshift of $z=6$ to our quasar sample and fit it again with our fiducial model. We denote this variation as \textit{bright end test} in Figure\,\ref{fig:qlf_fit_comp} and Table,\ref{tab:mlfit_results}.
The artificial inclusion of these two bright sources significantly changes all QLF parameters, which is reflected in the different shape of this variation (orange) to our fiducial model (grey) in  Figure\,\ref{fig:qlf_fit_comp}. Most notably the bright end slope changes to $\beta=-3.12$ and the break is now $0.8$\,mag fainter. This test highlights the strong influence individual sources at the extreme bright end can have on the QLF measurement.

With the conclusions from the different fit variations in mind, we now discuss our fiducial fit results in context with the current literature.
Our QLF model favors a break magnitude of $ M_{1450}=-26.38$. This value is brighter than previous work at $z\approx6$ \citep{Willott2010a, Jiang2016, Matsuoka2018c}, but also fainter than work at $z\approx5$ \citep{Yang2016, McGreer2018}. It is also significantly fainter than the study of \citet{Kulkarni2019}, which model the QLF from $z=0$ to $6$ and fit a break magnitude of $ M_{1450}\approx-29$ at $z=6$, effectively constraining only the faint end slope with the data at $z=6$.
The differences to the results of \citet{Jiang2016} and \citet{Matsuoka2018c} are at least in part due to the inclusion of bright quasars at $z>6.25$ in these samples as our bright-end test (Table\,\ref{tab:mlfit_results}, last row) highlights.\\

\begin{deluxetable*}{lccccc}
\tabletypesize{\footnotesize} 
\tablecaption{The binned PS1 distant quasar survey QLF at $5.7 \le z \le 6.2$\label{tab:binnedqlf}}
\tablehead{\colhead{$M_{1450}$ bin} &\colhead{$\rm{med}(M_{1450})$} &\colhead{$\rm{med}(z)$} &\colhead{$N$} &\colhead{$\Phi$ (uncorr.)} &\colhead{$\Phi$} \\ 
\colhead{(mag)} &\colhead{(mag)} &\nocolhead{} &\nocolhead{} &\colhead{$(\rm{Gpc}^{-3}\rm{mag}^{-1})$} &\colhead{$(\rm{Gpc}^{-3}\rm{mag}^{-1})$} 
} 
\startdata 
$-27.50\pm{0.25}$ & $-27.32$ & $5.83$ & $6$ & $0.110$ & $0.149_{-0.059}^{+0.089}$ \\ 
 $-27.00\pm{0.25}$ & $-27.06$ & $5.82$ & $15$ & $0.275$ & $0.381_{-0.097}^{+0.126}$ \\ 
 $-26.50\pm{0.25}$ & $-26.44$ & $5.90$ & $31$ & $0.567$ & $0.843_{-0.151}^{+0.180}$ \\ 
 $-26.00\pm{0.25}$ & $-25.98$ & $5.86$ & $37$ & $0.677$ & $1.700_{-0.278}^{+0.328}$ \\ 
 $-25.38\pm{0.38}$ & $-25.52$ & $5.84$ & $32$ & $0.391$ & $3.433_{-0.604}^{+0.721}$ \\ 
 \enddata 
\end{deluxetable*}

\begin{deluxetable*}{ccccccc}
\tabletypesize{\footnotesize} 
\tablecaption{PS1 QLF parameter values as constrained by the ML fit\label{tab:mlfit_results}}
\tablehead{\colhead{Model} &\colhead{Data} &\colhead{$\log\Phi^*(z=6)$} &\colhead{$M^*$} &\colhead{$\alpha$} &\colhead{$\beta$} &\colhead{$k$} \\ 
\nocolhead{} &\nocolhead{} &\colhead{$(\rm{Mpc}^{-3}\rm{mag}^{-1})$} &\colhead{(mag)} &\nocolhead{} &\nocolhead{} &\nocolhead{} 
} 
\startdata 
DPL &PS1 + SHELLQs &$-8.75^{+0.47}_{-0.41}$ &$-26.38^{+0.79}_{-0.60}$ &$-1.70^{+0.29}_{-0.19}$ &$-3.84^{+0.63}_{-1.21}$ &$-0.70$\tablenotemark{a} \\ \tableline 
DPL &PS1 + SHELLQs &$-8.74^{+0.38}_{-0.35}$ &$-26.45^{+0.64}_{-0.51}$ &$-1.66^{+0.24}_{-0.18}$ &$-3.95^{+0.62}_{-1.12}$ &$-0.20^{+0.20}_{-0.20}$  \\  
SPL &PS1 &$-8.85^{+0.04}_{-0.04}$ &$-26.00$\tablenotemark{a} &$-2.81^{+0.14}_{-0.14}$ & &$-0.70$\tablenotemark{a}  \\  
DPL &$\rm{PS1}\ (\rm{M}_{1450}<-26$)\ \rm{+\ SHELLQs}\  &$-9.09^{+0.72}_{-0.34}$ &$-26.84^{+1.09}_{-0.45}$ &$-1.85^{+0.36}_{-0.15}$ &$-4.41^{+1.09}_{-1.77}$ &$-0.70$\tablenotemark{a}  \\  
DPL &PS1{+}SHELLQs DPL; bright--end test &$-8.28^{+0.39}_{-0.44}$ &$-25.57^{+0.80}_{-0.77}$ &$-1.45^{+0.38}_{-0.26}$ &$-3.12^{+0.32}_{-0.45}$ &$-0.70$\tablenotemark{a}  \\  
\enddata 
\tablenotetext{a}{Parameters held fixed in QLF analysis.}\end{deluxetable*}

\begin{deluxetable*}{cccccccc}
\tabletypesize{\footnotesize} 
\tablecaption{Selected literature QLF measurements at $z>4.5$ \label{tab:litcomp}}
\tablehead{\colhead{Reference} & \colhead{Redshift range} &\colhead{Model} &\colhead{$\log\Phi^*(z=6)$} &\colhead{$M^*$} &\colhead{$\alpha$} &\colhead{$\beta$} &\colhead{$k$} \\ 
\nocolhead{} &\nocolhead{} &\nocolhead{} &\colhead{$(\rm{Mpc}^{-3}\rm{mag}^{-1})$} &\colhead{(mag)} &\nocolhead{} &\nocolhead{} &\nocolhead{} 
} 
\startdata 
\citet{Willott2010a} & $5.74 < z < 6.42$ & DPL & $-7.94$ & $-25.13$ & $-1.5$\tablenotemark{a} & $-2.81$ & $-0.47$\tablenotemark{a} \\
\citet{Jiang2016} & $5.7 < z \le 6.4$ & DPL & $-8.00$ & $-25.2_{-3.8}^{+1.2}$ & $-1.9_{-0.58}^{+0.44}$ &  $-2.8$\tablenotemark{a} & $-0.70$\tablenotemark{a} \\ 
\citet{Matsuoka2018c} & $5.7 \le z \le 6.5$ & DPL & $-7.96_{-0.42}^{+0.32}$ & $-24.9_{-0.9}^{+0.75}$ & $-1.23_{-0.34}^{+0.44}$ & $-2.73_{-0.31}^{+.23}$ & $-0.70$\tablenotemark{a} \\
\tableline
\citet{Yang2016} & $4.7 \le z \le 5.4$ & DPL & $-8.82_{-0.15}^{+0.15}$ & $-26.98_{-0.23}^{+0.23}$ & $-2.03$\tablenotemark{a} & $-3.58_{-0.24}^{+0.24}$ & $-0.47$\tablenotemark{a} \\
\citet{McGreer2018} & $4.7 \le z \le 5.4$ & DPL & $-8.97_{-0.18}^{+0.15}$ & $-27.47_{-0.26}^{+0.22}$ & $-1.97_{-0.09}^{+0.09}$ & $-4.0$\tablenotemark{a} & $-0.47$\tablenotemark{a} \\
\citet{WangFeige2019} & $6.45 \le z \le 7.05$ & DPL & $-8.49_{-0.14}^{+0.10}$ & $-25.2$\tablenotemark{a} & $-1.9$\tablenotemark{a} & $-2.54_{-0.29}^{+0.29}$ & $-0.78$\tablenotemark{a} 
\enddata 
\tablenotetext{a}{Parameters held fixed in QLF analysis.}
\end{deluxetable*}

Allowing for $k$ to vary or excluding the fainter PS1 sample data with the high spectroscopic incompleteness from our fit (see second and fourth row in Table\,\ref{tab:mlfit_results}) does not change our best--fit break magnitude significantly, given its uncertainties. Rather, it seems to be a robust result given the combined PS1 + SHELLQs quasar sample (also see Figure\,\ref{fig:mlfit_covariance}).

The break magnitude $ M_{1450}$ and density normalization $\Phi^{\star}(z=6)$ values are highly covariant in a broken double power law (Figure\,\ref{fig:mlfit_covariance}). As a natural consequence of our brighter break magnitude we measure a lower value of $\Phi^{\star}(z=6)$ than the previous studies at $z=6$. Taking this covariance into account, our best--fit QLF model agrees well with previous determinations especially.

Our fiducial best fit returns a bright--end slope with a relatively value of $\beta=-3.84$, significantly steeper than the literature data at $z=6$ \citep[$\beta\approx-2.8$][]{Willott2010a, Jiang2016, Matsuoka2018c}.
The exception is the study of \citet{Kulkarni2019}.
Following their global QLF fit, they find a very steep bright end slope of $\beta=-5.05_{-1.18}^{+0.76}$. However, their work does not yet include the faint SHELLQs quasars at $z\gtrsim6$. In consequence, they derive an extremely bright best-fit break magnitude at $z=6$, $M^*\approx-29$, such that the data they use only constrains the faint end slope at this redshift.
We emphasize that the PS1 quasar sample that determines $\beta$ covers a narrower redshift range. Therefore, our samples do not include some very luminous quasars at $z>6.25$, e.g., J1148+J1148+5251 or J0100+2802. Artificially including these sources in our sample with an assumed redshift of $z=6$ changes the best--fit model results significantly (Table\,\ref{tab:mlfit_results}, last row). We conclude that these few sources at very bright magnitudes are the main driver between differences between our fiducial measurements and the studies of \citet{Jiang2016} and \citet{Matsuoka2018c}.
%
%
On the other hand, a steep bright end slope is not uncommon. In fact, at $z\le 5$ a range of studies \citep[e.g.,][]{Richards2006, McGreer2013, Yang2016} find a bright end slope of $\beta\lesssim-3$, with some studies reporting an even steeper slope of $\beta\approx-4$ \citep{Schindler2019a, Boutsia2021, PanZhiwei2022}.
Viewed in this context, our results at $z\approx6$ indicate that the bright end slope is generally steep ($\beta\approx-4$) and does not evolve significantly with redshift.

The faint end slope measurement of our maximum likelihood fit is largely determined by the SHELLQs quasar sample. Our best fit value of $\alpha=-1.70$ lies between the value of $\alpha=-1.23$ measured by \citet{Matsuoka2018c} and the previous determination of $\alpha=-1.9$ by \citet{Jiang2016}.
The addition of the SHELLQs sample to our fit explains the flatter slope compared to \citet{Jiang2016} as data at these faint luminosities was not available at the time.
The $2\sigma$ differences to the result of \citet{Matsuoka2018c} is mostly driven by the influence of bright sources (Table\,\ref{tab:mlfit_results}, last row) on the QLF results, which are present in their sample but not included in ours. These sources significantly affect the resulting break magnitude, which is covariant with the faint end slope (see Figure\,\ref{fig:mlfit_covariance}).
Additionally, the QLF of \citet{Matsuoka2018c} includes data from \citet{Willott2010a} not present in our analysis.

Our fiducial QLF fit assumes an exponential density evolution with $k\approx-0.7$. As a test we remove this assumption and allow $k$ to vary, resulting in a best-fit value of $k=-0.20\pm0.2$. This value is in tension with our assumption, but does this mean that our assumption of $k\approx-0.7$ is not justified?
We have based the assumption on literature data, and the following discussion on the quasar density redshift evolution in Section\,\ref{sec:qsodensity} strongly supports our assumed value of $k$ for the fiducial fit.
Hence, we conclude that the redshift range, which is limited by our sample selection, is not large enough to probe the quasar density evolution sufficiently.

\begin{figure}[t]
    \centering
    \includegraphics[width=0.45\textwidth]{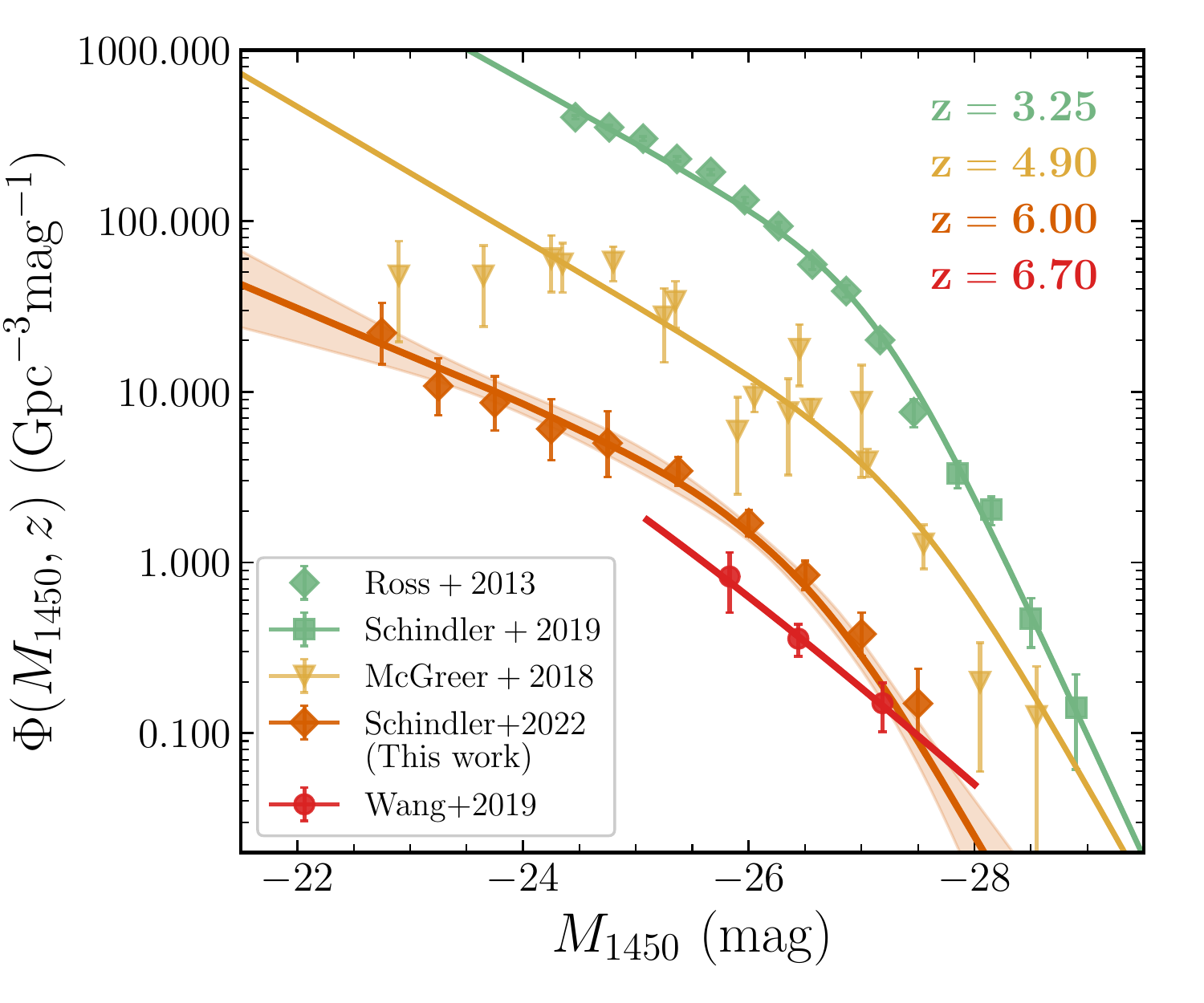}
    \caption{Redshift evolution of the QLF from $z\approx3-7$ colored by redshift. Filled symbols show the binned QLF at these redshifts from different studies in the literature \citep{Ross2013, McGreer2018, Schindler2019a, WangFeige2019}. This comparison visualizes the strong decline in quasar number densities across redshift $z\approx3$ to $7$.}
    \label{fig:qlf_redsh_comp}
\end{figure}

\section{Discussion}\label{sec:discussion}

\subsection[Evolution of the bright-end quasar density]{Evolution of the $M_{1450}<-26$ quasar density}\label{sec:qsodensity}

\begin{figure*}[ht]
    \centering
    \includegraphics[width=\textwidth]{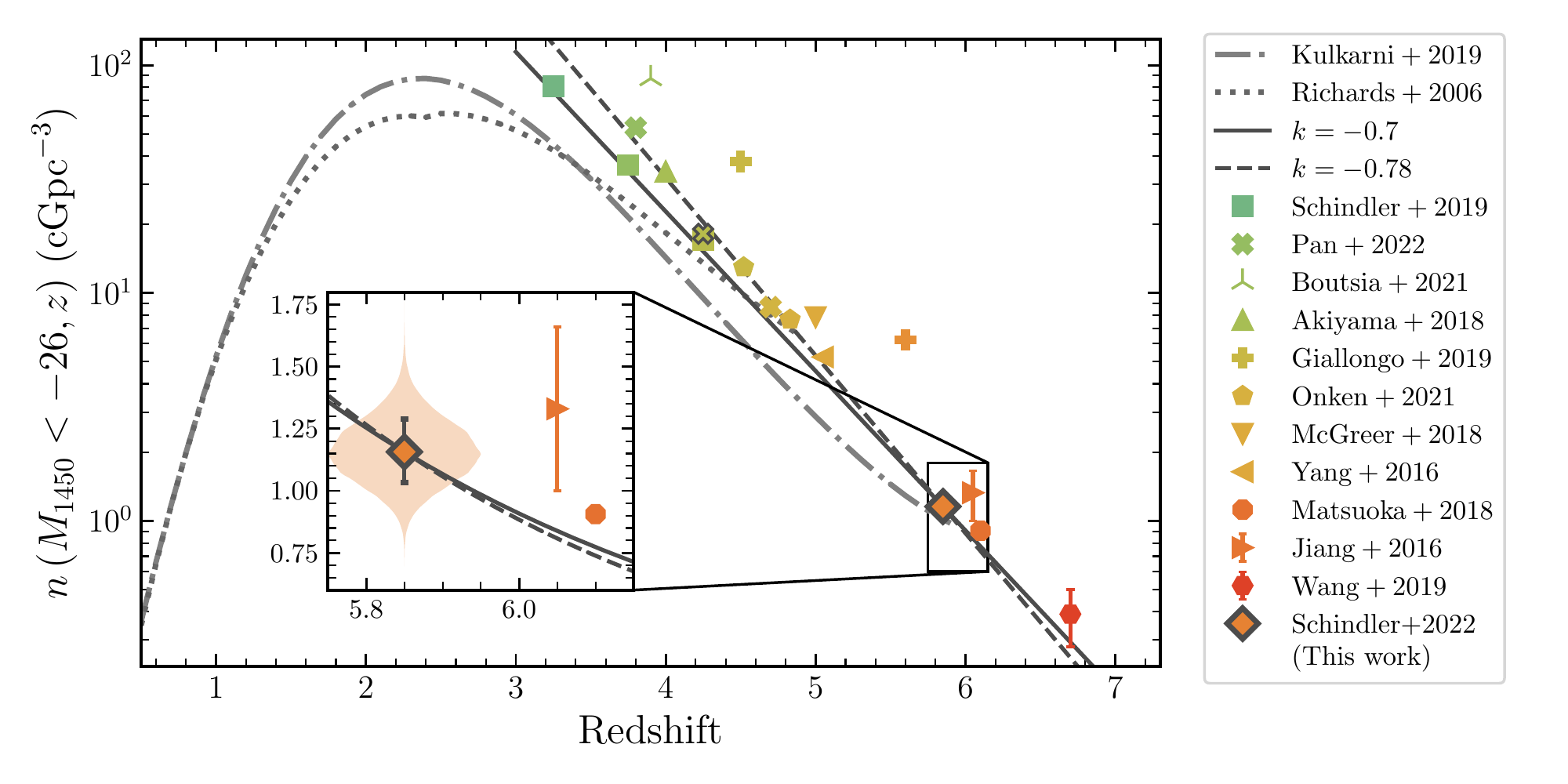}
    \caption{The density of luminous quasars $n( M_{1450}{<}-26)$ as a function of redshift. Our result, integrating our DPL QLF (first row in Table\,\ref{tab:mlfit_results}) at $z=5.85$ is shown as an orange diamond. The uncertainties (grey) shown in the inset indicate the 16 to 84 percentile region of the fit posteriors, which are shown with the violin plot (orange shaded region).
    The density evolution that we assumed for our fit, $n( M_{1450}{<}-26)\propto 10^{-k(z-6)}$ with $k=-0.7$, and the evolution assuming a value of $k=-0.78$ are shown as the solid and dashed dark grey lines. We compare our results to a large range of individual values from the literature indicated by different markers and colors \citep{Jiang2016, Yang2016, Akiyama2018, Matsuoka2018c, McGreer2018, Giallongo2019, Schindler2019a, WangFeige2019, Boutsia2020, Onken2022, PanZhiwei2022} and the density evolution from the QLFs of \citet[][]{Richards2006} and \citet{Kulkarni2019} (grey dotted and dot-dashed lines, respectively). All data points with the exception of the values from \citet{Jiang2016} and \citet{WangFeige2019} are calculated by integrating the QLF model of the respective work. The other two values are determined in \citet{WangFeige2019} using their Equation\,12.
    We note the discrepancy between the majority of points and the number densities from \citet{Giallongo2019} and \citet{Boutsia2021}, which are systematically higher.
    Given the most recent measurements at $z=3-5$, our assumed value of $k=-0.7$ describes the increase of the luminous quasar density  well from $z=7$ to $3$ without the need for a steeper ($k=-0.78$) evolution.}
    \label{fig:qso_density}
\end{figure*}

The bright--end quasar density $n( M_{1450}{<}-26)$ has been known to increase from $z\approx5$ to $z\approx2$ \citep{Schmidt1995}.
\citet{Fan2001c} modeled the quasar density up to $z\approx6$ with $n( M_{1450}{<}-26)\propto 10^{k(z-z_{\text{ref}})}$, finding a value of $k=-0.47$. \citet{McGreer2013} and \citet{Jiang2016} reported and even steeper increase of the luminous quasar density with $k=-0.7$, the factor assumed in our maximum likelihood analysis above.
From $z\approx7$ to $z\approx6$ the quasar density is reported to increase even more steeply with $k=-0.78$ \citep{WangFeige2019}.
This finding is supported by quasar searches from the VIKING survey (B. Venemans, private communication) that go beyond the first discoveries \citep{Venemans2013}.
%
Going backwards in cosmic time, this seemingly accelerating decrease in quasar density has important consequences for the predicted number of discoverable quasars at even higher redshifts, $z>8$.
In Figure\,\ref{fig:qlf_redsh_comp} we compare the QLFs of \citet{WangFeige2019} at $z=6.7$, of \citet{McGreer2018} at $z=4.9$, and of \citet{Schindler2019a} at $z=3.25$ with our results at $z=6.0$. The Figure shows the rise in quasar number counts from $z=6.7$ to $3.25$.

To determine the quasar density at the bright end, we integrate our best-fit QLF at $z=5.85$ down to $ M_{1450}=-26$. This results in a value of $n(M_{1450}<-26)=1.16_{-0.12}^{+0.13}\,\text{cGpc}^{-3}$, in line with the other measurements at $z\approx6$ \citep{Willott2010a, Jiang2016, Matsuoka2016}.
We chose a redshift of $z=5.85$ close to the median redshift of the PS1 quasar sample (see Table\,\ref{tab:binnedqlf}), which determines this magnitude range of the QLF.
We show our result in comparison with values from the literature in Figure\,\ref{fig:qso_density}. We include a range of studies that provide a significant re-evaluation of the QLF at $z=3-5$ \citep{Akiyama2018, Schindler2018, Giallongo2019, Schindler2019a, Boutsia2021, Onken2022} compared to the first results from SDSS \citep{Richards2006, Ross2013, Shen2012}.
%
%
We also show the redshift evolution from integrating the quasar density from the QLFs of \citet[][$z=0.3-5$]{Richards2006} and \citet[][$z=0.3-6$]{Kulkarni2019} as grey dotted and dot-dashed lines.
We note that \citet{Niida2020} and \citet{KimYongjung2020} also provide updated measurements on the $z\approx5$ QLF. Their integrated quasar densities are consistent with \citet{McGreer2018} and so we do not display them in Figure\,\ref{fig:qso_density} for the sake of its readability.

%
Quasar densities from the studies of \citet{Boutsia2021} and \citet{Giallongo2019} stand out in Figure\,\ref{fig:qso_density}, reporting significantly larger values than the remaining literature.
In the work of \citet{Boutsia2021} the $ M_{1450}{\sim}-26$ luminosities at $z\approx4$ are largely dominated by the assumed \citet{Fontanot2007} QLF with densities larger than reported in \citet{Richards2006}, as adopted in \citet{Schindler2019a}, and \citet{Akiyama2018}. The adoption of the binned \citet{Fontanot2007} QLF values for their analysis thus explains the large number densities compared to the other studies at the same redshift.
The analysis of \citet{Giallongo2019} on the other hand relies on photometric redshifts and does not include data that constrain their QLF model at $ M_{1450}{\sim}-27$ to $-25$, the critical range for determining $n(M_{1450}{<}-26)$. Therefore, the \citet{Giallongo2019} QLF based number densities shown in Figure\,\ref{fig:qso_density} should be interpreted with caution.

Furthermore, we include the exponential density evolution with $k=-0.78$ \citep{WangFeige2019a} and $k=-0.7$ anchored on our value as grey dashed and solid lines in Figure\,\ref{fig:qso_density}.
\citet{PanZhiwei2022} find a single value of $k=-0.7$ to describe the density evolution across $z=3.5$ to $z=5$. Based on our results we argue that this evolution continues to $z\approx7$, when excluding the discrepant data from \citet{Boutsia2021} and \citet{Giallongo2019}.
There is evidence that at $z<4$ the density evolution flattens ($k>-0.7$) before the turnover point at $z\approx2-2.5$ \citep{Richards2006, Kulkarni2019} as also discussed in \citet{Onken2022}.
In light of the recent literature and given the systematic uncertainties inherent in QLF estimates due to differing models for completeness correction, we conclude that the bright end density evolution at $z>4$ can be well described by an exponential decline with $k=-0.7$.
Comparing the work of \citet{Matsuoka2018c} and our new estimate of the $z=6$ quasar density with the value from \citet{WangFeige2019a}, we do not find  evidence for a more rapid decrease of the quasar density at $z>6.5$ as originally reported from the comparison with the \citet{Jiang2016} quasar density in \citet{WangFeige2019a}.

\subsection{Forecasting high redshift quasar detections} \label{sec:forecast}
We explore the impact of our new quasar luminosity function measurement on future high redshift quasar detections.
For this purpose we use the Euclid mission (expected launch 2023) as our main example. The Euclid wide-area survey will deliver \textit{Y}--, \textit{J}--, and \textit{H}--band photometry down to a $5\sigma$ limiting magnitude of $24.0\,\rm{mag}$ (for point-like sources) over $\sim15,000\,\rm{deg}^{2}$.
We use \texttt{simqso} to simulate quasars and their photometry in the Euclid bands at $7 \le z < 10$ according to our QLF  over the wide-area survey footprint.
The Lyman-$\alpha$ break at $1215\text{\AA}$ enters Euclid's \textit{J}--band at $z\gtrsim8.9$. Therefore, we require only $H<24.0$ for a detection in the Euclid wide-area survey.
We show the resulting detection number counts from our best-fit QLF model in comparison with the QLF models of  \citet{Jiang2016}, \citet{Matsuoka2018c}, and \citet{WangFeige2019} in Table\,\ref{tab:qlfpredictions}.


\begin{deluxetable*}{lcccc}
\tabletypesize{\scriptsize}
\tablecaption{Forecasting Euclid quasar detections. See Section~\ref{sec:forecast} for details. \label{tab:qlfpredictions}}
\tablehead{\colhead{QLF} & \colhead{k} &\colhead{$7{\le}z{<}8$} &\colhead{$8 {\le}z{<}9$} &\colhead{$9{\le}z{<}10$} \\
\nocolhead{} &
\nocolhead{} &\colhead{$M_{1450} {\lesssim}-22.4$} &\colhead{$M_{1450}{\lesssim}-22.7$} &\colhead{$M_{1450} {\lesssim}-23.0$}}
\startdata
\tableline
This work & -0.70& 312 & 61 & 8 \\
\citet{Matsuoka2018c} & -0.70 & 360 & 49 & 5  \\
\tableline
\citet{Jiang2016} & -0.70 & 809 & 123 & 18  \\
\citet{WangFeige2019} & -0.78 &668 & 71 & 5
 \enddata
\end{deluxetable*}

At $7 \le z < 8$ we expect to probe an absolute magnitude  of $M_{1450}\lesssim-22.4$. Therefore, the total number counts are strongly dependent on the faintest quasar population.
While the \citet{Matsuoka2018c} QLF and our measurement use the SHELLQs quasars to constrain the faint end, the \citet{Jiang2016} and \citet{WangFeige2019} need to extrapolate into this region.
Hence, the predicted number counts based on their QLFs should be treated with this caveat in mind.
Our work and the QLF from \citet{Matsuoka2018c} produce similar predictions, whereas the steep faint-end slopes of the \citet{Jiang2016} and \citet{WangFeige2019} QLFs lead to high quasar number counts in this redshift range.
The overall slightly lower number densities of our QLF measurement compared to \citet{Matsuoka2018c} make our predictions the least optimistic in this redshift range.
At $8 \le z < 9$ and $9 \le z < 10$ the Euclid \textit{H}--band magnitude probes $M_{1450}\lesssim-22.7$ and $M_{1450}\lesssim-23$, respectively.
Due to the extrapolation to the faint end the \citet{Jiang2016} QLF predicts the most optimistic number counts.
At $8 \le z < 9$ the expected quasar detections based on our measurement and the \citet{Matsuoka2018c} and \citet{WangFeige2019} QLF are around 50--70. The numbers drop to 5--8 at the highest redshift bin, where our new QLF measurement provides a more optimistic forecast.
This simple forecast does not claim to be a comprehensive prediction of the quasar yields based on quasar selection strategies as, for example, presented in \citet{EuclidPrepV2019}.
In this context our predicted detection numbers should be regarded as an upper limit to the number of quasars that could be discovered with Euclid depending on the selection strategy.
Our aim here is to simply  illustrate how our new QLF measurement impacts our expectations for quasar discoveries. As the \citet{EuclidPrepV2019} selection predictions are based on the \citet{Jiang2016} QLF, their resulting quasar yields are likely too optimistic in light of our new QLF measurement.

\subsection{Quasar contribution to hydrogen reionization}

\begin{figure*}
    \centering
    \includegraphics[width=\textwidth]{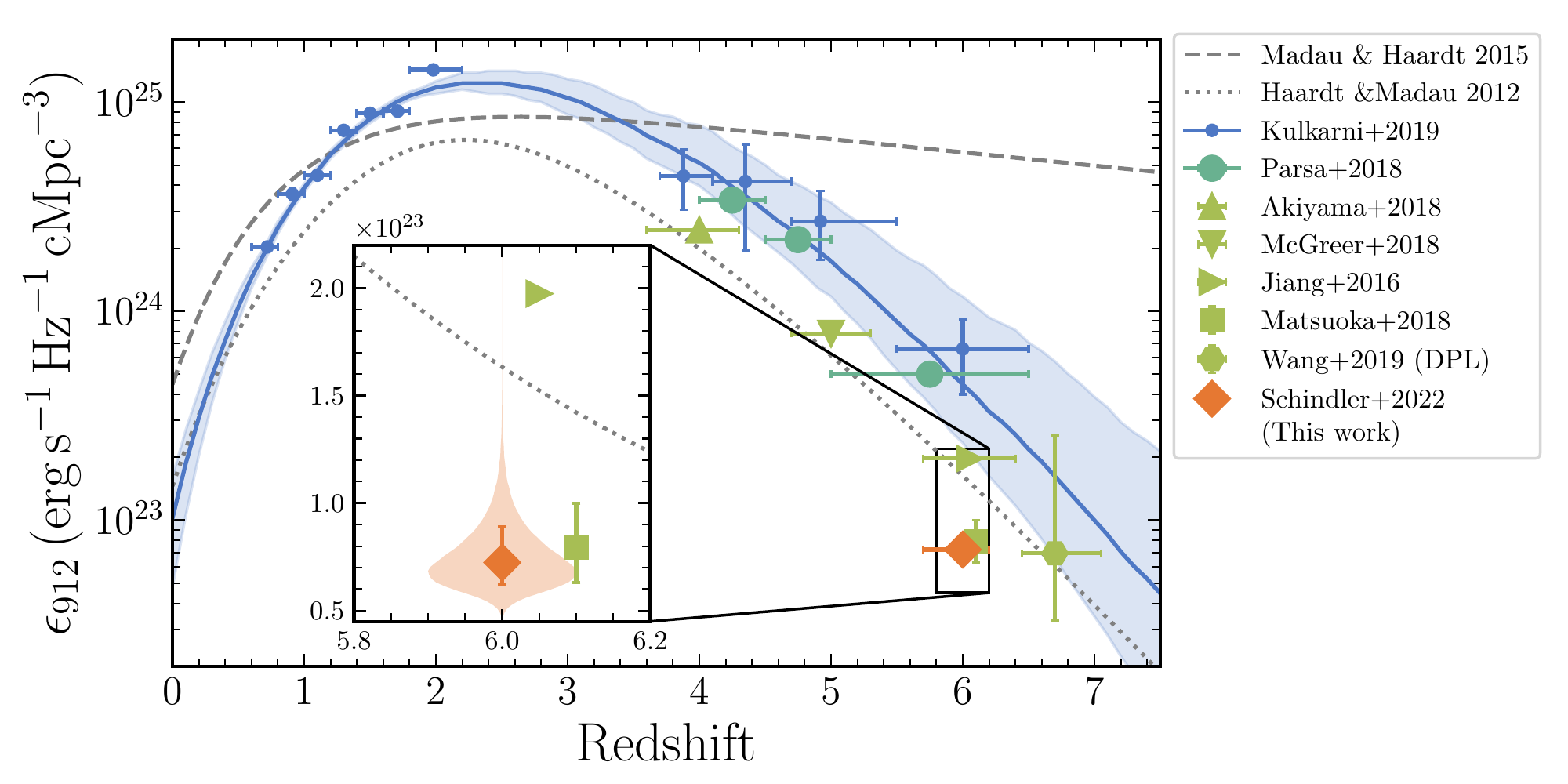}
    \caption{The quasar ionizing emissivity at $912\,$\AA\ as a function of redshift. The ionizing emissivity is derived from Equation\,\ref{eq:ionizing_emissivity} with a faint luminosity limit of $M_{1450}=-18$. We show our result as the orange diamond. The error bars in the inset indicate the 16 to 84 percentile region of the fit posteriors, which are shown with the violin plot (orange shaded region). We compare our data  to various results in the literature \citep{Jiang2016, Akiyama2018, Matsuoka2018c, McGreer2018, Parsa2018, Kulkarni2019, WangFeige2019a}. We include the individual data points from \citet{Kulkarni2019}, that were not affected by systematic errors as discussed by the authors.
    Horizontal error bars on indvidual data points indicate the redshift ranges of the different QLF samples. The blue solid line and the blue shaded area show the derived posterior median emissivity evolution model and $1\sigma$ uncertainties of \citet{Kulkarni2019}. We also display the models by \citet{Haardt2012} and \citet{Madau2015}. Our derived quasar emissivity falls well below other measurements at $z=6$ with the exception of the work from \citet{Matsuoka2018c} with which we share the SHELLQs quasar sample at the faint end.}
    \label{fig:ionizing_emissivity}
\end{figure*}

Based on our new measurement of the QLF at $z\sim6$ we calculate the quasar contribution to the HI photoionization rate of the UV background.
Following the literature \citep[e.g.,][]{Haardt1996, Haardt2012, FaucherGiguere2020} the HI photoionization rate is
\begin{equation}
    \Gamma_{\text{HI}}(z) = \int_{\nu_{912}}^{\infty} \dif{\nu} \sigma_{\text{HI}}(\nu) c n_{\nu}(\nu, z) \ , \label{eq:photoionization_rate}
\end{equation}
where $\sigma_{\text{HI}}(\nu)$ is the frequency dependent HI photoionization cross section and $n_{\nu}(\nu, z)$ is the number density of ionizing photons per unit frequency at redshift $z$. The lower boundary of the integral $\nu_{912}$ corresponds to the frequency at a wavelength of $912\text{\AA}$.
At $z=6$ we can assume that the optical depth of ionizing photons is smaller than unity, $\tau_{\text{eff}}\le 1$, allowing us to adopt the `local source' approximation \citep[e.g.,][]{ZuoLin1993, Madau1999}, simplifying $n_{\nu}(\nu, z)$ to
\begin{equation}
    n_{\nu}(\nu, z) \approx \frac{(1+z)^3}{h \nu} \frac{l(\nu,z)}{c} \epsilon(\nu, z) \ .
\end{equation}
In the equation above $l(\nu,z)$ is the mean free path of ionizing photons and $\epsilon(\nu, z)$ is the comoving emissivity of ionizing sources. We closely follow \citet{ShenXuejian2020} in adopting the frequency dependence on the mean free path based on the results of \citet{FaucherGiguere2008b}, $l(\nu,z) = l(\nu_{912},z) (\nu/\nu_{912})^{3(\beta-1)}$, where the power law index of the intergalactic HI column density distribution is assumed to be $\beta=1.5$ \citep{Madau1999}. Furthermore, we also assume a power law shape for the extreme UV quasar continuum with an index of $\alpha_{\text{UV}}=1.7$ \citep{Lusso2015},
\begin{equation}
   \epsilon(\nu, z) = \epsilon_{912}(z) \left(\frac{\nu}{\nu_{912}}\right)^{-\alpha_{\text{UV}}} \ .
\end{equation}

Assuming a frequency dependence of $\sigma_{\text{HI}}\propto\nu^{-3}$ we  analytically integrate Equation\,\ref{eq:photoionization_rate}, which yields
\begin{equation}
    \Gamma_{\text{HI}}(z) \approx \frac{(1+z)^3}{3+\alpha_{\text{UV}}-3(\beta-1)} \epsilon_{912}(z) l(\nu_{912},z) \sigma_{\text{HI}}(\nu_{912}) \ .
\end{equation}
For the HI photoionization cross-section we use a value of $\sigma_{\text{HI}}(\nu_{912})=6.35\times10^{-18}\,\text{cm}^{2}$ \citep{Verner1996, Becker2015b}.

Using our new measurement of the QLF we first calculate the ionizing emissivity of quasars at $1450\text{\AA}$,
\begin{equation}
    \epsilon_{1450}(z) = \int_{-\infty}^{-18} \Phi(M_{1450},z) L_{1450}(M_{1450}) \dif{M_{1450}} \ . \label{eq:ionizing_emissivity}
\end{equation}
Here we assume that the escape fraction of ionizing photons from the type-1 quasar population measured by the QLF is unity.
We adopt an upper integration boundary (faint limit) of $M_{1450}=-18$ for comparison with the recent literature \citep{Kulkarni2019, Matsuoka2018c, WangFeige2019}. Assuming a power-law SED for the quasars in the extreme UV \citep{Lusso2015},
\begin{gather}
     f_{\nu} \propto  \begin{cases}
      \nu^{-0.61} &  \forall \lambda > 912\text{\AA} \\
      \nu^{-1.70} & \forall \lambda < 912\text{\AA}  \
    \end{cases}
\end{gather}
We estimate the ionizing emissivity at $912\text{\AA}$ as
\begin{equation}
    \epsilon_{912}(z) = \epsilon_{1450}(z)\times (1450\,\text{\AA}/912\,\text{\AA})^{-0.6} \ .
\end{equation}
Based on our fiducial double power law fit to the combined PS1+SHELLQs quasar sample, we calculate a value of $\epsilon_{912}(z\!=\!6) =7.23_{-1.02}^{+1.65}\times 10^{22}\,\text{erg}\text{s}^{-1}\text{Hz}^{-1}\text{cMpc}^{-3}$. The errors reflect the statistical fit uncertainty on the QLF as corresponding to the $16\%$ to $84\%$ percentile range.
Figure\,\ref{fig:ionizing_emissivity} shows our result in comparison with values from the recent literature.
%
Our quasar sample is dominated by the SHELLQs quasar sample at the faint end, which strongly affects the ionizing emissivity. In comparison with other studies of the type-1 UV QLF at $z=6$, this fact largely explains the disagreement with the values of \citet{Jiang2016}, \citet{Parsa2018} and \citet{Kulkarni2019} and the agreement with the work of \citet{Matsuoka2018c}.
Furthermore, our best-fit QLF model has a steeper bright-end slope than all previous measurements, reducing the integrated emissivity of the luminous quasar contribution as well.

In order to calculate the photoionization rate based on our QLF measurement we need to adopt a value for the mean free path of ionizing photons at $z\approx6$. \citet{Becker2021} recently measured the mean free path of ionizing photons and find a value of $l(\nu_{912},z\!=\!6)=0.75_{-0.45}^{+0.65}\,\text{pMpc}$ at $z=6$, which falls below extrapolations from lower redshift. We adopt this value, noting that these measurements are in agreement with the independently calculated lower limits reported by \citet{Bosman2021arxivb}.

With these assumptions we calculate a quasar photoionization rate of $\Gamma_{\text{HI}}(z{=}6) = 5.86_{-3.51}^{+5.08}  {}_{-0.83}^{+1.33}\times10^{-16}\,\text{s}^{-1}$. The first errors reflect the $1\sigma$ uncertainties of the mean free path \citep{Becker2021}, and the second errors are due to the  statistical $1\sigma$ uncertainty in the QLF double power law fit.
The total photoionization rate at $z\approx6$ has been measured to be $\Gamma_{\text{HI}}(z=6)\approx 10^{-13}\text{s}^{-1}$ based on quasar near-zone sizes \citep{Wyithe2011c, Calverley2011} and the mean transmitted Ly$\alpha$ flux from quasar spectra \citep{dAloisio2018, Davies2018a}.
As a combination of our low emissivity values based on our new QLF measurement and the short mean free path of \citet{Becker2021}, the quasar contribution to the photoionization rate is roughly two orders of magnitude lower than the total value. This result strongly disfavors a dominant contribution of quasars to cosmic hydrogen reionization at high redshifts, in line with other recent studies of the high redshift QLF \citep{Jiang2016, Parsa2018, Matsuoka2018c, McGreer2018, Kulkarni2019, WangFeige2019a, JiangLinhua2022}.

\subsection{Quasar lensing and the QLF}
None of the \nqso{} quasars in our sample is known to be gravitationally lensed by a foreground galaxy.
This is a direct consequence of our selection criteria (Section\,\ref{sec:qso_selection}. Our morphology selection criterion aims at selection point sources and thus naturally excludes lensed quasars, which appear extended if they consist of multiple sources images or for which a foreground lens galaxy is detected.
In addition, the required color criteria are designed to select quasars using the Lyman-$\alpha$ break. In the case that a foreground lens galaxy contaminates the bluer bands, the criteria bias against the selection of such sources.

Following the serendipitous discovery of the highest-redshift lensed quasar, J043947.08+163415.7, at $z=6.51$ \citep{Fan2019}, \citet{Pacucci2019} re-evaluated the theoretical consequences of this discovery and conclude that a large fraction of quasars at $z>6$ are missed by current surveys. For a bright end slope of $\beta=-3.6$, the authors expect about half of the $z>6$ population to be lensed.
The steep bright-end slope of $\beta=-3.84$ of our new QLF measurement would lead to an even larger lensed fraction of the $z>6$ quasar population.
In consequence, our bright-end slope measurement would strongly suggest that we are missing a large fraction of lensed quasars.
Given that only one lensed quasar at $z>6$ has been discovered \citep{Fan2019}, our bright-end slope measurement and the resulting lensed fraction according to \citet{Pacucci2019} is in strong tension with observations.

The recent study by \citet{YueMinghao2022} revisits the predicted fraction of high-redshift lensed quasars. By adopting recent galaxy velocity dispersion functions that affect the lensing optical depth, they conclude that the lensed fraction for bright quasars at $z\sim6$ can reach $2\%$ to $6\%$ depending on the QLF bright-end slope.
Following their Figure\,5 and adopting $M_{\rm{lim}}-M^*\approx1\,\rm{mag}$ for our the PS1 quasar selection, results in a lensed fraction of $\lesssim1\%$ for $\beta\approx-4$.
In line with observations \citep{Fan2019}, this result suggests that even with our steep bright-end slope we would have found only $\sim1$ lensed quasar with our selection, hadn't it been biased against these sources.

\section{Conclusions}\label{sec:conclusions}

In this work we present the most precise measurement of the $z\approx6$ QLF at $-28\lesssim M_{1450}\lesssim-22$ to date, based on the combined sample of \nqsobinned{} quasars from the Pan-STARRS\,1 $z>5.6$ quasar survey (PS1) and 48 quasars from SHELLQs. We determine the full PS1 quasar survey completeness taking into account the different components of the PS1 quasar selection strategy and the state of the spectroscopic observations.
We use a maximum likelihood approach (see Appendix\,\ref{app:qlfanalysis}) sampled via MCMC in order to fit a double power law QLF model to the quasar data. Our fiducial model (Table\,\ref{tab:mlfit_results}, first row) is determined on the combined quasar sample and assumes an exponential evolution of the quasar density with $k=-0.7$. The four best fit parameters are $\log(\Phi^*(z=6)/\text{Mpc}^{-3}\text{mag}^{-1})=-8.75_{-0.41}^{+0.47}$, $M^*=-26.38_{-0.60}^{+0.79}\,\text{mag}$, $\alpha=-1.70_{-0.19}^{+0.29}$, and $\beta=-3.84_{-1.21}^{+0.63}$.
The combination of the PS1 and SHELLQs quasar sample constrains the break magnitude to be $\sim1\,\text{mag}$ brighter and the bright-end slope to be significantly steeper than previous studies at this redshift \citep{Willott2010a, Jiang2016, Matsuoka2018c}.

Using our fiducial QLF model we calculate the bright-end quasar density, $n( M_{1450}<-26,z=5.85)=1.16_{-0.12}^{+0.13}\,\text{cGpc}^{-3}$, and put it in perspective with its redshift evolution at $z\approx4-7$. We find that an exponential density evolution model with an exponent of $k=-0.7$, as assumed in our QLF fit, describes the literature data over this redshift range well without the need for an accelerating decline of the quasar density at $z>6.5$ as proposed by \citet{WangFeige2019}.

With our fiducial QLF model we derive the ionizing emissivity of the quasar population and their contribution to cosmic hydrogen reionization.
Using standard assumption, we calculate the ionizing emissivity to be $\epsilon_{912}(z=6) =7.23_{-1.02}^{+1.65}\times 10^{22}\,\text{erg}\text{s}^{-1}\text{Hz}^{-1}\text{cMpc}^{-3}$. This result is lower than some previous results \citep[e.g.,][]{Jiang2016}, but shows good agreement with \citet{Matsuoka2018c}, the most recent estimate of the $z\approx6$ QLF. Adopting the mean free path of \citet{Becker2021}, the only measurement at $z\approx6$, we estimate a HI quasar photoionization rate two order of magnitudes below estimates of its total value, strongly disfavoring quasars as a dominant driver of hydrogen reionization at $z\approx6$. \\


\begin{acknowledgments}
The authors would like to thank the anonymous referee for their insightful comments that helped to improve the manuscript. We would like to further thank Sarah E. I. Bosman, Joseph F. Hennawi, and Romain Meyer for fruitful discussions and comments on the manuscript.
JTS and RN acknowledge funding from the European Research Council (ERC) Advanced Grant program under the European Union’s Horizon 2020 research and innovation programme (Grant agreement No. 885301).
The work of T.C. and D.S. was carried out at the Jet Propulsion Laboratory, California Institute of Technology, under a contract with the National Aeronautics and Space Administration (80NM0018D0004).
T.C.'s research was supported by an appointment to the NASA Postdoctoral Program at the Jet Propulsion Laboratory, California Institute of Technology, administered by Universities Space Research Association under contract with NASA.
EPF is supported by the international Gemini Observatory, a program of NSF’s NOIRLab, which is managed by the Association of Universities for Research in Astronomy (AURA) under a cooperative agreement with the National Science Foundation, on behalf of the Gemini partnership of Argentina, Brazil, Canada, Chile, the Republic of Korea, and the United States of America.
Some of the results in this paper have been derived using the healpy and HEALPix package.
The Pan-STARRS1 Surveys (PS1) and the PS1 public science archive have been made possible through contributions by the Institute for Astronomy, the University of Hawaii, the Pan-STARRS Project Office, the Max-Planck Society and its participating institutes, the Max Planck Institute for Astronomy, Heidelberg and the Max Planck Institute for Extraterrestrial Physics, Garching, The Johns Hopkins University, Durham University, the University of Edinburgh, the Queen's University Belfast, the Harvard-Smithsonian Center for Astrophysics, the Las Cumbres Observatory Global Telescope Network Incorporated, the National Central University of Taiwan, the Space Telescope Science Institute, the National Aeronautics and Space Administration under Grant No. NNX08AR22G issued through the Planetary Science Division of the NASA Science Mission Directorate, the National Science Foundation Grant No. AST-1238877, the University of Maryland, Eotvos Lorand University (ELTE), the Los Alamos National Laboratory, and the Gordon and Betty Moore Foundation.
\end{acknowledgments}

%


\vspace{5mm}
\software{
Astropy \citep{astropy1, astropy2},
chainconsumer \citep{chainconsumer},
Numpy \citep{numpy, numpy2020},
healpy \citep{Zonca2019},
Pandas \citep{pandas_software, pandas_paper},
SciPy \citep{scipy},
emcee \citep{emcee}
}



\appendix

\section{Mathematical formulation of the QLF analysis}\label{app:qlfanalysis}

The QLF $\Phi(M_{1450},z)$ describes the number density of quasars as a function of their absolute magnitude at $1450\,\text{\AA}$ $M_{1450}$ and redshift $z$.
%
To derive the QLF it is imperative to take into account selection effects that are inherent in the parent catalog data and are imposed by the selection criteria for the quasar discovery survey (see Section\,\ref{sec:qso_selection}).
The selection function $S(\vect{q})$ describes the probability of a source with attributes $\vect{q}$ to be within the given sample.
In this section we explicitly derive the mathematical formulation of our MCMC maximum likelihood analysis of the QLF.

\subsection{The quasar incidence as predicted by the QLF}
We begin our discussion by closely following \citet[][their Equation\,1]{Rix2021arxiv} to describe the expected catalog incidence $\dif\Lambda(\vect{q})$ of quasars in our sample through the selection function $S(\vect{q})$ multiplied with a model family for quasars $\mathcal{M}(\vect{q}|\vect{\Theta}_{\text{mod}})$, parameterized by $\vect{\Theta}_{\text{mod}}$:
\begin{gather} \label{eq:incidence_rate}
    \dif\Lambda(\vect{q}) = \mathcal{M}(\vect{q}|\vect{\Theta}_{\text{mod}})\ S(\vect{q}) \dif\vect{q}
\end{gather}
In our case the QLF $\Phi$ forms the basis of the model family. In its most general form the QLF can be written as a function of the luminosity (in our case the absolute magnitude at $1450\text{\AA}$), the three dimensional position of quasars $\vect{x}=(r, \theta, \phi)$ (in spherical coordinates) and the QLF parameters $\vect{\Theta}_{\text{QLF}}$.
The distance to quasars $r$ is not a direct observable. Therefore, it is much more practical (and common) to formulate the quasar luminosity function as a function of redshift $z$. Furthermore, we have good reason to assume that our universe is isotropic on large scales. In this case the QLF is independent of sky position $(\theta, \phi)$. However, the selection function may depend on the sky position and thus we separate the sky dependence from the QLF using the unit normal vector $\hat{n}(\theta, \phi)$.
\begin{gather} \label{eq:model_family}
  \mathcal{M}(\vect{q}|\vect{\Theta}_{\text{mod}}) = \Phi(M_{1450}, z, \theta, \phi | \vect{\Theta}_{\text{QLF}}) = \Phi(M_{1450}, z | \vect{\Theta}_{\text{QLF}})\ \hat{n}(\theta, \phi)
\end{gather}
For clarity of the mathematical expressions we omit the subscript to the absolute magnitude at $1450\text{\AA}$, $M_{1450}$, in the following and simply denote it with $M$.
We now substitute Equation\,\ref{eq:model_family} into Equation\,\ref{eq:incidence_rate} and integrate both sides over volume ($\dif V$) and absolute magnitude ($\dif M$) to retain the total expected number of quasars as observed given the model and the selection function:
\begin{gather} \label{eq:incidence}
    \Lambda(M, z, \theta, \phi) =
    \int_{M}^{M+\Delta M} \int_V \Phi(M, z| \vect{\Theta}_{\text{QLF}})\ \hat{n}(\theta, \phi)\ S(\vect{q}(M, z, \theta, \phi)) \dif{V}(z|\vect{\Theta}_{\text{Cos}}) \dif{M}
\end{gather}
Due to the redshift dependency of the QLF executing the volume integral requires a cosmological model with its own range of parameters $\vect{\Theta}_{\text{Cos}}$. We will now rewrite the volume integral in terms of the differential comoving solid volume element ($\dif{V}/\dif{z}/\dif{\Omega}$), a standard quantity in any cosmological model:
\begin{gather}
    \dif{V}(z|\vect{\Theta}_{\text{Cos}}) =  \frac{\dif{V}}{\dif{z}\dif{\Omega}}(z|\vect{\Theta}_{\text{Cos}})\ \dif{z} \dif{\Omega} \label{eq:difvolelement}
\end{gather}
Substituting Equation\,\ref{eq:difvolelement} in Equation\,\ref{eq:incidence} allows us to separate the volume integration into integrals over redshift and solid angle:
\begin{gather} \label{eq:incidence_2}
   \Lambda(M, z, \theta, \phi) =  \int_{\Omega}
    \int_{M}^{M+\Delta M} \int_{z}^{z+\Delta z} \Phi(M, z| \vect{\Theta}_{\text{QLF}})   \frac{\dif{V}}{\dif{z}\dif{\Omega}}(z|\vect{\Theta}_{\text{Cos}})\  \hat{n}(\theta, \phi)\ S(\vect{q}(M, z, \theta, \phi)) \dif{z} \dif{M}  \dif{\Omega}
\end{gather}
For surveys of inhomogeneous depth it can be often difficult to find an analytic expression for the sky position dependence of the selection function. In this work we take the inhomogeneity into account when modeling the observed quasar properties by sampling from the depth distribution. Therefore, we continue as one would with a survey of homogeneous depth and drop the sky position dependence. Now the integral over they survey footprint simply yields the total footprint area $\Omega$.
\begin{gather} \label{eq:incidence_3}
   \Lambda(M, \Delta M, z, \Delta z) =  \Omega
    \int_{M}^{M+\Delta M} \int_{z}^{z+\Delta z} \Phi(M, z| \vect{\Theta}_{\text{QLF}})   \frac{\dif{V}}{\dif{z}\dif{\Omega}}(z|\vect{\Theta}_{\text{Cos}})\ S(\vect{q}(M, z)) \dif{z} \dif{M}
\end{gather}
Given a model for the quasar luminosity function $\Phi(M, z| \vect{\Theta}_{\text{QLF}})$, a cosmological model $\vect{\Theta}_{\text{Cos}}$ and a model of the observed properties given absolute magnitude and redshift $\vect{q}(M, z)$, we can now calculate the total expected number of observed quasars as a function of absolute magnitude and redshift.

%

\subsection{Formulating the likelihood function}
We derive the likelihood function for the QLF analysis following \citet{Marshall1983} and \citet{Fan2001b}.
The probability of detecting $n_{lm}$ quasars given the QLF $\Phi(M, z| \vect{\Theta}_{\text{QLF}})$ in an absolute magnitude bin $(\Delta M)_l$ and redshift bin $(\Delta z)_m$ can be written in terms of the Poisson distribution function using the incidence rate $\Lambda_{lm}(M,z)$ of Equation\,\ref{eq:incidence_3}:
\begin{gather}
    P(n_{lm}|\Phi(M, z| \vect{\Theta}_{\text{QLF}})) = \frac{\Lambda_{lm}(M,z)^{n_{lm}} e^{-\Lambda_{lm}(M,z)}}{n_{lm}!}
\end{gather}
The probability of finding $N(M, z)$ quasars in the entire survey, as characterized by the full absolute magnitude and redshift range, can then be written as the product of the probabilities over all absolute magnitude and redshift bins.
\begin{equation}\label{eq:probability_n}
\begin{split}
    P(N(M, z)|\Phi(M, z| \vect{\Theta}_{\text{QLF}})) & = \prod_{lm} \frac{\Lambda_{lm}(M,z)^{n_{lm}} e^{-\Lambda_{lm}(M,z)}}{n_{lm}!} \\
    & = \prod_{lm}^{n_{lm}=1} \Lambda_{lm}(M,z) e^{-\Lambda_{lm}(M,z)} \times \prod_{lm}^{n_{lm}=0} e^{-\Lambda_{lm}(M,z)}\\
    & = \prod_{lm}^{n_{lm}=1} \Lambda_{lm}(M,z) \times \prod_{lm} e^{-\Lambda_{lm}(M,z)}
\end{split}
\end{equation}
If the absolute magnitude and redshift bins are infitesimally small, then either $n_{lm}=1$ or $n_{lm}=0$ quasars can be found in each bin. We split the product into two terms for these two cases, simplifying the equation. We can furthermore, rearrange the terms to arrive at the final version of the probability $P(N(M, z)|\Phi(M, z| \vect{\Theta}_{\text{QLF}}))$ in Equation\,\ref{eq:probability_n}.
Our main goal is to constrain the QLF $\Phi(M, z| \vect{\Theta}_{\text{QLF}})$ and its parameters $\vect{\Theta}_{\text{QLF}}$ based on the observed distribution of high redshift quasars $N(M, z)$.
We are basically asking what the probability of $\Phi(M, z| \vect{\Theta}_{\text{QLF}})$ is given $N(M, z)$ observed quasars in the interval $\Delta M \Delta z$ and in the survey area $\Omega$.
\begin{gather}
    P(\Phi(M, z| \vect{\Theta}_{\text{QLF}})|N(M, z)) = P(N(M, z)|\Phi(M, z| \vect{\Theta}_{\text{QLF}})) \frac{P(\Phi(M, z| \vect{\Theta}_{\text{QLF}}))}{P(N(M, z))}
\end{gather}
Assuming flat priors for $N(M, z)$ and $\Phi(M, z| \vect{\Theta}_{\text{QLF}})$, we can simplify the probability of $\Phi(M, z| \vect{\Theta}_{\text{QLF}})$ given $N(M, z)$ to
\begin{gather}
    P(\Phi(M, z| \vect{\Theta}_{\text{QLF}})|N(M, z)) \propto P(N(M, z)|\Phi(M, z| \vect{\Theta}_{\text{QLF}}))
\end{gather}
We have already discussed how to express the second term in this equation via the incidence rate in Equation\,\ref{eq:probability_n}. We now formulate the logarithmic probability equivalent to the logarithmic likelihood:
\begin{equation}\label{eq:log_probability}
\begin{split}
    \ln\left[P(\Phi(M, z| \vect{\Theta}_{\text{QLF}})|N(M, z))\right] & \propto \ln \left[P(N(M, z)|\Phi(M, z| \vect{\Theta}_{\text{QLF}}))\right] \\
    & \propto \sum_{lm}^{n_{lm}=1} \ln \left[\Lambda_{lm}(M,z)\right] - \sum_{lm} \Lambda_{lm}(M,z) \ .
\end{split}
\end{equation}

To evaluate this equation further we will take a look at the incidence rate $\Lambda_{lm, j}(M,z)$ for a single quasar $j$. The quasar $j$ has an absolute magnitude $M_j$ and redshift $z_j$ in the bin centers of width $(\Delta M)_l$ and $(\Delta z)_m$. With these boundary conditions we can write Equation\,\ref{eq:incidence_3} for a single quasar as
\begin{gather}
     \Lambda_{lm,j}(M, z) =  \int_{M_j-(\Delta M)_l/2}^{M_j+(\Delta M)_l/2} \int_{z_j-(\Delta z)_m/2}^{z_j+(\Delta z)_m/2} \Phi(M, z| \vect{\Theta}_{\text{QLF}})   \frac{\dif{V}}{\dif{z}\dif{\Omega}}(z|\vect{\Theta}_{\text{Cos}})\ S(\vect{q}(M, z))\ \dif{z} \dif{M}\ \Omega
\end{gather}
In the limit of infinitesimal bin sizes around $M_j$ and $z_j$ the integrals can be trivially evaluated and we can drop the indices $l$ and $m$:
\begin{gather} \label{eq:single_qso_incidence}
     \Lambda_{j}(M, z) =\Phi(M_j, z_j| \vect{\Theta}_{\text{QLF}})   \frac{\dif{V}}{\dif{z}\dif{\Omega}}(z_j|\vect{\Theta}_{\text{Cos}})\ S_j(\vect{q}(M_j, z_j))\ (\Delta z)_m (\Delta M)_l\ \Omega \ .
\end{gather}
Starting from the first term of the right hand side of Equation\,\ref{eq:log_probability} we first rewrite the sum over all bins $l$ and $m$ for which $n_{lm}=1$ as a sum over all $N(M,z)$ quasars in the data set. Then we apply the natural logarithm to Equation\,\ref{eq:single_qso_incidence} and equivalent to \citet{Marshall1983} drop all terms independent of the QLF and the selection function, which would only add constant values to the logarithmic probability.
\begin{gather}
    \sum_{lm}^{n_{lm}=1} \ln \left[\Lambda_{lm}(M,z)\right] = \sum_{j=1}^{N(M,z)} \ln \left[\Lambda_{j}(M,z)\right] \propto  \sum_{j=1}^{N(M,z)} \ln \left[\Phi(M_j, z_j| \vect{\Theta}_{\text{QLF}}) S_j(\vect{q}(M_j, z_j))\right]
\end{gather}
The second term in Equation\,\ref{eq:log_probability} normalizes the logarithmic probability by summing over the full interval in absolute magnitude $\Delta M$ and redshift $\Delta z$ for which we aim to evaluate the QLF. By choosing the appropriate integration boundaries for the QLF evaluation in Equation\,\ref{eq:incidence_3}, we can simplify the expression to
\begin{gather}
    \sum_{lm} \Lambda_{lm}(M,z) = \Lambda(M,z) \ .
\end{gather}
The logarithmic probability of $\Phi(M, z| \vect{\Theta}_{\text{QLF}})$ given $N(M, z)$, our likelihood function, can then be approximated by
\begin{gather}
    \boxed{\ln\left[P(\Phi(M, z| \vect{\Theta}_{\text{QLF}})|N(M, z))\right] \propto \sum_{j=1}^{N(M,z)} \ln \left[\Phi(M_j, z_j| \vect{\Theta}_{\text{QLF}}) S_j(\vect{q}(M_j, z_j))\right] - \Lambda(M,z)} \label{eq:qlflogprob}
\end{gather}
with $\Lambda(M,z)$ as in Equation\,\ref{eq:incidence_3}.



\section{Extended discussion on the quasar model}\label{app:qsomodel}

%

To test the robustness of our \texttt{simqso} quasar model we compare its synthetic photometry to the recently  published \texttt{qsogen} model by \citet{Temple2021}. The latter has been used in the prediction of quasar yields for the Euclid mission \citep{EuclidPrepV2019}.
At the core of \texttt{qsogen} is a parametric quasar spectral model, which has been fit to optical to near-infrared colors of a subsample of SDSS DR16 quasars \citep{Lyke2020}. Equivalent to the \texttt{simqso} quasar model, the \texttt{qsogen} quasar models is explicitly designed to reproduce the median quasar colors of real sources.
A particular novel feature of \texttt{qsogen} are the emission line quasar templates, that allow to capture the diversity in quasar line strength and velocity shifts and can be modified via the \texttt{emline\_type} argument, which has a default value of $-0.9936$.

\begin{figure}
    \centering
    \includegraphics[width=\textwidth]{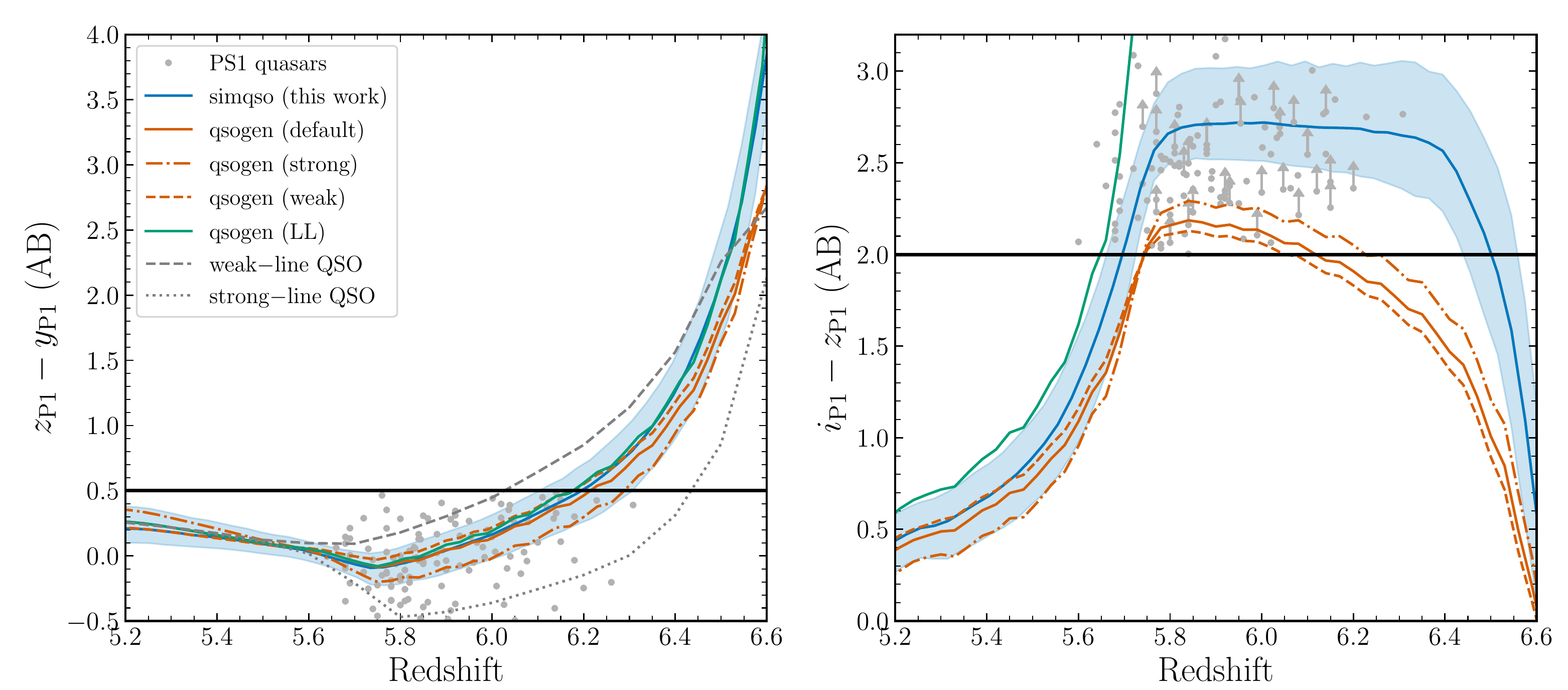}
    \caption{\textit{Left panel:} Synthetic quasar $z_{\rm{P1}}-y_{\rm{P1}}$ color as a function of redshift. The median quasar color from the \texttt{simqso} sample used to derive the selection function is shown in blue with a blue shading indicating the $5\%$ to $95\%$ percentile range. Quasar colors modeled with \texttt{qsogen} \citep{Temple2021} are shown in orange for the default model (solid line), a model with weak line emission (dashed line), and a model with strong line emission (dashed-dotted line). The green line depicts an alternative to the \texttt{qsogen} default model, where a Lyman limit system at the redshift of the quasar absorbs all flux blueward of $1215\textrm{\AA}$, mimicking full Gunn-Peterson absorption blueward of the Lyman-$\alpha$ line. PS1 dereddened colors of the PS1 quasar sample are shown as grey points. The black solid line shows the dividing line for the $z_{\rm{P1}}-y_{\rm{P1}} < 0.5$ color criterion.
    In addition, we show the color redshift evolution of a weak-line and a strong-line quasar template from \citet[][blue and red spectra in their Figure\,10]{Banados2016}. At $z\gtrsim6.3$ the color is biased blue by the non-evolving IGM transmission in the templates.
    \textit{Right panel:} Synthetic quasar $i_{\rm{P1}}-z_{\rm{P1}}$ color as a function of redshift. The symbols are the same with the addition of the black solid line that depicts the $i_{\rm{P1}}-z_{\rm{P1}} > 2$ color criterion used in this work.
    }
    \label{fig:qsogen_simqso}
\end{figure}

The quasar selection criteria on which our new QLF measurements is based rely on color cuts in the reddest PS1 filter bands, $i_{\rm{P1}}$, $z_{\rm{P1}}$, $y_{\rm{P1}}$. We use \texttt{qsogen} to calculate the magnitudes in these bands and then compare the resulting $i_{\rm{P1}}-z_{\rm{P1}}$, and $z_{\rm{P1}}-y_{\rm{P1}}$ colors between the \texttt{qsogen} and our quasar model based on \texttt{simqso}.
To capture the quasar diversity in terms of emission line properties with \texttt{qsogen}, we produce three models: the \textit{default} model uses the default value of \texttt{emline\_type}. The \textit{strong} emission line model uses \texttt{emline\_type}$=1$ and the \textit{weak} emission line model uses \texttt{emline\_type}$=-1$.
We compare the resulting colors as a function of redshift in Figure\,\ref{fig:qsogen_simqso}.
The $z_{\rm{P1}}-y_{\rm{P1}}$ color of  our quasar model and the \texttt{qsogen} models is consistent over $z=5.2-5.8$.
The diversity in quasar colors due to the different \texttt{qsogen} emission line models is well captured by the random sampling of continuum and emission line properties used in \texttt{simqso} (Section\,\ref{sec:simqso}).
Small differences the $z_{\rm{P1}}-y_{\rm{P1}}$ color appear at $z>5.8$, which increase towards $z=6.6$. We attribute the difference to the different prescriptions of absorption of neutral hydrogen due to the IGM, which become more pronounced at $z>5.8$ as the Lyman-$\alpha$ line moves to the red side of the $z_{\rm{P1}}$ filter band.
The selection of $z>5.7$ quasars is based on a color dropout selection requiring $i_{\rm{P1}}-z_{\rm{P1}} > 2$. The right panel of Figure\,\ref{fig:qsogen_simqso} shows $i_{\rm{P1}}-z_{\rm{P1}}$ as a function of redshift and highlights the dropout criterion with the black solid line. Quasars with colors above the black line would be selected.
With our \texttt{simqso} quasar model we start to select quasars at $z>5.6$ (blue solid line and blue shaded region). The downturn of the $i_{\rm{P1}}-z_{\rm{P1}}$ color at $z\gtrsim6.4$ is due to the Lyman-$\alpha$ line moving out of the $z_{\rm{P1}}$ filter band. As a consequence little to no quasar flux is measured in both the $i_{\rm{P1}}$- and $z_{\rm{P1}}$ band.
Our \texttt{simqso} quasar model mostly agrees with the \texttt{qsogen} models at $z\lesssim 5.6$. However, once the Lyman-$\alpha$ line leaves the $i_{\rm{P1}}$ band ($z\gtrsim5.6$) color differences become evident. We argue that these are mainly caused by the different prescriptions of IGM absorption in the spectral models. While \texttt{simqso} uses the stochastic Lyman-$\alpha$ forest model of \citet{McGreer2013} (see Section\,\ref{sec:simqso}), \texttt{qsogen} uses a prescription based on \citet{Becker2013}. Compared to our \texttt{simqso} model, the \texttt{qsogen} spectra have more flux blueward of Lyman-$\alpha$, resulting in a generally bluer color. When compared to quasars selected in this work (grey points) the \texttt{simqso} IGM model provides a better representation of the empirical data points\footnote{One might be concerned that the \texttt{simqso} synthethic colors do not encompass the full $i_{\rm{P1}}-z_{\rm{P1}}$ color scatter of PS1 quasars. We note that the synthethic color range is broadened significantly by the the PS1 photometric uncertainties (Section\,\ref{sec:modelps1phot}) before the selection function is assessed}. The \texttt{qsogen} models show significantly bluer $i_{\rm{P1}}-z_{\rm{P1}}$, especially at $z>5.6$.
Introducing a Lyman limit system at the redshift of the quasar (i.e. at rest-frame wavelength of $1215\textrm{\AA}$) to the \texttt{qsogen} default model, effectively reduces all blueward flux to zero. At $z\lesssim 6.0$ this approach is unphysical as the flux blueward of Lyman-$\alpha$ can be transmitted in ionized patches of the IGM. However, it is valid assumption at higher redshifts  \citep[see e.g. at $z>7$,][]{EuclidPrepV2019}.
We show this \texttt{qsogen} Lyman-limit model (\textit{LL}) as the green line in Figure\,\ref{fig:qsogen_simqso}. The \texttt{qsogen} \textit{LL} model has a significantly redder $i_{\rm{P1}}-z_{\rm{P1}}$ color at $z<5.7$, which diverges further at $z\approx5.7$. Compared to the PS1 quasar photometry at $z\approx5.6-6.4$ it is not a good representation of the data. However, the resulting $z_{\rm{P1}}-y_{\rm{P1}}$ color of our quasar model and \texttt{qsogen} is now fully consistent over the entire redshift range.
This agreement in the $z_{\rm{P1}}-y_{\rm{P1}}$ color highlights that, barring the differences in the prescription of IGM absorption, both our \texttt{simqso} quasar model and the \texttt{qsogen} model both produce consistent median colors as expected from their design goals.

\section{Supplemental figures}\label{app:suppfigures}
\begin{figure*}
\centering
\includegraphics[width=0.49\textwidth]{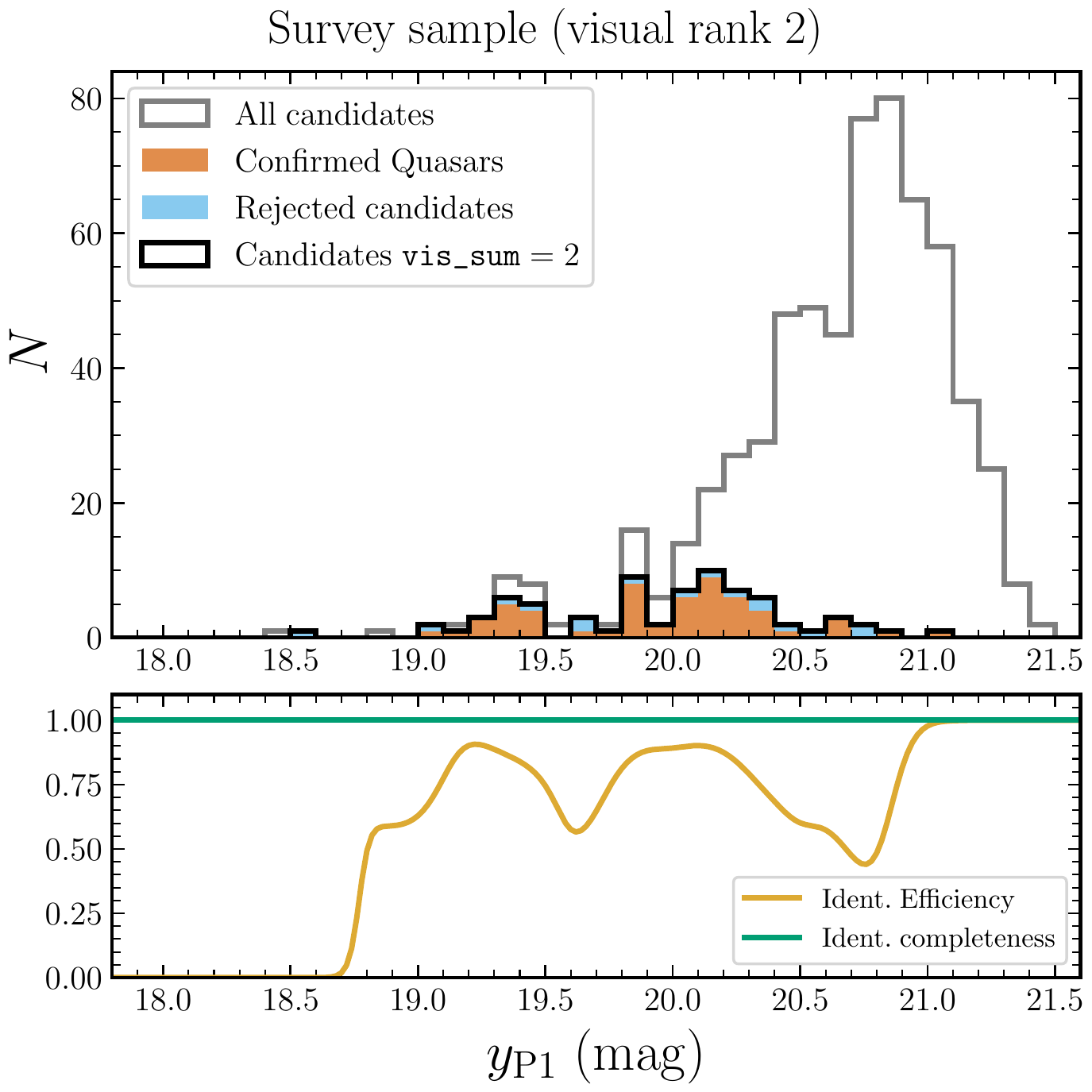}
\includegraphics[width=0.49\textwidth]{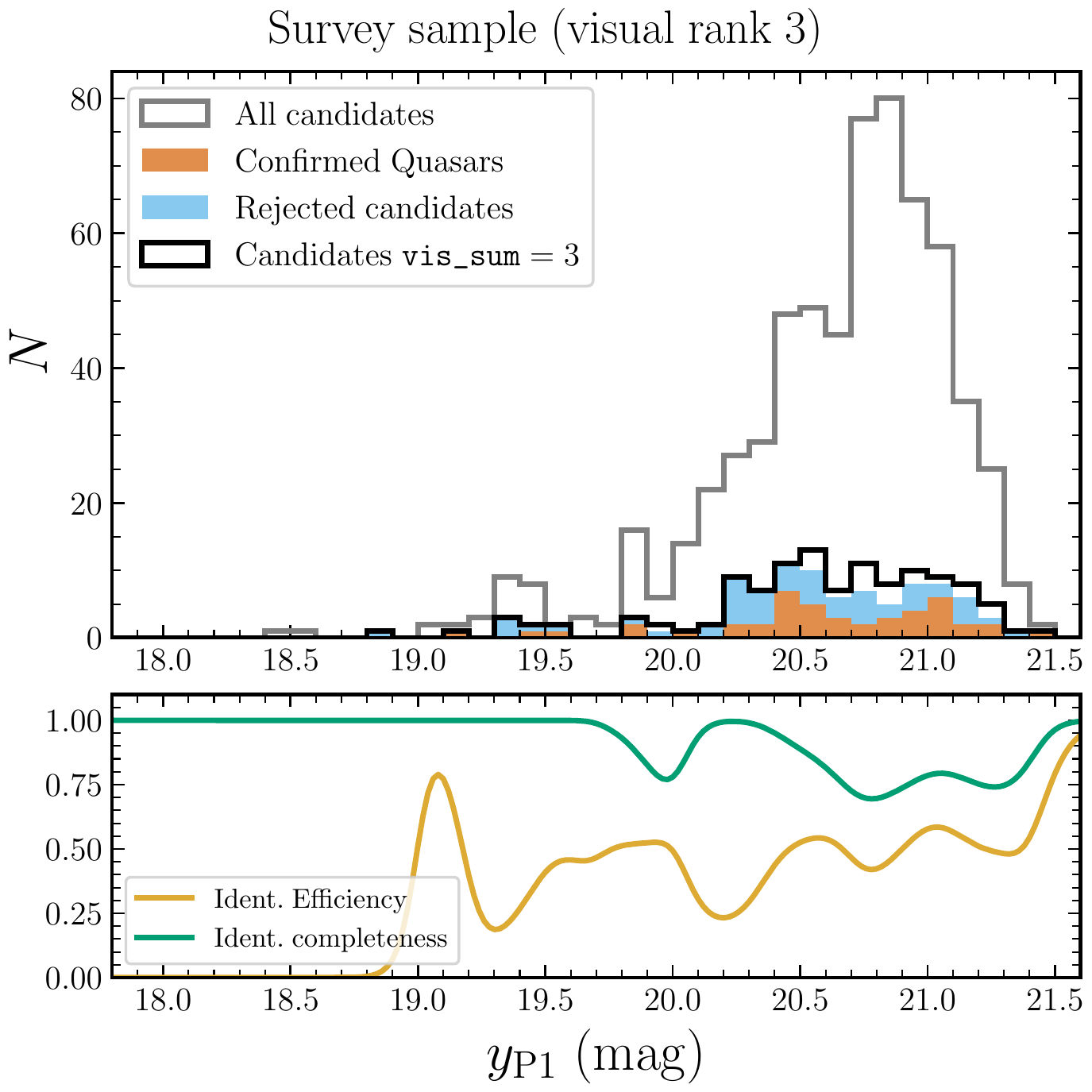}
\includegraphics[width=0.49\textwidth]{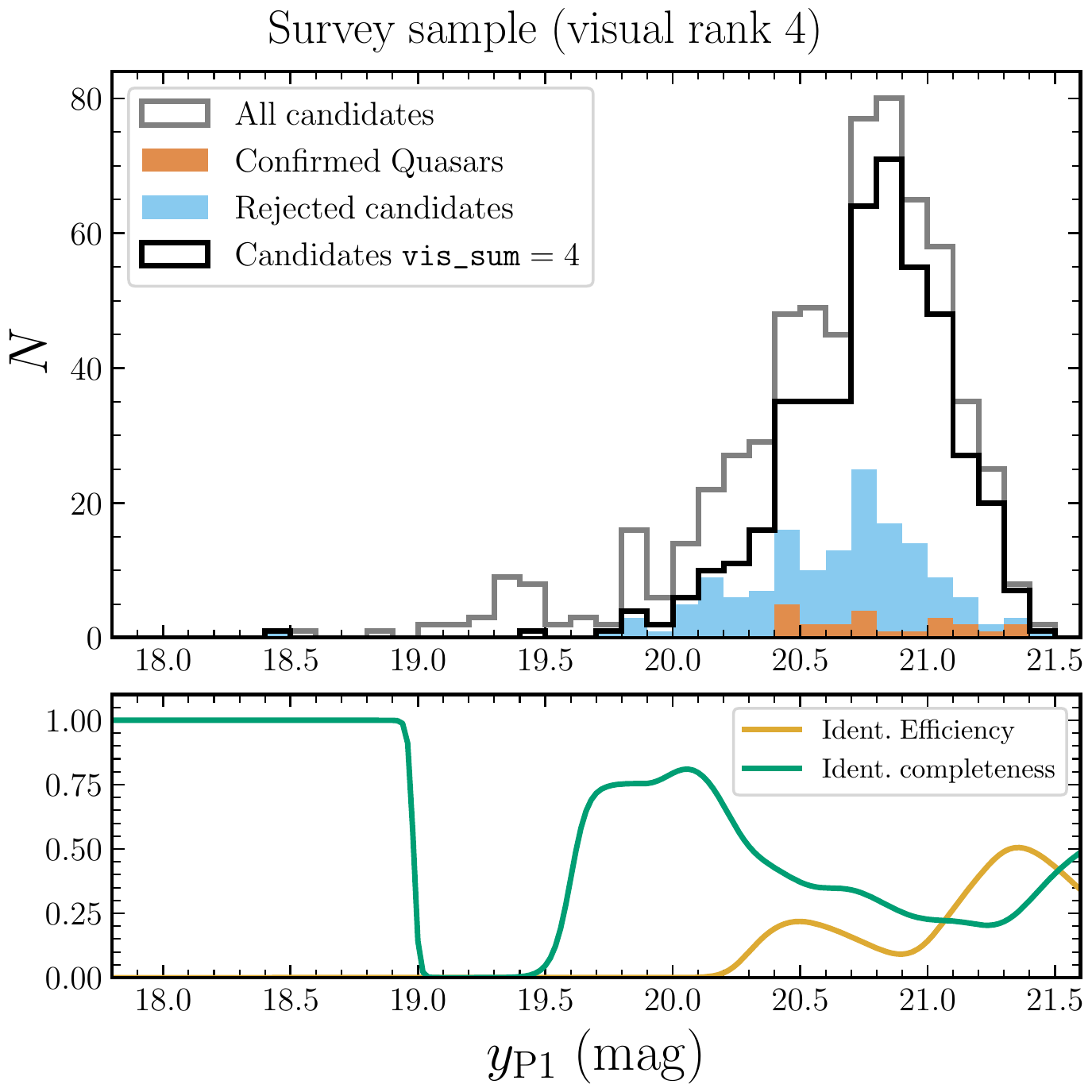}
\caption{In these three panels we show the observed survey samples, their selection efficiency and identification completeness for different visual ranks. Comparing the samples highlights that the majority of remaining candidates have the worst visual rank (\texttt{vis\_sum}$=4$) and the lowest selection efficiency. Our identification completeness (Equation\,\ref{eq:completeness2}) takes these differences into account.}
\label{fig:spec_completeness_ranks}
\end{figure*}

\begin{figure*}
    \centering
    \includegraphics[width=0.8\textwidth]{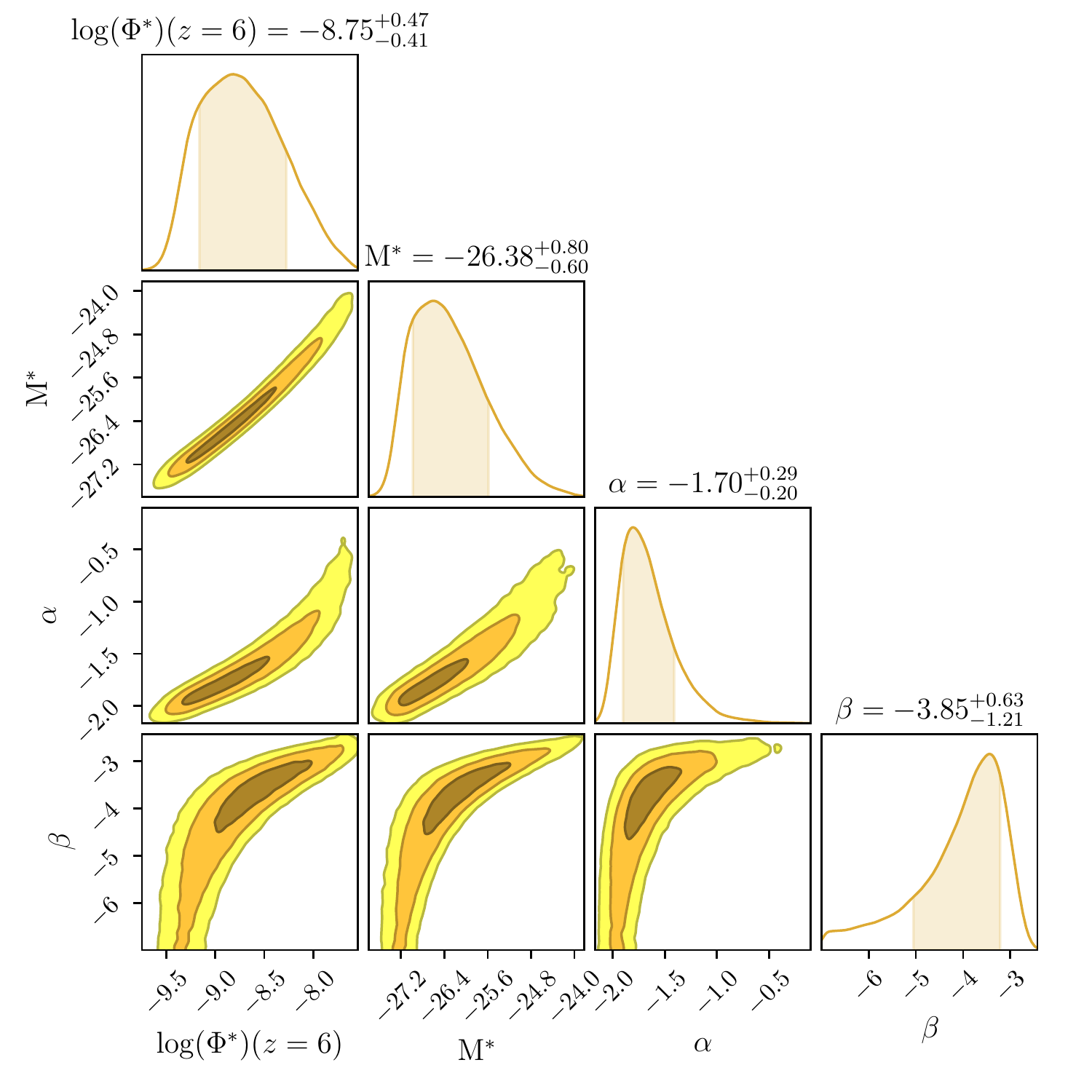}
    \caption{Covariance matrix of the fiducial double power law QLF fit to the combined PS1 + SHELLQs quasar sample (Table\,\ref{tab:mlfit_results}, first row). The contours highlight the $1\sigma,2\sigma$, and $3\sigma$ confidence bounds.}
    \label{fig:mlfit_covariance}
\end{figure*}

\begin{figure}
    \centering
    \includegraphics[width=0.8\textwidth]{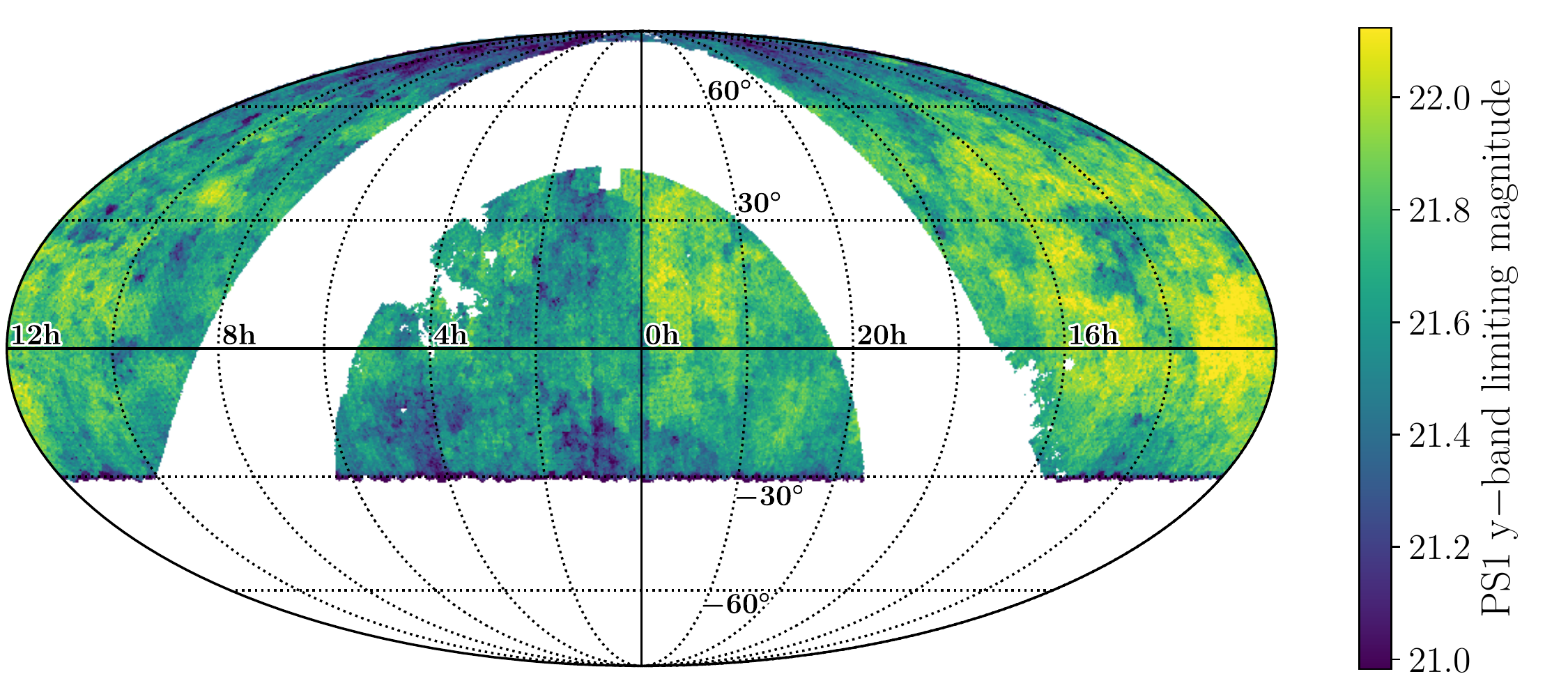}
    \caption{The PS1 \textit{y}--band limiting magnitude for our quasar selection survey area. The Figure details are analogous to Figure\,\ref{fig:zbanddepth} only for the PS1 \textit{y}--band.}
    \label{fig:ybanddepth}
\end{figure}

We provide a number of figures that provide supplemental information for a few sections of the paper. Figure\,\ref{fig:spec_completeness_ranks} shows the spectroscopic identification completeness for quasar samples with visual ranks 2,3, and 4 separately. For context of the covariance of the different QLF parameters in our ML fit we provide the full covariance matrix in Figure\,\ref{fig:mlfit_covariance}.
Analogous to Figure\,\ref{fig:zbanddepth}, we also show the PS1 \textit{y}--band limiting magnitude as a function of survey area in Figure\,\ref{fig:ybanddepth}.

\clearpage

\section{The PS1 quasar sample data table}
We present the full PS1 quasar sample in Table\,\ref{tab:qsos} below. The columns include the official PS1 designation, the dereddened PS1 \textit{y}-band magnitudes, the spectroscopic redshift, calculated absolute magnitude $ M_{1450}$, the discovery references and the reference for the redshift measurement.

\startlongtable
\begin{deluxetable*}{cccccc}
\tabletypesize{\footnotesize} 
\tablecaption{Quasars and their properties used in the PS1 QLF analysis \label{tab:qsos}}
\tablehead{\colhead{PS1 designation} &\colhead{$y_{\rm{P1}}$} &\colhead{$z$} &\colhead{$M_{1450}$} &\colhead{Ref} &\colhead{$z$ Ref} \\ 
\nocolhead{} &\colhead{(mag)} &\nocolhead{} &\colhead{(mag)} &\nocolhead{} &\nocolhead{}  
} 
\startdata 
$\text{PSOJ157.61297+05.41528}$ & $19.86$ & $6.31$ & $-26.80$ & \citet{Fan2001c} & \citet{Kurk2007}  \\ 
$\text{PSOJ245.88256+31.20014}$ & $20.25$ & $6.26$ & $-26.43$ & \citet{Fan2004} & \citet{WangRan2011b}  \\ 
$\text{PSOJ162.18777+46.62181}$ & $19.67$ & $6.23$ & $-27.01$ & \citet{Fan2003} & \citet{Carilli2010}  \\ 
$\text{PSOJ184.33893+01.52846}$ & $21.47$ & $6.20$ & $-25.19$ & \citet{Banados2016},\citet{WangFeige2017} &\citet{Banados2016}  \\ 
$\text{PSOJ060.55290+24.85678}$ & $19.92$ & $6.18$ & $-26.74$ & \citet{Banados2016} & \citet{Banados2016}  \\ 
$\text{PSOJ281.33614+53.76314}$ & $20.21$ & $6.18$ & $-26.44$ & \citet{Banados2022} & \citet{Banados2022}  \\ 
$\text{PSOJ359.13521-06.38312}$ & $20.07$ & $6.17$ & $-26.59$ & \citet{Banados2016},\citet{WangRan2016} &\citet{Eilers2020a_Err}  \\ 
$\text{PSOJ242.40529+30.69660}$ & $20.46$ & $6.16$ & $-26.19$ & Warren, S. J. et al., in prep.,\citet{Jiang2016} & Warren, S. J. et al., in prep.  \\ 
$\text{PSOJ130.76568+29.18709}$ & $21.00$ & $6.15$ & $-25.64$ & \citet{WangFeige2019} & \citet{WangFeige2019}  \\ 
$\text{PSOJ334.01815-05.00488}$ & $21.09$ & $6.15$ & $-25.56$ & \citet{Banados2022} & \citet{Banados2022}  \\ 
$\text{PSOJ127.05586+26.56541}$ & $20.42$ & $6.14$ & $-26.22$ & Warren, S. J. et al., in prep.,\citet{Banados2016} &\citet{Banados2022}  \\ 
$\text{PSOJ218.77147+04.81896}$ & $20.47$ & $6.14$ & $-26.18$ & \citet{Banados2022} & \citet{Banados2022}  \\ 
$\text{PSOJ192.71635+31.50608}$ & $20.33$ & $6.14$ & $-26.30$ & \citet{Fan2006} & \citet{Shen2019a}  \\ 
$\text{PSOJ199.79708+09.84763}$ & $19.82$ & $6.13$ & $-26.82$ & \citet{Mortlock2009} & \citet{Venemans2020}  \\ 
$\text{PSOJ065.50414-19.45796}$ & $20.27$ & $6.12$ & $-26.36$ & \citet{Banados2016} & \citet{Decarli2018}  \\ 
$\text{PSOJ160.32976+20.13997}$ & $20.66$ & $6.12$ & $-25.97$ & \citet{WangFeige2018} & \citet{WangFeige2019}  \\ 
$\text{PSOJ239.71246-07.40266}$ & $19.33$ & $6.11$ & $-27.30$ & \citet{Banados2016} & \citet{Eilers2020a_Err}  \\ 
$\text{PSOJ002.37869+32.87025}$ & $21.10$ & $6.10$ & $-25.52$ & \citet{Banados2016} & \citet{Banados2016}  \\ 
$\text{PSOJ315.22789-17.25606}$ & $21.23$ & $6.08$ & $-25.39$ & \citet{Willott2010a} & \citet{Venemans2020}  \\ 
$\text{PSOJ240.72490+42.47359}$ & $19.82$ & $6.08$ & $-26.80$ & \citet{Fan2004} & \citet{Shen2019a}  \\ 
$\text{PSOJ130.62263+12.31404}$ & $19.86$ & $6.08$ & $-26.75$ & \citet{DeRosa2011},\citet{Jiang2015} &\citet{Decarli2018}  \\ 
$\text{PSOJ239.78789+22.20398}$ & $20.88$ & $6.07$ & $-25.73$ & \citet{WangFeige2017} & \citet{WangFeige2017}  \\ 
$\text{PSOJ247.64126+40.20269}$ & $20.62$ & $6.06$ & $-25.98$ & \citet{Fan2003} & \citet{Shen2019a}  \\ 
$\text{PSOJ293.03178+71.65233}$ & $20.11$ & $6.05$ & $-26.50$ & \citet{Banados2016} & \citet{Banados2016}  \\ 
$\text{PSOJ250.34052+37.92226}$ & $21.09$ & $6.05$ & $-25.51$ & \citet{Willott2007} & \citet{Willott2010a}  \\ 
$\text{PSOJ210.72777+40.40088}$ & $21.00$ & $6.04$ & $-25.59$ & \citet{Banados2016} & \citet{Banados2016}  \\ 
$\text{PSOJ313.52710-00.08733}$ & $20.57$ & $6.04$ & $-26.03$ & \citet{Jiang2008} & \citet{Venemans2020}  \\ 
$\text{PSOJ181.90597+06.50283}$ & $20.15$ & $6.04$ & $-26.46$ & \citet{Jiang2015} & \citet{Decarli2018}  \\ 
$\text{PSOJ196.53444+03.94065}$ & $20.07$ & $6.03$ & $-26.53$ & \citet{Fan2001c} & \citet{Venemans2020}  \\ 
$\text{PSOJ265.92982+41.41395}$ & $21.32$ & $6.03$ & $-25.27$ & \citet{Banados2022} & \citet{Eilers2020a}  \\ 
$\text{PSOJ124.61417+17.38111}$ & $19.26$ & $6.02$ & $-27.34$ & \citet{Fan2006} & \citet{Carilli2010}  \\ 
$\text{PSOJ194.48949+63.82699}$ & $20.42$ & $6.02$ & $-26.18$ & \citet{Jiang2015} & \citet{Jiang2015}  \\ 
$\text{PSOJ340.20404-18.66219}$ & $20.39$ & $6.01$ & $-26.21$ & \citet{Banados2014} & \citet{Banados2016}  \\ 
$\text{PSOJ174.32386+35.83245}$ & $19.41$ & $6.01$ & $-27.18$ & \citet{Fan2006} & \citet{Shen2019a}  \\ 
$\text{PSOJ007.02733+04.95712}$ & $20.15$ & $6.00$ & $-26.43$ & \citet{Banados2014},\citet{Jiang2015} &\citet{Venemans2020}  \\ 
$\text{PSOJ037.97064-28.83892}$ & $20.58$ & $6.00$ & $-26.01$ & \citet{Banados2014} & \citet{Banados2016}  \\ 
$\text{PSOJ182.31219+53.46335}$ & $21.05$ & $5.99$ & $-25.53$ & \citet{Banados2022} & \citet{Banados2022}  \\ 
$\text{PSOJ029.51725-29.08868}$ & $19.37$ & $5.98$ & $-27.21$ & \citet{Banados2016} & \citet{Banados2022}  \\ 
$\text{PSOJ027.15681+06.00556}$ & $19.35$ & $5.98$ & $-27.23$ & \citet{Jiang2015} & \citet{Becker2015}  \\ 
$\text{PSOJ056.71684-16.47693}$ & $20.07$ & $5.97$ & $-26.50$ & \citet{Banados2016} & \citet{Eilers2020a}  \\ 
$\text{PSOJ288.64769+63.24792}$ & $20.76$ & $5.96$ & $-25.81$ & \citet{Banados2022} & \citet{Banados2022}  \\ 
$\text{PSOJ130.33131+29.08460}$ & $20.19$ & $5.95$ & $-26.37$ & \citet{Goto2006} & \citet{Shen2019a}  \\ 
$\text{PSOJ167.47265+56.95211}$ & $20.62$ & $5.95$ & $-25.95$ & \citet{Banados2016} & \citet{Banados2016}  \\ 
$\text{PSOJ267.00210+22.78120}$ & $21.04$ & $5.95$ & $-25.52$ & \citet{Banados2016} & \citet{Banados2016}  \\ 
$\text{PSOJ002.10738-06.43457}$ & $20.29$ & $5.93$ & $-26.26$ & \citet{Banados2016},\citet{Jiang2015} &\citet{Banados2016}  \\ 
$\text{PSOJ076.23444-10.88780}$ & $20.61$ & $5.93$ & $-25.95$ & \citet{Banados2022} & \citet{Banados2022}  \\ 
$\text{PSOJ139.61940+19.67919}$ & $20.71$ & $5.92$ & $-25.83$ & \citet{WangFeige2018} & \citet{WangFeige2018}  \\ 
$\text{PSOJ213.36298-22.56173}$ & $19.89$ & $5.92$ & $-26.65$ & \citet{Banados2014} & \citet{Banados2016}  \\ 
$\text{PSOJ228.68712+21.23882}$ & $20.60$ & $5.92$ & $-25.95$ & \citet{Banados2016} & \citet{Banados2016}  \\ 
$\text{PSOJ328.73399-09.50762}$ & $20.46$ & $5.92$ & $-26.08$ & \citet{Banados2016} & \citet{Banados2016}  \\ 
$\text{PSOJ135.87045-13.83368}$ & $21.01$ & $5.91$ & $-25.53$ & \citet{Banados2016} & \citet{Banados2016}  \\ 
$\text{PSOJ194.12902+25.54761}$ & $21.07$ & $5.91$ & $-25.47$ & \citet{Banados2016} & \citet{Banados2016}  \\ 
$\text{PSOJ218.39674+28.33067}$ & $20.45$ & $5.91$ & $-26.09$ & \citet{Banados2022} & \citet{Banados2022}  \\ 
$\text{PSOJ212.79703+12.29368}$ & $20.10$ & $5.90$ & $-26.44$ & \citet{Fan2004} & \citet{Kurk2007}  \\ 
$\text{PSOJ023.00711-02.26753}$ & $20.20$ & $5.90$ & $-26.35$ & \citet{Banados2016} & \citet{Banados2016}  \\ 
$\text{PSOJ319.60403-10.93263}$ & $20.05$ & $5.90$ & $-26.49$ & \citet{Banados2016} & \citet{Banados2016}  \\ 
$\text{PSOJ216.31805+32.90265}$ & $20.37$ & $5.89$ & $-26.16$ & \citet{Cool2006} & \citet{Carilli2010}  \\ 
$\text{PSOJ042.66908-02.91745}$ & $20.02$ & $5.89$ & $-26.51$ & \citet{Banados2016} & \citet{Banados2016}  \\ 
$\text{PSOJ187.30502+04.32436}$ & $21.09$ & $5.89$ & $-25.45$ & \citet{Banados2014} & \citet{Banados2014}  \\ 
$\text{PSOJ075.93563-07.50613}$ & $20.13$ & $5.88$ & $-26.40$ & \citet{Banados2016} & \citet{Banados2016}  \\ 
$\text{PSOJ157.90703-02.65990}$ & $20.26$ & $5.88$ & $-26.27$ & \citet{Banados2016} & \citet{Banados2016}  \\ 
$\text{PSOJ210.82969+09.04750}$ & $20.33$ & $5.88$ & $-26.20$ & \citet{Banados2016},\citet{Jiang2015} &\citet{Banados2016}  \\ 
$\text{PSOJ175.42940+71.32363}$ & $20.62$ & $5.86$ & $-25.90$ & \citet{Banados2022} & \citet{Banados2022}  \\ 
$\text{PSOJ183.29919-12.76762}$ & $19.22$ & $5.86$ & $-27.30$ & \citet{Banados2014} & \citet{Banados2014}  \\ 
$\text{PSOJ025.23764-11.68319}$ & $19.86$ & $5.85$ & $-26.65$ & \citet{Banados2016} & \citet{Banados2016}  \\ 
$\text{PSOJ127.28174+03.06571}$ & $20.73$ & $5.85$ & $-25.78$ & \citet{Banados2016} & \citet{Banados2016}  \\ 
$\text{PSOJ190.92005+25.48995}$ & $20.61$ & $5.85$ & $-25.90$ & \citet{Banados2014},\citet{Jiang2016} &\citet{Jiang2016}  \\ 
$\text{PSOJ219.04890+50.11865}$ & $20.20$ & $5.85$ & $-26.32$ & \citet{Fan2006} & \citet{Carilli2010}  \\ 
$\text{PSOJ130.14626+56.40561}$ & $19.56$ & $5.84$ & $-26.95$ & \citet{Fan2006} & \citet{WangRan2010}  \\ 
$\text{PSOJ001.46807-00.11546}$ & $20.91$ & $5.84$ & $-25.59$ & \citet{Fan2004} & \citet{DeRosa2011}  \\ 
$\text{PSOJ148.48293+69.18128}$ & $20.23$ & $5.84$ & $-26.28$ & \citet{Banados2022} & \citet{Banados2022}  \\ 
$\text{PSOJ197.71978+25.53518}$ & $20.76$ & $5.84$ & $-25.74$ & \citet{Banados2016} & \citet{Banados2016}  \\ 
$\text{PSOJ210.87224-12.00948}$ & $21.11$ & $5.84$ & $-25.40$ & \citet{Banados2014} & \citet{Banados2014}  \\ 
$\text{PSOJ306.35128-04.82274}$ & $20.57$ & $5.84$ & $-25.94$ & \citet{Banados2022} & \citet{Banados2022}  \\ 
$\text{PSOJ175.90977+38.14133}$ & $20.08$ & $5.84$ & $-26.43$ & \citet{Jiang2016} & \citet{Eilers2020a}  \\ 
$\text{PSOJ352.40341-15.33732}$ & $21.20$ & $5.83$ & $-25.30$ & \citet{Banados2018c} & \citet{RojasRuiz2021}  \\ 
$\text{PSOJ212.29742-15.98660}$ & $20.89$ & $5.83$ & $-25.61$ & \citet{Banados2016} & \citet{Banados2016}  \\ 
$\text{PSOJ122.72630+51.09447}$ & $19.86$ & $5.82$ & $-26.64$ & \citet{Jiang2016} & \citet{Banados2016}  \\ 
$\text{PSOJ235.94506+17.00789}$ & $20.20$ & $5.82$ & $-26.30$ & \citet{Banados2016} & \citet{Banados2016}  \\ 
$\text{PSOJ236.29124+16.60886}$ & $20.83$ & $5.82$ & $-25.67$ & \citet{Banados2016} & \citet{Banados2016}  \\ 
$\text{PSOJ000.66411+25.84303}$ & $19.42$ & $5.82$ & $-27.08$ & \citet{Fan2004} & \citet{Shen2019a}  \\ 
$\text{PSOJ129.18276+00.91479}$ & $19.03$ & $5.81$ & $-27.47$ & \citet{Fan2001c} & \citet{Kurk2007}  \\ 
$\text{PSOJ174.79204-12.28454}$ & $20.23$ & $5.81$ & $-26.26$ & \citet{Banados2016} & \citet{Banados2016}  \\ 
$\text{PSOJ238.85104-06.89765}$ & $20.63$ & $5.81$ & $-25.86$ & \citet{Banados2016} & \citet{Banados2016}  \\ 
$\text{PSOJ357.82898+06.40193}$ & $21.37$ & $5.81$ & $-25.12$ & \citet{Banados2016} & \citet{Banados2016}  \\ 
$\text{PSOJ017.06916-11.99193}$ & $20.66$ & $5.80$ & $-25.82$ & \citet{Banados2022} & \citet{Banados2022}  \\ 
$\text{PSOJ124.00326+12.99894}$ & $21.05$ & $5.80$ & $-25.43$ & \citet{Banados2022} & \citet{Banados2022}  \\ 
$\text{PSOJ193.39924-02.78203}$ & $21.10$ & $5.80$ & $-25.39$ & \citet{Banados2022} & \citet{Banados2022}  \\ 
$\text{PSOJ307.76357-05.19589}$ & $20.95$ & $5.80$ & $-25.53$ & \citet{Banados2022} & \citet{Banados2022}  \\ 
$\text{PSOJ308.48295-27.64850}$ & $19.90$ & $5.80$ & $-26.59$ & \citet{Banados2016} & \citet{Banados2016}  \\ 
$\text{PSOJ021.42133-25.88228}$ & $19.71$ & $5.79$ & $-26.78$ & \citet{Banados2016} & \citet{Banados2016}  \\ 
$\text{PSOJ161.13770-01.41721}$ & $19.32$ & $5.78$ & $-27.16$ & \citet{Fan2000} & \citet{Venemans2020}  \\ 
$\text{PSOJ169.84022+01.21906}$ & $21.20$ & $5.78$ & $-25.27$ & \citet{WangFeige2018} & \citet{WangFeige2018}  \\ 
$\text{PSOJ215.15142-16.04173}$ & $19.25$ & $5.78$ & $-27.23$ & \citet{Morganson2012} & Li, J. et al. 2022, in prep.  \\ 
$\text{PSOJ236.46704+60.47332}$ & $19.13$ & $5.78$ & $-27.35$ & \citet{WangFeige2016} & \citet{WangFeige2016}  \\ 
$\text{PSOJ243.64732+01.24579}$ & $19.96$ & $5.78$ & $-26.52$ & \citet{YangJinyi2019b} & \citet{YangJinyi2019b}  \\ 
$\text{PSOJ000.04163-04.27391}$ & $20.46$ & $5.77$ & $-26.01$ & \citet{Banados2022} & \citet{Banados2022}  \\ 
$\text{PSOJ172.17701+26.88666}$ & $20.73$ & $5.77$ & $-25.73$ & \citet{Banados2016} & \citet{Banados2016}  \\ 
$\text{PSOJ187.10477-02.56090}$ & $20.83$ & $5.77$ & $-25.63$ & \citet{Banados2016} & \citet{Banados2016}  \\ 
$\text{PSOJ209.38256-08.71714}$ & $21.25$ & $5.77$ & $-25.21$ & \citet{Banados2022} & \citet{Banados2022}  \\ 
$\text{PSOJ351.30935+26.47995}$ & $19.42$ & $5.77$ & $-27.06$ & \citet{WangFeige2016} & \citet{WangFeige2016}  \\ 
$\text{PSOJ249.46742+02.69955}$ & $19.37$ & $5.76$ & $-27.10$ & \citet{Wenzl2021} & \citet{Wenzl2021}  \\ 
$\text{PSOJ261.12470+37.30605}$ & $20.49$ & $5.76$ & $-25.97$ & \citet{Banados2022} & \citet{Banados2022}  \\ 
$\text{PSOJ072.58253-07.89183}$ & $20.96$ & $5.75$ & $-25.49$ & \citet{Banados2022} & \citet{Banados2022}  \\ 
$\text{PSOJ156.44661+38.95732}$ & $20.13$ & $5.75$ & $-26.34$ & \citet{Banados2022} & \citet{Banados2022}  \\ 
$\text{PSOJ201.92220+57.54400}$ & $20.56$ & $5.74$ & $-25.89$ & \citet{Banados2016} & \citet{Banados2016}  \\ 
$\text{PSOJ271.44556+49.30671}$ & $20.59$ & $5.74$ & $-25.86$ & \citet{Banados2022} & \citet{Banados2022}  \\ 
$\text{PSOJ169.14063+58.88944}$ & $20.44$ & $5.73$ & $-26.01$ & \citet{Banados2022} & \citet{Banados2022}  \\ 
$\text{PSOJ320.87027-24.36041}$ & $20.37$ & $5.73$ & $-26.07$ & \citet{Banados2016} & \citet{Banados2016}  \\ 
$\text{PSOJ030.88490+00.20813}$ & $20.41$ & $5.72$ & $-26.03$ & \citet{Venemans2007} & \citet{Mortlock2009}  \\ 
$\text{PSOJ209.20588-26.70839}$ & $19.45$ & $5.72$ & $-26.99$ & \citet{Banados2016} & \citet{Banados2016}  \\ 
$\text{PSOJ071.45075-02.33330}$ & $19.13$ & $5.69$ & $-27.30$ & \citet{Banados2016} & \citet{Banados2016}  \\ 
$\text{PSOJ175.40916-20.26547}$ & $20.37$ & $5.69$ & $-26.05$ & \citet{Banados2016} & \citet{Banados2016}  \\ 
$\text{PSOJ196.34762+15.38990}$ & $20.44$ & $5.69$ & $-25.98$ & \citet{Banados2022} & \citet{Banados2022}  \\ 
$\text{PSOJ207.59836+37.80990}$ & $20.59$ & $5.69$ & $-25.82$ & \citet{Banados2022},\citet{Gloudemans2022} &\citet{Banados2022}  \\ 
$\text{PSOJ004.81406-24.29917}$ & $19.41$ & $5.68$ & $-27.02$ & \citet{Banados2016} & \citet{Banados2016}  \\ 
$\text{PSOJ038.19141-18.57350}$ & $20.47$ & $5.68$ & $-25.95$ & \citet{Banados2022} & \citet{Banados2022}  \\ 
$\text{PSOJ040.01591+17.54581}$ & $20.88$ & $5.68$ & $-25.52$ & \citet{Banados2016} & \citet{Banados2016}  \\ 
$\text{PSOJ055.42440-00.80355}$ & $20.18$ & $5.68$ & $-26.24$ & \citet{Banados2015a} & \citet{Banados2015}  \\ 
$\text{PSOJ178.37330+28.50753}$ & $19.86$ & $5.68$ & $-26.56$ & \citet{Banados2022} & \citet{Banados2022}  \\ 
$\text{PSOJ245.06367-00.19786}$ & $21.21$ & $5.68$ & $-25.20$ & \citet{Banados2016} & \citet{Banados2016}  \\ 
$\text{PSOJ197.86749+45.80408}$ & $20.75$ & $5.66$ & $-25.65$ & \citet{Banados2022} & \citet{Banados2022}  \\ 
$\text{PSOJ002.54292+03.06321}$ & $20.93$ & $5.64$ & $-25.46$ & \citet{Banados2022} & \citet{Banados2022}  \\ 
$\text{PSOJ224.65067+10.21379}$ & $19.84$ & $5.60$ & $-26.55$ & \citet{Banados2022} & \citet{Banados2022}  \\ 
$\text{PSOJ330.28172+23.64389}$ & $20.49$ & $5.83$ & $-26.01$ & \citet{Gloudemans2022},\citet{Banados2022} &\citet{Gloudemans2022}  \\ 
\enddata 
\end{deluxetable*}


\bibliography{all}{}

\begin{thebibliography}{}
\expandafter\ifx\csname natexlab\endcsname\relax\def\natexlab#1{#1}\fi
\providecommand{\url}[1]{\href{#1}{#1}}
\providecommand{\dodoi}[1]{doi:~\href{http://doi.org/#1}{\nolinkurl{#1}}}
\providecommand{\doeprint}[1]{\href{http://ascl.net/#1}{\nolinkurl{http://ascl.net/#1}}}
\providecommand{\doarXiv}[1]{\href{https://arxiv.org/abs/#1}{\nolinkurl{https://arxiv.org/abs/#1}}}

\bibitem[{{Akiyama} {et~al.}(2018){Akiyama}, {He}, {Ikeda}, {Niida}, {Nagao},
  {Bosch}, {Coupon}, {Enoki}, {Imanishi}, {Kashikawa}, {Kawaguchi}, {Komiyama},
  {Lee}, {Matsuoka}, {Miyazaki}, {Nishizawa}, {Oguri}, {Ono}, {Onoue}, {Ouchi},
  {Schulze}, {Silverman}, {Tanaka}, {Tanaka}, {Terashima}, {Toba}, \&
  {Ueda}}]{Akiyama2018}
{Akiyama}, M., {He}, W., {Ikeda}, H., {et~al.} 2018, \pasj, 70, S34,
  \dodoi{10.1093/pasj/psx091}

\bibitem[{{Alam} {et~al.}(2015){Alam}, {Albareti}, {Allende Prieto}, {Anders},
  {Anderson}, {Anderton}, {Andrews}, {Armengaud}, {Aubourg}, {Bailey}, {Basu},
  {Bautista}, {Beaton}, {Beers}, {Bender}, {Berlind}, {Beutler}, {Bhardwaj},
  {Bird}, {Bizyaev}, {Blake}, {Blanton}, {Blomqvist}, {Bochanski}, {Bolton},
  {Bovy}, {Shelden Bradley}, {Brandt}, {Brauer}, {Brinkmann}, {Brown},
  {Brownstein}, {Burden}, {Burtin}, {Busca}, {Cai}, {Capozzi}, {Carnero
  Rosell}, {Carr}, {Carrera}, {Chambers}, {Chaplin}, {Chen}, {Chiappini},
  {Chojnowski}, {Chuang}, {Clerc}, {Comparat}, {Covey}, {Croft}, {Cuesta},
  {Cunha}, {da Costa}, {Da Rio}, {Davenport}, {Dawson}, {De Lee}, {Delubac},
  {Deshpande}, {Dhital}, {Dutra-Ferreira}, {Dwelly}, {Ealet}, {Ebelke},
  {Edmondson}, {Eisenstein}, {Ellsworth}, {Elsworth}, {Epstein}, {Eracleous},
  {Escoffier}, {Esposito}, {Evans}, {Fan}, {Fern{\'a}ndez-Alvar}, {Feuillet},
  {Filiz Ak}, {Finley}, {Finoguenov}, {Flaherty}, {Fleming}, {Font-Ribera},
  {Foster}, {Frinchaboy}, {Galbraith-Frew}, {Garc{\'\i}a},
  {Garc{\'\i}a-Hern{\'a}ndez}, {Garc{\'\i}a P{\'e}rez}, {Gaulme}, {Ge},
  {G{\'e}nova-Santos}, {Georgakakis}, {Ghezzi}, {Gillespie}, {Girardi},
  {Goddard}, {Gontcho}, {Gonz{\'a}lez Hern{\'a}ndez}, {Grebel}, {Green},
  {Grieb}, {Grieves}, {Gunn}, {Guo}, {Harding}, {Hasselquist}, {Hawley},
  {Hayden}, {Hearty}, {Hekker}, {Ho}, {Hogg}, {Holley-Bockelmann}, {Holtzman},
  {Honscheid}, {Huber}, {Huehnerhoff}, {Ivans}, {Jiang}, {Johnson},
  {Kinemuchi}, {Kirkby}, {Kitaura}, {Klaene}, {Knapp}, {Kneib}, {Koenig},
  {Lam}, {Lan}, {Lang}, {Laurent}, {Le Goff}, {Leauthaud}, {Lee}, {Lee},
  {Licquia}, {Liu}, {Long}, {L{\'o}pez-Corredoira}, {Lorenzo-Oliveira},
  {Lucatello}, {Lundgren}, {Lupton}, {Mack}, {Mahadevan}, {Maia}, {Majewski},
  {Malanushenko}, {Malanushenko}, {Manchado}, {Manera}, {Mao}, {Maraston},
  {Marchwinski}, {Margala}, {Martell}, {Martig}, {Masters}, {Mathur},
  {McBride}, {McGehee}, {McGreer}, {McMahon}, {M{\'e}nard}, {Menzel},
  {Merloni}, {M{\'e}sz{\'a}ros}, {Miller}, {Miralda-Escud{\'e}}, {Miyatake},
  {Montero-Dorta}, {More}, {Morganson}, {Morice-Atkinson}, {Morrison},
  {Mosser}, {Muna}, {Myers}, {Nandra}, {Newman}, {Neyrinck}, {Nguyen},
  {Nichol}, {Nidever}, {Noterdaeme}, {Nuza}, {O'Connell}, {O'Connell},
  {O'Connell}, {Ogando}, {Olmstead}, {Oravetz}, {Oravetz}, {Osumi}, {Owen},
  {Padgett}, {Padmanabhan}, {Paegert}, {Palanque-Delabrouille}, {Pan},
  {Parejko}, {P{\^a}ris}, {Park}, {Pattarakijwanich}, {Pellejero-Ibanez},
  {Pepper}, {Percival}, {P{\'e}rez-Fournon}, {P{\'e}rez-R{\`a}fols},
  {Petitjean}, {Pieri}, {Pinsonneault}, {Porto de Mello}, {Prada}, {Prakash},
  {Price-Whelan}, {Protopapas}, {Raddick}, {Rahman}, {Reid}, {Rich}, {Rix},
  {Robin}, {Rockosi}, {Rodrigues}, {Rodr{\'\i}guez-Torres}, {Roe}, {Ross},
  {Ross}, {Rossi}, {Ruan}, {Rubi{\~n}o-Mart{\'\i}n}, {Rykoff},
  {Salazar-Albornoz}, {Salvato}, {Samushia}, {S{\'a}nchez}, {Santiago},
  {Sayres}, {Schiavon}, {Schlegel}, {Schmidt}, {Schneider}, {Schultheis},
  {Schwope}, {Sc{\'o}ccola}, {Scott}, {Sellgren}, {Seo}, {Serenelli}, {Shane},
  {Shen}, {Shetrone}, {Shu}, {Silva Aguirre}, {Sivarani}, {Skrutskie},
  {Slosar}, {Smith}, {Sobreira}, {Souto}, {Stassun}, {Steinmetz}, {Stello},
  {Strauss}, {Streblyanska}, {Suzuki}, {Swanson}, {Tan}, {Tayar}, {Terrien},
  {Thakar}, {Thomas}, {Thomas}, {Thompson}, {Tinker}, {Tojeiro}, {Troup},
  {Vargas-Maga{\~n}a}, {Vazquez}, {Verde}, {Viel}, {Vogt}, {Wake}, {Wang},
  {Weaver}, {Weinberg}, {Weiner}, {White}, {Wilson}, {Wisniewski},
  {Wood-Vasey}, {Ye`che}, {York}, {Zakamska}, {Zamora}, {Zasowski}, {Zehavi},
  {Zhao}, {Zheng}, {Zhou}, {Zhou}, {Zou}, \& {Zhu}}]{Alam2015}
{Alam}, S., {Albareti}, F.~D., {Allende Prieto}, C., {et~al.} 2015, \apjs, 219,
  12, \dodoi{10.1088/0067-0049/219/1/12}

\bibitem[{{Astropy Collaboration} {et~al.}(2013){Astropy Collaboration},
  {Robitaille}, {Tollerud}, {Greenfield}, {Droettboom}, {Bray}, {Aldcroft},
  {Davis}, {Ginsburg}, {Price-Whelan}, {Kerzendorf}, {Conley}, {Crighton},
  {Barbary}, {Muna}, {Ferguson}, {Grollier}, {Parikh}, {Nair}, {Unther},
  {Deil}, {Woillez}, {Conseil}, {Kramer}, {Turner}, {Singer}, {Fox}, {Weaver},
  {Zabalza}, {Edwards}, {Azalee Bostroem}, {Burke}, {Casey}, {Crawford},
  {Dencheva}, {Ely}, {Jenness}, {Labrie}, {Lim}, {Pierfederici}, {Pontzen},
  {Ptak}, {Refsdal}, {Servillat}, \& {Streicher}}]{astropy1}
{Astropy Collaboration}, {Robitaille}, T.~P., {Tollerud}, E.~J., {et~al.} 2013,
  \aap, 558, A33, \dodoi{10.1051/0004-6361/201322068}

\bibitem[{{Astropy Collaboration} {et~al.}(2018){Astropy Collaboration},
  {Price-Whelan}, {Sip{\H{o}}cz}, {G{\"u}nther}, {Lim}, {Crawford}, {Conseil},
  {Shupe}, {Craig}, {Dencheva}, {Ginsburg}, {Vand erPlas}, {Bradley},
  {P{\'e}rez-Su{\'a}rez}, {de Val-Borro}, {Aldcroft}, {Cruz}, {Robitaille},
  {Tollerud}, {Ardelean}, {Babej}, {Bach}, {Bachetti}, {Bakanov}, {Bamford},
  {Barentsen}, {Barmby}, {Baumbach}, {Berry}, {Biscani}, {Boquien}, {Bostroem},
  {Bouma}, {Brammer}, {Bray}, {Breytenbach}, {Buddelmeijer}, {Burke},
  {Calderone}, {Cano Rodr{\'\i}guez}, {Cara}, {Cardoso}, {Cheedella}, {Copin},
  {Corrales}, {Crichton}, {D'Avella}, {Deil}, {Depagne}, {Dietrich}, {Donath},
  {Droettboom}, {Earl}, {Erben}, {Fabbro}, {Ferreira}, {Finethy}, {Fox},
  {Garrison}, {Gibbons}, {Goldstein}, {Gommers}, {Greco}, {Greenfield},
  {Groener}, {Grollier}, {Hagen}, {Hirst}, {Homeier}, {Horton}, {Hosseinzadeh},
  {Hu}, {Hunkeler}, {Ivezi{\'c}}, {Jain}, {Jenness}, {Kanarek}, {Kendrew},
  {Kern}, {Kerzendorf}, {Khvalko}, {King}, {Kirkby}, {Kulkarni}, {Kumar},
  {Lee}, {Lenz}, {Littlefair}, {Ma}, {Macleod}, {Mastropietro}, {McCully},
  {Montagnac}, {Morris}, {Mueller}, {Mumford}, {Muna}, {Murphy}, {Nelson},
  {Nguyen}, {Ninan}, {N{\"o}the}, {Ogaz}, {Oh}, {Parejko}, {Parley}, {Pascual},
  {Patil}, {Patil}, {Plunkett}, {Prochaska}, {Rastogi}, {Reddy Janga},
  {Sabater}, {Sakurikar}, {Seifert}, {Sherbert}, {Sherwood-Taylor}, {Shih},
  {Sick}, {Silbiger}, {Singanamalla}, {Singer}, {Sladen}, {Sooley},
  {Sornarajah}, {Streicher}, {Teuben}, {Thomas}, {Tremblay}, {Turner},
  {Terr{\'o}n}, {van Kerkwijk}, {de la Vega}, {Watkins}, {Weaver}, {Whitmore},
  {Woillez}, {Zabalza}, \& {Astropy Contributors}}]{astropy2}
{Astropy Collaboration}, {Price-Whelan}, A.~M., {Sip{\H{o}}cz}, B.~M., {et~al.}
  2018, \aj, 156, 123, \dodoi{10.3847/1538-3881/aabc4f}

\bibitem[{{Avni} \& {Bahcall}(1980)}]{Avni1980}
{Avni}, Y., \& {Bahcall}, J.~N. 1980, \apj, 235, 694, \dodoi{10.1086/157673}

\bibitem[{{Ba{\~n}ados} {et~al.}(2018{\natexlab{a}}){Ba{\~n}ados}, {Carilli},
  {Walter}, {Momjian}, {Decarli}, {Farina}, {Mazzucchelli}, \&
  {Venemans}}]{Banados2018c}
{Ba{\~n}ados}, E., {Carilli}, C., {Walter}, F., {et~al.} 2018{\natexlab{a}},
  \apjl, 861, L14, \dodoi{10.3847/2041-8213/aac511}

\bibitem[{{Ba{\~n}ados} {et~al.}(2015{\natexlab{a}}){Ba{\~n}ados}, {Decarli},
  {Walter}, {Venemans}, {Farina}, \& {Fan}}]{Banados2015}
{Ba{\~n}ados}, E., {Decarli}, R., {Walter}, F., {et~al.} 2015{\natexlab{a}},
  \apjl, 805, L8, \dodoi{10.1088/2041-8205/805/1/L8}

\bibitem[{{Ba{\~n}ados} {et~al.}(2014){Ba{\~n}ados}, {Venemans}, {Morganson},
  {Decarli}, {Walter}, {Chambers}, {Rix}, {Farina}, {Fan}, {Jiang}, {McGreer},
  {De Rosa}, {Simcoe}, {Wei{\ss}}, {Price}, {Morgan}, {Burgett}, {Greiner},
  {Kaiser}, {Kudritzki}, {Magnier}, {Metcalfe}, {Stubbs}, {Sweeney}, {Tonry},
  {Wainscoat}, \& {Waters}}]{Banados2014}
{Ba{\~n}ados}, E., {Venemans}, B.~P., {Morganson}, E., {et~al.} 2014, \aj, 148,
  14, \dodoi{10.1088/0004-6256/148/1/14}

\bibitem[{{Ba{\~n}ados} {et~al.}(2015{\natexlab{b}}){Ba{\~n}ados}, {Venemans},
  {Morganson}, {Hodge}, {Decarli}, {Walter}, {Stern}, {Schlafly}, {Farina},
  {Greiner}, {Chambers}, {Fan}, {Rix}, {Burgett}, {Draper}, {Flewelling},
  {Kaiser}, {Metcalfe}, {Morgan}, {Tonry}, \& {Wainscoat}}]{Banados2015a}
---. 2015{\natexlab{b}}, \apj, 804, 118, \dodoi{10.1088/0004-637X/804/2/118}

\bibitem[{{Ba{\~n}ados} {et~al.}(2016){Ba{\~n}ados}, {Venemans}, {Decarli},
  {Farina}, {Mazzucchelli}, {Walter}, {Fan}, {Stern}, {Schlafly}, {Chambers},
  {Rix}, {Jiang}, {McGreer}, {Simcoe}, {Wang}, {Yang}, {Morganson}, {De Rosa},
  {Greiner}, {Balokovi{\'c}}, {Burgett}, {Cooper}, {Draper}, {Flewelling},
  {Hodapp}, {Jun}, {Kaiser}, {Kudritzki}, {Magnier}, {Metcalfe}, {Miller},
  {Schindler}, {Tonry}, {Wainscoat}, {Waters}, \& {Yang}}]{Banados2016}
{Ba{\~n}ados}, E., {Venemans}, B.~P., {Decarli}, R., {et~al.} 2016, \apjs, 227,
  11, \dodoi{10.3847/0067-0049/227/1/11}

\bibitem[{{Ba{\~n}ados} {et~al.}(2018{\natexlab{b}}){Ba{\~n}ados}, {Venemans},
  {Mazzucchelli}, {Farina}, {Walter}, {Wang}, {Decarli}, {Stern}, {Fan},
  {Davies}, {Hennawi}, {Simcoe}, {Turner}, {Rix}, {Yang}, {Kelson}, {Rudie}, \&
  {Winters}}]{Banados2018}
{Ba{\~n}ados}, E., {Venemans}, B.~P., {Mazzucchelli}, C., {et~al.}
  2018{\natexlab{b}}, \nat, 553, 473, \dodoi{10.1038/nature25180}

\bibitem[{{Baldwin}(1977)}]{Baldwin1977}
{Baldwin}, J.~A. 1977, \apj, 214, 679, \dodoi{10.1086/155294}

\bibitem[{{Banados} {et~al.}(2022){Banados}, {Schindler}, {Venemans}, {Connor},
  {Decarli}, {Farina}, {Mazzucchelli}, {Meyer}, {Stern}, {Walter}, {Fan},
  {Hennawi}, {Khusanova}, {Morrell}, {Nanni}, {Noirot}, {Pensabene}, {Rix},
  {Simon}, {Verdoes Kleijn}, {Xie}, {Yang}, \& {Connor}}]{Banados2022}
{Banados}, E., {Schindler}, J.-T., {Venemans}, B.~P., {et~al.} 2022, arXiv
  e-prints, arXiv:2212.04452.
\newblock \doarXiv{2212.04452}

\bibitem[{{Becker} \& {Bolton}(2013)}]{Becker2013}
{Becker}, G.~D., \& {Bolton}, J.~S. 2013, \mnras, 436, 1023,
  \dodoi{10.1093/mnras/stt1610}

\bibitem[{{Becker} {et~al.}(2015{\natexlab{a}}){Becker}, {Bolton}, \&
  {Lidz}}]{Becker2015b}
{Becker}, G.~D., {Bolton}, J.~S., \& {Lidz}, A. 2015{\natexlab{a}}, \pasa, 32,
  e045, \dodoi{10.1017/pasa.2015.45}

\bibitem[{{Becker} {et~al.}(2015{\natexlab{b}}){Becker}, {Bolton}, {Madau},
  {Pettini}, {Ryan-Weber}, \& {Venemans}}]{Becker2015}
{Becker}, G.~D., {Bolton}, J.~S., {Madau}, P., {et~al.} 2015{\natexlab{b}},
  \mnras, 447, 3402, \dodoi{10.1093/mnras/stu2646}

\bibitem[{{Becker} {et~al.}(2021){Becker}, {D'Aloisio}, {Christenson}, {Zhu},
  {Worseck}, \& {Bolton}}]{Becker2021}
{Becker}, G.~D., {D'Aloisio}, A., {Christenson}, H.~M., {et~al.} 2021, \mnras,
  508, 1853, \dodoi{10.1093/mnras/stab2696}

\bibitem[{{Bischetti} {et~al.}(2022){Bischetti}, {Feruglio}, {D'Odorico},
  {Arav}, {Ba{\~n}ados}, {Becker}, {Bosman}, {Carniani}, {Cristiani}, {Cupani},
  {Davies}, {Eilers}, {Farina}, {Ferrara}, {Maiolino}, {Mazzucchelli},
  {Mesinger}, {Meyer}, {Onoue}, {Piconcelli}, {Ryan-Weber}, {Schindler},
  {Wang}, {Yang}, {Zhu}, \& {Fiore}}]{Bischetti2022}
{Bischetti}, M., {Feruglio}, C., {D'Odorico}, V., {et~al.} 2022, \nat, 605,
  244, \dodoi{10.1038/s41586-022-04608-1}

\bibitem[{{Boroson} \& {Green}(1992)}]{Boroson1992a}
{Boroson}, T.~A., \& {Green}, R.~F. 1992, \apjs, 80, 109,
  \dodoi{10.1086/191661}

\bibitem[{{Bosman}(2021)}]{Bosman2021arxivb}
{Bosman}, S. E.~I. 2021, arXiv e-prints, arXiv:2108.12446.
\newblock \doarXiv{2108.12446}

\bibitem[{{Boutsia} {et~al.}(2020){Boutsia}, {Grazian}, {Calderone},
  {Cristiani}, {Cupani}, {Guarneri}, {Fontanot}, {Amorin}, {D'Odorico},
  {Giallongo}, {Salvato}, {Omizzolo}, {Romano}, \& {Menci}}]{Boutsia2020}
{Boutsia}, K., {Grazian}, A., {Calderone}, G., {et~al.} 2020, arXiv e-prints,
  arXiv:2008.03865.
\newblock \doarXiv{2008.03865}

\bibitem[{{Boutsia} {et~al.}(2021){Boutsia}, {Grazian}, {Fontanot},
  {Giallongo}, {Menci}, {Calderone}, {Cristiani}, {D'Odorico}, {Cupani},
  {Guarneri}, \& {Omizzolo}}]{Boutsia2021}
{Boutsia}, K., {Grazian}, A., {Fontanot}, F., {et~al.} 2021, \apj, 912, 111,
  \dodoi{10.3847/1538-4357/abedb5}

\bibitem[{{Boyle} {et~al.}(2000){Boyle}, {Shanks}, {Croom}, {Smith}, {Miller},
  {Loaring}, \& {Heymans}}]{Boyle2000}
{Boyle}, B.~J., {Shanks}, T., {Croom}, S.~M., {et~al.} 2000, \mnras, 317, 1014,
  \dodoi{10.1046/j.1365-8711.2000.03730.x}

\bibitem[{{Boyle} {et~al.}(1988){Boyle}, {Shanks}, \& {Peterson}}]{Boyle1988}
{Boyle}, B.~J., {Shanks}, T., \& {Peterson}, B.~A. 1988, \mnras, 235, 935,
  \dodoi{10.1093/mnras/235.3.935}

\bibitem[{{Calderone} {et~al.}(2019){Calderone}, {Boutsia}, {Cristiani},
  {Grazian}, {Amorin}, {D{\textquoteright}Odorico}, {Cupani}, {Fontanot}, \&
  {Salvato}}]{Calderone2019}
{Calderone}, G., {Boutsia}, K., {Cristiani}, S., {et~al.} 2019, \apj, 887, 268,
  \dodoi{10.3847/1538-4357/ab510a}

\bibitem[{{Calverley} {et~al.}(2011){Calverley}, {Becker}, {Haehnelt}, \&
  {Bolton}}]{Calverley2011}
{Calverley}, A.~P., {Becker}, G.~D., {Haehnelt}, M.~G., \& {Bolton}, J.~S.
  2011, \mnras, 412, 2543, \dodoi{10.1111/j.1365-2966.2010.18072.x}

\bibitem[{{Cappelluti} {et~al.}(2016){Cappelluti}, {Comastri}, {Fontana},
  {Zamorani}, {Amorin}, {Castellano}, {Merlin}, {Santini}, {Elbaz},
  {Schreiber}, {Shu}, {Wang}, {Dunlop}, {Bourne}, {Bruce}, {Buitrago},
  {Micha{\l}owski}, {Derriere}, {Ferguson}, {Faber}, \&
  {Vito}}]{Cappelluti2016}
{Cappelluti}, N., {Comastri}, A., {Fontana}, A., {et~al.} 2016, \apj, 823, 95,
  \dodoi{10.3847/0004-637X/823/2/95}

\bibitem[{{Carilli} {et~al.}(2010){Carilli}, {Wang}, {Fan}, {Walter}, {Kurk},
  {Riechers}, {Wagg}, {Hennawi}, {Jiang}, {Menten}, {Bertoldi}, {Strauss}, \&
  {Cox}}]{Carilli2010}
{Carilli}, C.~L., {Wang}, R., {Fan}, X., {et~al.} 2010, \apj, 714, 834,
  \dodoi{10.1088/0004-637X/714/1/834}

\bibitem[{{Chambers} {et~al.}(2016){Chambers}, {Magnier}, {Metcalfe},
  {Flewelling}, {Huber}, {Waters}, {Denneau}, {Draper}, {Farrow}, {Finkbeiner},
  {Holmberg}, {Koppenhoefer}, {Price}, {Rest}, {Saglia}, {Schlafly}, {Smartt},
  {Sweeney}, {Wainscoat}, {Burgett}, {Chastel}, {Grav}, {Heasley}, {Hodapp},
  {Jedicke}, {Kaiser}, {Kudritzki}, {Luppino}, {Lupton}, {Monet}, {Morgan},
  {Onaka}, {Shiao}, {Stubbs}, {Tonry}, {White}, {Ba{\~n}ados}, {Bell},
  {Bender}, {Bernard}, {Boegner}, {Boffi}, {Botticella}, {Calamida},
  {Casertano}, {Chen}, {Chen}, {Cole}, {Deacon}, {Frenk}, {Fitzsimmons},
  {Gezari}, {Gibbs}, {Goessl}, {Goggia}, {Gourgue}, {Goldman}, {Grant},
  {Grebel}, {Hambly}, {Hasinger}, {Heavens}, {Heckman}, {Henderson}, {Henning},
  {Holman}, {Hopp}, {Ip}, {Isani}, {Jackson}, {Keyes}, {Koekemoer}, {Kotak},
  {Le}, {Liska}, {Long}, {Lucey}, {Liu}, {Martin}, {Masci}, {McLean}, {Mindel},
  {Misra}, {Morganson}, {Murphy}, {Obaika}, {Narayan}, {Nieto-Santisteban},
  {Norberg}, {Peacock}, {Pier}, {Postman}, {Primak}, {Rae}, {Rai}, {Riess},
  {Riffeser}, {Rix}, {R{\"o}ser}, {Russel}, {Rutz}, {Schilbach}, {Schultz},
  {Scolnic}, {Strolger}, {Szalay}, {Seitz}, {Small}, {Smith}, {Soderblom},
  {Taylor}, {Thomson}, {Taylor}, {Thakar}, {Thiel}, {Thilker}, {Unger},
  {Urata}, {Valenti}, {Wagner}, {Walder}, {Walter}, {Watters}, {Werner},
  {Wood-Vasey}, \& {Wyse}}]{Chambers2016}
{Chambers}, K.~C., {Magnier}, E.~A., {Metcalfe}, N., {et~al.} 2016, arXiv
  e-prints, arXiv:1612.05560.
\newblock \doarXiv{1612.05560}

\bibitem[{{Cool} {et~al.}(2006){Cool}, {Kochanek}, {Eisenstein}, {Stern},
  {Brand}, {Brown}, {Dey}, {Eisenhardt}, {Fan}, {Gonzalez}, {Green}, {Jannuzi},
  {McKenzie}, {Rieke}, {Rieke}, {Soifer}, {Spinrad}, \& {Elston}}]{Cool2006}
{Cool}, R.~J., {Kochanek}, C.~S., {Eisenstein}, D.~J., {et~al.} 2006, \aj, 132,
  823, \dodoi{10.1086/505535}

\bibitem[{{Croom} {et~al.}(2009){Croom}, {Richards}, {Shanks}, {Boyle},
  {Strauss}, {Myers}, {Nichol}, {Pimbblet}, {Ross}, {Schneider}, {Sharp}, \&
  {Wake}}]{Croom2009}
{Croom}, S.~M., {Richards}, G.~T., {Shanks}, T., {et~al.} 2009, \mnras, 399,
  1755, \dodoi{10.1111/j.1365-2966.2009.15398.x}

\bibitem[{{D'Aloisio} {et~al.}(2018){D'Aloisio}, {McQuinn}, {Davies}, \&
  {Furlanetto}}]{dAloisio2018}
{D'Aloisio}, A., {McQuinn}, M., {Davies}, F.~B., \& {Furlanetto}, S.~R. 2018,
  \mnras, 473, 560, \dodoi{10.1093/mnras/stx2341}

\bibitem[{{Davies} {et~al.}(2018){Davies}, {Hennawi}, {Eilers}, \&
  {Luki{\'c}}}]{Davies2018a}
{Davies}, F.~B., {Hennawi}, J.~F., {Eilers}, A.-C., \& {Luki{\'c}}, Z. 2018,
  \apj, 855, 106, \dodoi{10.3847/1538-4357/aaaf70}

\bibitem[{{De Rosa} {et~al.}(2011){De Rosa}, {Decarli}, {Walter}, {Fan},
  {Jiang}, {Kurk}, {Pasquali}, \& {Rix}}]{DeRosa2011}
{De Rosa}, G., {Decarli}, R., {Walter}, F., {et~al.} 2011, \apj, 739, 56,
  \dodoi{10.1088/0004-637X/739/2/56}

\bibitem[{{Decarli} {et~al.}(2018){Decarli}, {Walter}, {Venemans},
  {Ba{\~n}ados}, {Bertoldi}, {Carilli}, {Fan}, {Farina}, {Mazzucchelli},
  {Riechers}, {Rix}, {Strauss}, {Wang}, \& {Yang}}]{Decarli2018}
{Decarli}, R., {Walter}, F., {Venemans}, B.~P., {et~al.} 2018, \apj, 854, 97,
  \dodoi{10.3847/1538-4357/aaa5aa}

\bibitem[{{Di Matteo} {et~al.}(2005){Di Matteo}, {Springel}, \&
  {Hernquist}}]{DiMatteo2005}
{Di Matteo}, T., {Springel}, V., \& {Hernquist}, L. 2005, \nat, 433, 604,
  \dodoi{10.1038/nature03335}

\bibitem[{{Diamond-Stanic} {et~al.}(2009){Diamond-Stanic}, {Fan}, {Brandt},
  {Shemmer}, {Strauss}, {Anderson}, {Carilli}, {Gibson}, {Jiang}, {Kim},
  {Richards}, {Schmidt}, {Schneider}, {Shen}, {Smith}, {Vestergaard}, \&
  {Young}}]{DiamondStanic2009}
{Diamond-Stanic}, A.~M., {Fan}, X., {Brandt}, W.~N., {et~al.} 2009, \apj, 699,
  782, \dodoi{10.1088/0004-637X/699/1/782}

\bibitem[{{Eilers} {et~al.}(2020){Eilers}, {Hennawi}, {Decarli}, {Davies},
  {Venemans}, {Walter}, {Ba{\~n}ados}, {Fan}, {Farina}, {Mazzucchelli},
  {Novak}, {Schindler}, {Simcoe}, {Wang}, \& {Yang}}]{Eilers2020a}
{Eilers}, A.-C., {Hennawi}, J.~F., {Decarli}, R., {et~al.} 2020, \apj, 900, 37,
  \dodoi{10.3847/1538-4357/aba52e}

\bibitem[{{Eilers} {et~al.}(2021){Eilers}, {Hennawi}, {Decarli}, {Davies},
  {Venemans}, {Walter}, {Ba{\~n}ados}, {Fan}, {Farina}, {Mazzucchelli},
  {Novak}, {Schindler}, {Simcoe}, {Wang}, \& {Yang}}]{Eilers2020a_Err}
---. 2021, \apj, 914, 74, \dodoi{10.3847/1538-4357/ac05c3}

\bibitem[{{Euclid Collaboration} {et~al.}(2019){Euclid Collaboration},
  {Barnett}, {Warren}, {Mortlock}, {Cuby}, {Conselice}, {Hewett}, {Willott},
  {Auricchio}, {Balaguera-Antol{\'\i}nez}, {Baldi}, {Bardelli}, {Bellagamba},
  {Bender}, {Biviano}, {Bonino}, {Bozzo}, {Branchini}, {Brescia}, {Brinchmann},
  {Burigana}, {Camera}, {Capobianco}, {Carbone}, {Carretero}, {Carvalho},
  {Castander}, {Castellano}, {Cavuoti}, {Cimatti}, {Cl{\'e}dassou}, {Congedo},
  {Conversi}, {Copin}, {Corcione}, {Coupon}, {Courtois}, {Cropper}, {Da Silva},
  {Duncan}, {Dusini}, {Ealet}, {Farrens}, {Fosalba}, {Fotopoulou},
  {Fourmanoit}, {Frailis}, {Fumana}, {Galeotta}, {Garilli}, {Gillard},
  {Gillis}, {Graci{\'a}-Carpio}, {Grupp}, {Hoekstra}, {Hormuth}, {Israel},
  {Jahnke}, {Kermiche}, {Kilbinger}, {Kirkpatrick}, {Kitching}, {Kohley},
  {Kubik}, {Kunz}, {Kurki-Suonio}, {Laureijs}, {Ligori}, {Lilje}, {Lloro},
  {Maiorano}, {Mansutti}, {Marggraf}, {Martinet}, {Marulli}, {Massey}, {Mauri},
  {Medinaceli}, {Mei}, {Mellier}, {Metcalf}, {Metge}, {Meylan}, {Moresco},
  {Moscardini}, {Munari}, {Neissner}, {Niemi}, {Nutma}, {Padilla}, {Paltani},
  {Pasian}, {Paykari}, {Percival}, {Pettorino}, {Polenta}, {Poncet},
  {Pozzetti}, {Raison}, {Renzi}, {Rhodes}, {Rix}, {Romelli}, {Roncarelli},
  {Rossetti}, {Saglia}, {Sapone}, {Scaramella}, {Schneider}, {Scottez},
  {Secroun}, {Serrano}, {Sirri}, {Stanco}, {Sureau}, {Tallada-Cresp{\'\i}},
  {Tavagnacco}, {Taylor}, {Tenti}, {Tereno}, {Toledo-Moreo}, {Torradeflot},
  {Valenziano}, {Vassallo}, {Wang}, {Zacchei}, {Zamorani}, {Zoubian}, \&
  {Zucca}}]{EuclidPrepV2019}
{Euclid Collaboration}, {Barnett}, R., {Warren}, S.~J., {et~al.} 2019, \aap,
  631, A85, \dodoi{10.1051/0004-6361/201936427}

\bibitem[{{Fan}(1999)}]{Fan1999a}
{Fan}, X. 1999, \aj, 117, 2528, \dodoi{10.1086/300848}

\bibitem[{{Fan} {et~al.}(2000){Fan}, {White}, {Davis}, {Becker}, {Strauss},
  {Haiman}, {Schneider}, {Gregg}, {Gunn}, {Knapp}, {Lupton}, {Anderson},
  {Anderson}, {Annis}, {Bahcall}, {Boroski}, {Brunner}, {Chen}, {Connolly},
  {Csabai}, {Doi}, {Fukugita}, {Hennessy}, {Hindsley}, {Ichikawa},
  {Ivezi{\'c}}, {Loveday}, {Meiksin}, {McKay}, {Munn}, {Newberg}, {Nichol},
  {Okamura}, {Pier}, {Sekiguchi}, {Shimasaku}, {Stoughton}, {Szalay},
  {Szokoly}, {Thakar}, {Vogeley}, \& {York}}]{Fan2000}
{Fan}, X., {White}, R.~L., {Davis}, M., {et~al.} 2000, \aj, 120, 1167,
  \dodoi{10.1086/301534}

\bibitem[{{Fan} {et~al.}(2001{\natexlab{a}}){Fan}, {Narayanan}, {Lupton},
  {Strauss}, {Knapp}, {Becker}, {White}, {Pentericci}, {Leggett}, {Haiman},
  {Gunn}, {Ivezi{\'c}}, {Schneider}, {Anderson}, {Brinkmann}, {Bahcall},
  {Connolly}, {Csabai}, {Doi}, {Fukugita}, {Geballe}, {Grebel}, {Harbeck},
  {Hennessy}, {Lamb}, {Miknaitis}, {Munn}, {Nichol}, {Okamura}, {Pier},
  {Prada}, {Richards}, {Szalay}, \& {York}}]{Fan2001c}
{Fan}, X., {Narayanan}, V.~K., {Lupton}, R.~H., {et~al.} 2001{\natexlab{a}},
  \aj, 122, 2833, \dodoi{10.1086/324111}

\bibitem[{{Fan} {et~al.}(2001{\natexlab{b}}){Fan}, {Strauss}, {Schneider},
  {Gunn}, {Lupton}, {Becker}, {Davis}, {Newman}, {Richards}, {White},
  {Anderson}, {Annis}, {Bahcall}, {Brunner}, {Csabai}, {Hennessy}, {Hindsley},
  {Fukugita}, {Kunszt}, {Ivezi{\'c}}, {Knapp}, {McKay}, {Munn}, {Pier},
  {Szalay}, \& {York}}]{Fan2001b}
{Fan}, X., {Strauss}, M.~A., {Schneider}, D.~P., {et~al.} 2001{\natexlab{b}},
  \aj, 121, 54, \dodoi{10.1086/318033}

\bibitem[{{Fan} {et~al.}(2003){Fan}, {Strauss}, {Schneider}, {Becker}, {White},
  {Haiman}, {Gregg}, {Pentericci}, {Grebel}, {Narayanan}, {Loh}, {Richards},
  {Gunn}, {Lupton}, {Knapp}, {Ivezi{\'c}}, {Brandt}, {Collinge}, {Hao},
  {Harbeck}, {Prada}, {Schaye}, {Strateva}, {Zakamska}, {Anderson},
  {Brinkmann}, {Bahcall}, {Lamb}, {Okamura}, {Szalay}, \& {York}}]{Fan2003}
---. 2003, \aj, 125, 1649, \dodoi{10.1086/368246}

\bibitem[{{Fan} {et~al.}(2004){Fan}, {Hennawi}, {Richards}, {Strauss},
  {Schneider}, {Donley}, {Young}, {Annis}, {Lin}, {Lampeitl}, {Lupton}, {Gunn},
  {Knapp}, {Brandt}, {Anderson}, {Bahcall}, {Brinkmann}, {Brunner}, {Fukugita},
  {Szalay}, {Szokoly}, \& {York}}]{Fan2004}
{Fan}, X., {Hennawi}, J.~F., {Richards}, G.~T., {et~al.} 2004, \aj, 128, 515,
  \dodoi{10.1086/422434}

\bibitem[{{Fan} {et~al.}(2006){Fan}, {Strauss}, {Richards}, {Hennawi},
  {Becker}, {White}, {Diamond-Stanic}, {Donley}, {Jiang}, {Kim}, {Vestergaard},
  {Young}, {Gunn}, {Lupton}, {Knapp}, {Schneider}, {Brandt}, {Bahcall},
  {Barentine}, {Brinkmann}, {Brewington}, {Fukugita}, {Harvanek}, {Kleinman},
  {Krzesinski}, {Long}, {Neilsen}, {Nitta}, {Snedden}, \& {Voges}}]{Fan2006}
{Fan}, X., {Strauss}, M.~A., {Richards}, G.~T., {et~al.} 2006, \aj, 131, 1203,
  \dodoi{10.1086/500296}

\bibitem[{{Fan} {et~al.}(2019){Fan}, {Wang}, {Yang}, {Keeton}, {Yue},
  {Zabludoff}, {Bian}, {Bonaglia}, {Georgiev}, {Hennawi}, {Li}, {McGreer},
  {Naidu}, {Pacucci}, {Rabien}, {Thompson}, {Venemans}, {Walter}, {Wang}, \&
  {Wu}}]{Fan2019}
{Fan}, X., {Wang}, F., {Yang}, J., {et~al.} 2019, \apjl, 870, L11,
  \dodoi{10.3847/2041-8213/aaeffe}

\bibitem[{{Faucher-Gigu{\`e}re}(2020)}]{FaucherGiguere2020}
{Faucher-Gigu{\`e}re}, C.-A. 2020, \mnras, 493, 1614,
  \dodoi{10.1093/mnras/staa302}

\bibitem[{{Faucher-Gigu{\`e}re} {et~al.}(2008){Faucher-Gigu{\`e}re}, {Lidz},
  {Hernquist}, \& {Zaldarriaga}}]{FaucherGiguere2008b}
{Faucher-Gigu{\`e}re}, C.-A., {Lidz}, A., {Hernquist}, L., \& {Zaldarriaga}, M.
  2008, \apj, 688, 85, \dodoi{10.1086/592289}

\bibitem[{{Fontanot} {et~al.}(2007){Fontanot}, {Cristiani}, {Monaco}, {Nonino},
  {Vanzella}, {Brandt}, {Grazian}, \& {Mao}}]{Fontanot2007}
{Fontanot}, F., {Cristiani}, S., {Monaco}, P., {et~al.} 2007, \aap, 461, 39,
  \dodoi{10.1051/0004-6361:20066073}

\bibitem[{{Foreman-Mackey} {et~al.}(2013){Foreman-Mackey}, {Hogg}, {Lang}, \&
  {Goodman}}]{emcee}
{Foreman-Mackey}, D., {Hogg}, D.~W., {Lang}, D., \& {Goodman}, J. 2013, \pasp,
  125, 306, \dodoi{10.1086/670067}

\bibitem[{{Gehrels}(1986)}]{Gehrels1986}
{Gehrels}, N. 1986, \apj, 303, 336, \dodoi{10.1086/164079}

\bibitem[{{Giallongo} {et~al.}(2015){Giallongo}, {Grazian}, {Fiore}, {Fontana},
  {Pentericci}, {Vanzella}, {Dickinson}, {Kocevski}, {Castellano}, {Cristiani},
  {Ferguson}, {Finkelstein}, {Grogin}, {Hathi}, {Koekemoer}, {Newman}, \&
  {Salvato}}]{Giallongo2015}
{Giallongo}, E., {Grazian}, A., {Fiore}, F., {et~al.} 2015, \aap, 578, A83,
  \dodoi{10.1051/0004-6361/201425334}

\bibitem[{{Giallongo} {et~al.}(2019){Giallongo}, {Grazian}, {Fiore}, {Kodra},
  {Urrutia}, {Castellano}, {Cristiani}, {Dickinson}, {Fontana}, {Menci},
  {Pentericci}, {Boutsia}, {Newman}, \& {Puccetti}}]{Giallongo2019}
---. 2019, \apj, 884, 19, \dodoi{10.3847/1538-4357/ab39e1}

\bibitem[{{Glikman} {et~al.}(2011){Glikman}, {Djorgovski}, {Stern}, {Dey},
  {Jannuzi}, \& {Lee}}]{Glikman2011}
{Glikman}, E., {Djorgovski}, S.~G., {Stern}, D., {et~al.} 2011, \apjl, 728,
  L26, \dodoi{10.1088/2041-8205/728/2/L26}

\bibitem[{{Gloudemans} {et~al.}(2022){Gloudemans}, {Duncan}, {Saxena},
  {Harikane}, {Hill}, {Zeimann}, {Rottgering}, {Yang}, {Best}, {Banados},
  {Drabent}, {Hardcastle}, {Hennawi}, {Lansbury}, {Magliocchetti}, {Miley},
  {Nanni}, {Shimwell}, {Smith}, {Venemans}, \& {Wagenveld}}]{Gloudemans2022}
{Gloudemans}, A.~J., {Duncan}, K.~J., {Saxena}, A., {et~al.} 2022, arXiv
  e-prints, arXiv:2210.01811.
\newblock \doarXiv{2210.01811}

\bibitem[{{G{\'o}rski} {et~al.}(2005){G{\'o}rski}, {Hivon}, {Banday},
  {Wandelt}, {Hansen}, {Reinecke}, \& {Bartelmann}}]{Gorski2005}
{G{\'o}rski}, K.~M., {Hivon}, E., {Banday}, A.~J., {et~al.} 2005, \apj, 622,
  759, \dodoi{10.1086/427976}

\bibitem[{{Goto}(2006)}]{Goto2006}
{Goto}, T. 2006, \mnras, 371, 769, \dodoi{10.1111/j.1365-2966.2006.10702.x}

\bibitem[{{Grazian} {et~al.}(2020){Grazian}, {Giallongo}, {Fiore}, {Boutsia},
  {Civano}, {Cristiani}, {Cupani}, {Dickinson}, {Fontanot}, {Menci}, \&
  {Romano}}]{Grazian2020}
{Grazian}, A., {Giallongo}, E., {Fiore}, F., {et~al.} 2020, \apj, 897, 94,
  \dodoi{10.3847/1538-4357/ab99a3}

\bibitem[{{Grazian} {et~al.}(2022){Grazian}, {Giallongo}, {Boutsia},
  {Calderone}, {Cristiani}, {Cupani}, {Fontanot}, {Guarneri}, \&
  {Ozdalkiran}}]{Grazian2022}
{Grazian}, A., {Giallongo}, E., {Boutsia}, K., {et~al.} 2022, \apj, 924, 62,
  \dodoi{10.3847/1538-4357/ac33a4}

\bibitem[{{Haardt} \& {Madau}(1996)}]{Haardt1996}
{Haardt}, F., \& {Madau}, P. 1996, \apj, 461, 20, \dodoi{10.1086/177035}

\bibitem[{{Haardt} \& {Madau}(2012)}]{Haardt2012}
---. 2012, \apj, 746, 125, \dodoi{10.1088/0004-637X/746/2/125}

\bibitem[{{Harris} {et~al.}(2020){Harris}, {Jarrod Millman}, {van der Walt},
  {Gommers}, {Virtanen}, {Cournapeau}, {Wieser}, {Taylor}, {Berg}, {Smith},
  {Kern}, {Picus}, {Hoyer}, {van Kerkwijk}, {Brett}, {Haldane}, {Fern{\'a}ndez
  del R{\'\i}o}, {Wiebe}, {Peterson}, {G{\'e}rard-Marchant}, {Sheppard},
  {Reddy}, {Weckesser}, {Abbasi}, {Gohlke}, \& {Oliphant}}]{numpy2020}
{Harris}, C.~R., {Jarrod Millman}, K., {van der Walt}, S.~J., {et~al.} 2020,
  arXiv e-prints, arXiv:2006.10256.
\newblock \doarXiv{2006.10256}

\bibitem[{{Hinton}(2016)}]{chainconsumer}
{Hinton}, S.~R. 2016, The Journal of Open Source Software, 1, 00045,
  \dodoi{10.21105/joss.00045}

\bibitem[{{Inayoshi} {et~al.}(2020){Inayoshi}, {Visbal}, \&
  {Haiman}}]{Inayoshi2020}
{Inayoshi}, K., {Visbal}, E., \& {Haiman}, Z. 2020, \araa, 58, 27,
  \dodoi{10.1146/annurev-astro-120419-014455}

\bibitem[{{Jiang} {et~al.}(2015){Jiang}, {McGreer}, {Fan}, {Bian}, {Cai},
  {Cl{\'e}ment}, {Wang}, \& {Fan}}]{Jiang2015}
{Jiang}, L., {McGreer}, I.~D., {Fan}, X., {et~al.} 2015, \aj, 149, 188,
  \dodoi{10.1088/0004-6256/149/6/188}

\bibitem[{{Jiang} {et~al.}(2006){Jiang}, {Fan}, {Hines}, {Shi}, {Vestergaard},
  {Bertoldi}, {Brandt}, {Carilli}, {Cox}, {Le Floc'h}, {Pentericci},
  {Richards}, {Rieke}, {Schneider}, {Strauss}, {Walter}, \&
  {Brinkmann}}]{Jiang2006}
{Jiang}, L., {Fan}, X., {Hines}, D.~C., {et~al.} 2006, \aj, 132, 2127,
  \dodoi{10.1086/508209}

\bibitem[{{Jiang} {et~al.}(2008){Jiang}, {Fan}, {Annis}, {Becker}, {White},
  {Chiu}, {Lin}, {Lupton}, {Richards}, {Strauss}, {Jester}, \&
  {Schneider}}]{Jiang2008}
{Jiang}, L., {Fan}, X., {Annis}, J., {et~al.} 2008, \aj, 135, 1057,
  \dodoi{10.1088/0004-6256/135/3/1057}

\bibitem[{{Jiang} {et~al.}(2016){Jiang}, {McGreer}, {Fan}, {Strauss},
  {Ba{\~n}ados}, {Becker}, {Bian}, {Farnsworth}, {Shen}, {Wang}, {Wang},
  {Wang}, {White}, {Wu}, {Wu}, {Yang}, \& {Yang}}]{Jiang2016}
{Jiang}, L., {McGreer}, I.~D., {Fan}, X., {et~al.} 2016, \apj, 833, 222,
  \dodoi{10.3847/1538-4357/833/2/222}

\bibitem[{{Jiang} {et~al.}(2022){Jiang}, {Ning}, {Fan}, {Ho}, {Luo}, {Wang},
  {Wu}, {Wu}, {Yang}, \& {Zheng}}]{JiangLinhua2022}
{Jiang}, L., {Ning}, Y., {Fan}, X., {et~al.} 2022, Nature Astronomy, 6, 850,
  \dodoi{10.1038/s41550-022-01708-w}

\bibitem[{{Kaiser} {et~al.}(2002){Kaiser}, {Aussel}, {Burke}, {Boesgaard},
  {Chambers}, {Chun}, {Heasley}, {Hodapp}, {Hunt}, {Jedicke}, {Jewitt},
  {Kudritzki}, {Luppino}, {Maberry}, {Magnier}, {Monet}, {Onaka}, {Pickles},
  {Rhoads}, {Simon}, {Szalay}, {Szapudi}, {Tholen}, {Tonry}, {Waterson}, \&
  {Wick}}]{Kaiser2002}
{Kaiser}, N., {Aussel}, H., {Burke}, B.~E., {et~al.} 2002, in \procspie, Vol.
  4836, Survey and Other Telescope Technologies and Discoveries, ed. J.~A.
  {Tyson} \& S.~{Wolff}, 154--164, \dodoi{10.1117/12.457365}

\bibitem[{{Kaiser} {et~al.}(2010){Kaiser}, {Burgett}, {Chambers}, {Denneau},
  {Heasley}, {Jedicke}, {Magnier}, {Morgan}, {Onaka}, \& {Tonry}}]{Kaiser2010}
{Kaiser}, N., {Burgett}, W., {Chambers}, K., {et~al.} 2010, in \procspie, Vol.
  7733, Ground-based and Airborne Telescopes III, 77330E,
  \dodoi{10.1117/12.859188}

\bibitem[{{Kim} {et~al.}(2015){Kim}, {Im}, {Jeon}, {Kim}, {Choi}, {Hong},
  {Hyun}, {Jun}, {Karouzos}, {Kim}, {Kim}, {Kim}, {Kim}, {Lee}, {Pak}, {Park},
  {Taak}, \& {Yoon}}]{KimYongjung2015}
{Kim}, Y., {Im}, M., {Jeon}, Y., {et~al.} 2015, \apjl, 813, L35,
  \dodoi{10.1088/2041-8205/813/2/L35}

\bibitem[{{Kim} {et~al.}(2020){Kim}, {Im}, {Jeon}, {Kim}, {Pak}, {Hyun},
  {Taak}, {Shin}, {Lim}, {Paek}, {Paek}, {Jiang}, {Choi}, {Hong}, {Ji}, {Jun},
  {Karouzos}, {Kim}, {Kim}, {Kim}, {Kim}, {Lee}, {Lee}, {Park}, {Yoon},
  {Byeon}, {Hwang}, {Kim}, {Kim}, \& {Park}}]{KimYongjung2020}
---. 2020, \apj, 904, 111, \dodoi{10.3847/1538-4357/abc0ea}

\bibitem[{{Kormendy} \& {Ho}(2013)}]{Kormendy2013}
{Kormendy}, J., \& {Ho}, L.~C. 2013, \araa, 51, 511,
  \dodoi{10.1146/annurev-astro-082708-101811}

\bibitem[{{Kulkarni} {et~al.}(2019){Kulkarni}, {Worseck}, \&
  {Hennawi}}]{Kulkarni2019}
{Kulkarni}, G., {Worseck}, G., \& {Hennawi}, J.~F. 2019, \mnras, 488, 1035,
  \dodoi{10.1093/mnras/stz1493}

\bibitem[{{Kurk} {et~al.}(2007){Kurk}, {Walter}, {Fan}, {Jiang}, {Riechers},
  {Rix}, {Pentericci}, {Strauss}, {Carilli}, \& {Wagner}}]{Kurk2007}
{Kurk}, J.~D., {Walter}, F., {Fan}, X., {et~al.} 2007, \apj, 669, 32,
  \dodoi{10.1086/521596}

\bibitem[{{Lusso} {et~al.}(2015){Lusso}, {Worseck}, {Hennawi}, {Prochaska},
  {Vignali}, {Stern}, \& {O'Meara}}]{Lusso2015}
{Lusso}, E., {Worseck}, G., {Hennawi}, J.~F., {et~al.} 2015, \mnras, 449, 4204,
  \dodoi{10.1093/mnras/stv516}

\bibitem[{{Lyke} {et~al.}(2020){Lyke}, {Higley}, {McLane}, {Schurhammer},
  {Myers}, {Ross}, {Dawson}, {Chabanier}, {Martini}, {Busca}, {Mas des
  Bourboux}, {Salvato}, {Streblyanska}, {Zarrouk}, {Burtin}, {Anderson},
  {Bautista}, {Bizyaev}, {Brandt}, {Brinkmann}, {Brownstein}, {Comparat},
  {Green}, {de la Macorra}, {Mu{\~n}oz Guti{\'e}rrez}, {Hou}, {Newman},
  {Palanque-Delabrouille}, {P{\^a}ris}, {Percival}, {Petitjean}, {Rich},
  {Rossi}, {Schneider}, {Smith}, {Vivek}, \& {Weaver}}]{Lyke2020}
{Lyke}, B.~W., {Higley}, A.~N., {McLane}, J.~N., {et~al.} 2020, \apjs, 250, 8,
  \dodoi{10.3847/1538-4365/aba623}

\bibitem[{{Lyu} \& {Rieke}(2017)}]{LyuJianwei2017b}
{Lyu}, J., \& {Rieke}, G.~H. 2017, \apj, 841, 76,
  \dodoi{10.3847/1538-4357/aa7051}

\bibitem[{{Madau} \& {Haardt}(2015)}]{Madau2015}
{Madau}, P., \& {Haardt}, F. 2015, \apjl, 813, L8,
  \dodoi{10.1088/2041-8205/813/1/L8}

\bibitem[{{Madau} {et~al.}(1999){Madau}, {Haardt}, \& {Rees}}]{Madau1999}
{Madau}, P., {Haardt}, F., \& {Rees}, M.~J. 1999, \apj, 514, 648,
  \dodoi{10.1086/306975}

\bibitem[{{Magnier} {et~al.}(2020{\natexlab{a}}){Magnier}, {Chambers},
  {Flewelling}, {Hoblitt}, {Huber}, {Price}, {Sweeney}, {Waters}, {Denneau},
  {Draper}, {Hodapp}, {Jedicke}, {Kaiser}, {Kudritzki}, {Metcalfe}, {Stubbs},
  \& {Wainscoat}}]{Magnier2020DataProcessing}
{Magnier}, E.~A., {Chambers}, K.~C., {Flewelling}, H.~A., {et~al.}
  2020{\natexlab{a}}, \apjs, 251, 3, \dodoi{10.3847/1538-4365/abb829}

\bibitem[{{Magnier} {et~al.}(2020{\natexlab{b}}){Magnier}, {Sweeney},
  {Chambers}, {Flewelling}, {Huber}, {Price}, {Waters}, {Denneau}, {Draper},
  {Farrow}, {Jedicke}, {Hodapp}, {Kaiser}, {Kudritzki}, {Metcalfe}, {Stubbs},
  \& {Wainscoat}}]{Magnier2020PixelAnalysis}
{Magnier}, E.~A., {Sweeney}, W.~E., {Chambers}, K.~C., {et~al.}
  2020{\natexlab{b}}, \apjs, 251, 5, \dodoi{10.3847/1538-4365/abb82c}

\bibitem[{{Marshall} {et~al.}(1983){Marshall}, {Tananbaum}, {Avni}, \&
  {Zamorani}}]{Marshall1983}
{Marshall}, H.~L., {Tananbaum}, H., {Avni}, Y., \& {Zamorani}, G. 1983, \apj,
  269, 35, \dodoi{10.1086/161016}

\bibitem[{{Matsuoka} {et~al.}(2016){Matsuoka}, {Onoue}, {Kashikawa}, {Iwasawa},
  {Strauss}, {Nagao}, {Imanishi}, {Niida}, {Toba}, {Akiyama}, {Asami}, {Bosch},
  {Foucaud}, {Furusawa}, {Goto}, {Gunn}, {Harikane}, {Ikeda}, {Kawaguchi},
  {Kikuta}, {Komiyama}, {Lupton}, {Minezaki}, {Miyazaki}, {Morokuma},
  {Murayama}, {Nishizawa}, {Ono}, {Ouchi}, {Price}, {Sameshima}, {Silverman},
  {Sugiyama}, {Tait}, {Takada}, {Takata}, {Tanaka}, {Tang}, \&
  {Utsumi}}]{Matsuoka2016}
{Matsuoka}, Y., {Onoue}, M., {Kashikawa}, N., {et~al.} 2016, \apj, 828, 26,
  \dodoi{10.3847/0004-637X/828/1/26}

\bibitem[{{Matsuoka} {et~al.}(2018){Matsuoka}, {Strauss}, {Kashikawa}, {Onoue},
  {Iwasawa}, {Tang}, {Lee}, {Imanishi}, {Nagao}, {Akiyama}, {Asami}, {Bosch},
  {Furusawa}, {Goto}, {Gunn}, {Harikane}, {Ikeda}, {Izumi}, {Kawaguchi},
  {Kato}, {Kikuta}, {Kohno}, {Komiyama}, {Lupton}, {Minezaki}, {Miyazaki},
  {Murayama}, {Niida}, {Nishizawa}, {Noboriguchi}, {Oguri}, {Ono}, {Ouchi},
  {Price}, {Sameshima}, {Schulze}, {Shirakata}, {Silverman}, {Sugiyama},
  {Tait}, {Takada}, {Takata}, {Tanaka}, {Toba}, {Utsumi}, {Wang}, \&
  {Yamashita}}]{Matsuoka2018c}
{Matsuoka}, Y., {Strauss}, M.~A., {Kashikawa}, N., {et~al.} 2018, \apj, 869,
  150, \dodoi{10.3847/1538-4357/aaee7a}

\bibitem[{{McGreer} {et~al.}(2021){McGreer}, {Moustakas}, \&
  {Schindler}}]{McGreer2021}
{McGreer}, I., {Moustakas}, J., \& {Schindler}, J. 2021, {simqso: Simulated
  quasar spectra generator}, Astrophysics Source Code Library, record
  ascl:2106.008.
\newblock \doeprint{2106.008}

\bibitem[{{McGreer} {et~al.}(2018){McGreer}, {Fan}, {Jiang}, \&
  {Cai}}]{McGreer2018}
{McGreer}, I.~D., {Fan}, X., {Jiang}, L., \& {Cai}, Z. 2018, \aj, 155, 131,
  \dodoi{10.3847/1538-3881/aaaab4}

\bibitem[{{McGreer} {et~al.}(2013){McGreer}, {Jiang}, {Fan}, {Richards},
  {Strauss}, {Ross}, {White}, {Shen}, {Schneider}, {Myers}, {Brandt}, {DeGraf},
  {Glikman}, {Ge}, \& {Streblyanska}}]{McGreer2013}
{McGreer}, I.~D., {Jiang}, L., {Fan}, X., {et~al.} 2013, \apj, 768, 105,
  \dodoi{10.1088/0004-637X/768/2/105}

\bibitem[{{Morganson} {et~al.}(2012){Morganson}, {De Rosa}, {Decarli},
  {Walter}, {Chambers}, {McGreer}, {Fan}, {Burgett}, {Flewelling}, {Greiner},
  {Hodapp}, {Kaiser}, {Magnier}, {Price}, {Rix}, {Sweeney}, \&
  {Waters}}]{Morganson2012}
{Morganson}, E., {De Rosa}, G., {Decarli}, R., {et~al.} 2012, \aj, 143, 142,
  \dodoi{10.1088/0004-6256/143/6/142}

\bibitem[{{Mortlock} {et~al.}(2009){Mortlock}, {Patel}, {Warren}, {Venemans},
  {McMahon}, {Hewett}, {Simpson}, {Sharp}, {Burningham}, {Dye}, {Ellis},
  {Gonzales-Solares}, \& {Hu{\'e}lamo}}]{Mortlock2009}
{Mortlock}, D.~J., {Patel}, M., {Warren}, S.~J., {et~al.} 2009, \aap, 505, 97,
  \dodoi{10.1051/0004-6361/200811161}

\bibitem[{{Neeleman} {et~al.}(2021){Neeleman}, {Novak}, {Venemans}, {Walter},
  {Decarli}, {Kaasinen}, {Schindler}, {Ba{\~n}ados}, {Carilli}, {Drake}, {Fan},
  \& {Rix}}]{Neeleman2021}
{Neeleman}, M., {Novak}, M., {Venemans}, B.~P., {et~al.} 2021, \apj, 911, 141,
  \dodoi{10.3847/1538-4357/abe70f}

\bibitem[{Newville {et~al.}(2014)Newville, Stensitzki, Allen, \&
  Ingargiola}]{lmfit2014}
Newville, M., Stensitzki, T., Allen, D.~B., \& Ingargiola, A. 2014, {LMFIT:
  Non-Linear Least-Square Minimization and Curve-Fitting for Python}, 0.8.0,
  Zenodo, \dodoi{10.5281/zenodo.11813}

\bibitem[{{Niida} {et~al.}(2020){Niida}, {Nagao}, {Ikeda}, {Akiyama},
  {Matsuoka}, {He}, {Matsuoka}, {Toba}, {Onoue}, {Kobayashi}, {Taniguchi},
  {Furusawa}, {Harikane}, {Imanishi}, {Kashikawa}, {Kawaguchi}, {Komiyama},
  {Shirakata}, {Terashima}, \& {Ueda}}]{Niida2020}
{Niida}, M., {Nagao}, T., {Ikeda}, H., {et~al.} 2020, \apj, 904, 89,
  \dodoi{10.3847/1538-4357/abbe11}

\bibitem[{{Onken} {et~al.}(2022){Onken}, {Wolf}, {Bian}, {Fan}, {Hon},
  {Raithel}, {Tisserand}, \& {Lai}}]{Onken2022}
{Onken}, C.~A., {Wolf}, C., {Bian}, F., {et~al.} 2022, \mnras, 511, 572,
  \dodoi{10.1093/mnras/stac051}

\bibitem[{{Pacucci} \& {Loeb}(2019)}]{Pacucci2019}
{Pacucci}, F., \& {Loeb}, A. 2019, \apjl, 870, L12,
  \dodoi{10.3847/2041-8213/aaf86a}

\bibitem[{{Page} \& {Carrera}(2000)}]{Page2000}
{Page}, M.~J., \& {Carrera}, F.~J. 2000, \mnras, 311, 433,
  \dodoi{10.1046/j.1365-8711.2000.03105.x}

\bibitem[{{Pan} {et~al.}(2022){Pan}, {Jiang}, {Fan}, {Wu}, \&
  {Yang}}]{PanZhiwei2022}
{Pan}, Z., {Jiang}, L., {Fan}, X., {Wu}, J., \& {Yang}, J. 2022, \apj, 928,
  172, \dodoi{10.3847/1538-4357/ac5aab}

\bibitem[{pandas~development team(2020)}]{pandas_software}
pandas~development team, T. 2020, pandas-dev/pandas: Pandas, latest,  Zenodo,
  \dodoi{10.5281/zenodo.3509134}

\bibitem[{{P{\^a}ris} {et~al.}(2014){P{\^a}ris}, {Petitjean}, {Aubourg},
  {Ross}, {Myers}, {Streblyanska}, {Bailey}, {Hall}, {Strauss}, {Anderson},
  {Bizyaev}, {Borde}, {Brinkmann}, {Bovy}, {Brandt}, {Brewington},
  {Brownstein}, {Cook}, {Ebelke}, {Fan}, {Filiz Ak}, {Finley}, {Font-Ribera},
  {Ge}, {Hamann}, {Ho}, {Jiang}, {Kinemuchi}, {Malanushenko}, {Malanushenko},
  {Marchante}, {McGreer}, {McMahon}, {Miralda-Escud{\'e}}, {Muna},
  {Noterdaeme}, {Oravetz}, {Palanque-Delabrouille}, {Pan}, {Perez-Fournon},
  {Pieri}, {Riffel}, {Schlegel}, {Schneider}, {Simmons}, {Viel}, {Weaver},
  {Wood-Vasey}, {Y{\`e}che}, \& {York}}]{Paris2014}
{P{\^a}ris}, I., {Petitjean}, P., {Aubourg}, {\'E}., {et~al.} 2014, \aap, 563,
  A54, \dodoi{10.1051/0004-6361/201322691}

\bibitem[{{P{\^a}ris} {et~al.}(2017){P{\^a}ris}, {Petitjean}, {Ross}, {Myers},
  {Aubourg}, {Streblyanska}, {Bailey}, {Armengaud}, {Palanque-Delabrouille},
  {Y{\`e}che}, {Hamann}, {Strauss}, {Albareti}, {Bovy}, {Bizyaev}, {Niel
  Brandt}, {Brusa}, {Buchner}, {Comparat}, {Croft}, {Dwelly}, {Fan},
  {Font-Ribera}, {Ge}, {Georgakakis}, {Hall}, {Jiang}, {Kinemuchi},
  {Malanushenko}, {Malanushenko}, {McMahon}, {Menzel}, {Merloni}, {Nandra},
  {Noterdaeme}, {Oravetz}, {Pan}, {Pieri}, {Prada}, {Salvato}, {Schlegel},
  {Schneider}, {Simmons}, {Viel}, {Weinberg}, \& {Zhu}}]{Paris2017}
{P{\^a}ris}, I., {Petitjean}, P., {Ross}, N.~P., {et~al.} 2017, \aap, 597, A79,
  \dodoi{10.1051/0004-6361/201527999}

\bibitem[{{Parsa} {et~al.}(2018){Parsa}, {Dunlop}, \& {McLure}}]{Parsa2018}
{Parsa}, S., {Dunlop}, J.~S., \& {McLure}, R.~J. 2018, \mnras, 474, 2904,
  \dodoi{10.1093/mnras/stx2887}

\bibitem[{{Pei}(1995)}]{Pei1995}
{Pei}, Y.~C. 1995, \apj, 438, 623, \dodoi{10.1086/175105}

\bibitem[{{Ricci} {et~al.}(2017){Ricci}, {Marchesi}, {Shankar}, {La Franca}, \&
  {Civano}}]{RicciF2017}
{Ricci}, F., {Marchesi}, S., {Shankar}, F., {La Franca}, F., \& {Civano}, F.
  2017, \mnras, 465, 1915, \dodoi{10.1093/mnras/stw2909}

\bibitem[{{Richards} {et~al.}(2006){Richards}, {Strauss}, {Fan}, {Hall},
  {Jester}, {Schneider}, {Vanden Berk}, {Stoughton}, {Anderson}, {Brunner},
  {Gray}, {Gunn}, {Ivezi{\'c}}, {Kirkland}, {Knapp}, {Loveday}, {Meiksin},
  {Pope}, {Szalay}, {Thakar}, {Yanny}, {York}, {Barentine}, {Brewington},
  {Brinkmann}, {Fukugita}, {Harvanek}, {Kent}, {Kleinman}, {Krzesi{\'n}ski},
  {Long}, {Lupton}, {Nash}, {Neilsen}, {Nitta}, {Schlegel}, \&
  {Snedden}}]{Richards2006}
{Richards}, G.~T., {Strauss}, M.~A., {Fan}, X., {et~al.} 2006, \aj, 131, 2766,
  \dodoi{10.1086/503559}

\bibitem[{{Rix} {et~al.}(2021){Rix}, {Hogg}, {Boubert}, {Brown}, {Casey},
  {Drimmel}, {Everall}, {Fouesneau}, \& {Price-Whelan}}]{Rix2021arxiv}
{Rix}, H.-W., {Hogg}, D.~W., {Boubert}, D., {et~al.} 2021, arXiv e-prints,
  arXiv:2106.07653.
\newblock \doarXiv{2106.07653}

\bibitem[{{Rojas-Ruiz} {et~al.}(2021){Rojas-Ruiz}, {Ba{\~n}ados}, {Neeleman},
  {Connor}, {Eilers}, {Venemans}, {Khusanova}, {Carilli}, {Mazzucchelli},
  {Decarli}, {Momjian}, \& {Novak}}]{RojasRuiz2021}
{Rojas-Ruiz}, S., {Ba{\~n}ados}, E., {Neeleman}, M., {et~al.} 2021, \apj, 920,
  150, \dodoi{10.3847/1538-4357/ac1a13}

\bibitem[{{Ross} {et~al.}(2013){Ross}, {McGreer}, {White}, {Richards}, {Myers},
  {Palanque-Delabrouille}, {Strauss}, {Anderson}, {Shen}, {Brandt},
  {Y{\`e}che}, {Swanson}, {Aubourg}, {Bailey}, {Bizyaev}, {Bovy}, {Brewington},
  {Brinkmann}, {DeGraf}, {Di Matteo}, {Ebelke}, {Fan}, {Ge}, {Malanushenko},
  {Malanushenko}, {Mandelbaum}, {Maraston}, {Muna}, {Oravetz}, {Pan},
  {P{\^a}ris}, {Petitjean}, {Schawinski}, {Schlegel}, {Schneider}, {Silverman},
  {Simmons}, {Snedden}, {Streblyanska}, {Suzuki}, {Weinberg}, \&
  {York}}]{Ross2013}
{Ross}, N.~P., {McGreer}, I.~D., {White}, M., {et~al.} 2013, \apj, 773, 14,
  \dodoi{10.1088/0004-637X/773/1/14}

\bibitem[{{Scaramella} {et~al.}(2021){Scaramella}, {Amiaux}, {Mellier},
  {Burigana}, {Carvalho}, {Cuillandre}, {Da Silva}, {Derosa}, {Dinis},
  {Maiorano}, {Maris}, {Tereno}, {Laureijs}, {Boenke}, {Buenadicha}, {Dupac},
  {Gaspar Venancio}, {G{\'o}mez-{\'A}lvarez}, {Hoar}, {Alvarez}, {Racca},
  {Saavedra-Criado}, {Schwartz}, {Vavrek}, {Schirmer}, {Aussel}, {Azzollini},
  {Cardone}, {Cropper}, {Ealet}, {Garilli}, {Gillard}, {Granett}, {Guzzo},
  {Hoekstra}, {Jahnke}, {Kitching}, {Meneghetti}, {Miller}, {Nakajima},
  {Niemi}, {Pasian}, {Percival}, {Sauvage}, {Scodeggio}, {Wachter}, {Zacchei},
  {Aghanim}, {Amara}, {Auphan}, {Auricchio}, {Awan}, {Balestra}, {Bender},
  {Bodendorf}, {Bonino}, {Branchini}, {Brau-Nogue}, {Brescia}, {Candini},
  {Capobianco}, {Carbone}, {Carlberg}, {Carretero}, {Casas}, {Castander},
  {Castellano}, {Cavuoti}, {Cimatti}, {Cledassou}, {Congedo}, {Conselice},
  {Conversi}, {Copin}, {Corcione}, {Costille}, {Courbin}, {Degaudenzi},
  {Douspis}, {Dubath}, {Duncan}, {Dusini}, {Farrens}, {Ferriol}, {Fosalba},
  {Fourmanoit}, {Frailis}, {Franceschi}, {Franzetti}, {Fumana}, {Gillis},
  {Giocoli}, {Grazian}, {Grupp}, {Haugan}, {Holmes}, {Hormuth}, {Hudelot},
  {Kermiche}, {Kiessling}, {Kilbinger}, {Kohley}, {Kubik}, {K{\"u}mmel},
  {Kunz}, {Kurki-Suonio}, {Ligori}, {Lilje}, {Lloro}, {Mansutti}, {Marggraf},
  {Markovic}, {Marulli}, {Massey}, {Maurogordato}, {Melchior}, {Merlin},
  {Meylan}, {Mohr}, {Moresco}, {Morin}, {Moscardini}, {Munari}, {Nichol},
  {Padilla}, {Paltani}, {Peacock}, {Pedersen}, {Pettorino}, {Pires}, {Poncet},
  {Popa}, {Pozzetti}, {Raison}, {Rebolo}, {Rhodes}, {Rix}, {Roncarelli},
  {Rossetti}, {Saglia}, {Schneider}, {Schrabback}, {Secroun}, {Seidel},
  {Serrano}, {Sirignano}, {Sirri}, {Skottfelt}, {Stanco}, {Starck},
  {Tallada-Cresp{\'\i}}, {Tavagnacco}, {Taylor}, {Teplitz}, {Toledo-Moreo},
  {Torradeflot}, {Trifoglio}, {Valentijn}, {Valenziano}, {Verdoes Kleijn},
  {Wang}, {Welikala}, {Weller}, {Wetzstein}, {Zamorani}, {Zoubian}, {Andreon},
  {Baldi}, {Bardelli}, {Boucaud}, {Camera}, {Fabbian}, {Farinelli},
  {Graci{\'a}-Carpio}, {Maino}, {Medinaceli}, {Mei}, {Neissner}, {Polenta},
  {Renzi}, {Romelli}, {Rosset}, {Sureau}, {Tenti}, {Vassallo}, {Zucca},
  {Baccigalupi}, {Balaguera-Antol{\'\i}nez}, {Battaglia}, {Biviano}, {Borgani},
  {Bozzo}, {Cabanac}, {Cappi}, {Casas}, {Castignani}, {Colodro-Conde},
  {Coupon}, {Courtois}, {Cuby}, {de la Torre}, {Desai}, {Di Ferdinando},
  {Dole}, {Fabricius}, {Farina}, {Ferreira}, {Finelli}, {Flose-Reimberg},
  {Fotopoulou}, {Galeotta}, {Ganga}, {Gozaliasl}, {Hook}, {Keihanen},
  {Kirkpatrick}, {Liebing}, {Lindholm}, {Mainetti}, {Martinelli}, {Martinet},
  {Maturi}, {McCracken}, {Metcalf}, {Morgante}, {Nightingale}, {Nucita},
  {Patrizii}, {Potter}, {Riccio}, {S{\'a}nchez}, {Sapone}, {Schewtschenko},
  {Schultheis}, {Scottez}, {Teyssier}, {Tutusaus}, {Valiviita}, {Viel},
  {Vriend}, \& {Whittaker}}]{EuclidPrepI2021}
{Scaramella}, R., {Amiaux}, J., {Mellier}, Y., {et~al.} 2021, arXiv e-prints,
  arXiv:2108.01201.
\newblock \doarXiv{2108.01201}

\bibitem[{{Schindler} {et~al.}(2018){Schindler}, {Fan}, {McGreer}, {Yang},
  {Wang}, {Green}, {Garavito-Camargo}, {Huang}, {O{\textquoteright}Donnell},
  {Patej}, {Pucha}, {Rees}, \& {Spalding}}]{Schindler2018}
{Schindler}, J.-T., {Fan}, X., {McGreer}, I.~D., {et~al.} 2018, \apj, 863, 144,
  \dodoi{10.3847/1538-4357/aad2dd}

\bibitem[{{Schindler} {et~al.}(2019){Schindler}, {Fan}, {McGreer}, {Yang},
  {Wang}, {Green}, {Fynbo}, {Krogager}, {Green}, {Huang}, {Kadowaki}, {Patej},
  {Wu}, \& {Yue}}]{Schindler2019a}
---. 2019, \apj, 871, 258, \dodoi{10.3847/1538-4357/aaf86c}

\bibitem[{{Schlafly} \& {Finkbeiner}(2011)}]{Schlafly2011}
{Schlafly}, E.~F., \& {Finkbeiner}, D.~P. 2011, \apj, 737, 103,
  \dodoi{10.1088/0004-637X/737/2/103}

\bibitem[{{Schlegel} {et~al.}(1998){Schlegel}, {Finkbeiner}, \&
  {Davis}}]{Schlegel1998}
{Schlegel}, D.~J., {Finkbeiner}, D.~P., \& {Davis}, M. 1998, \apj, 500, 525,
  \dodoi{10.1086/305772}

\bibitem[{{Schmidt}(1968)}]{Schmidt1968}
{Schmidt}, M. 1968, \apj, 151, 393, \dodoi{10.1086/149446}

\bibitem[{{Schmidt} {et~al.}(1995){Schmidt}, {Schneider}, \&
  {Gunn}}]{Schmidt1995}
{Schmidt}, M., {Schneider}, D.~P., \& {Gunn}, J.~E. 1995, \aj, 110, 68,
  \dodoi{10.1086/117497}

\bibitem[{{Schneider} {et~al.}(2010){Schneider}, {Richards}, {Hall}, {Strauss},
  {Anderson}, {Boroson}, {Ross}, {Shen}, {Brandt}, {Fan}, {Inada}, {Jester},
  {Knapp}, {Krawczyk}, {Thakar}, {Vanden Berk}, {Voges}, {Yanny}, {York},
  {Bahcall}, {Bizyaev}, {Blanton}, {Brewington}, {Brinkmann}, {Eisenstein},
  {Frieman}, {Fukugita}, {Gray}, {Gunn}, {Hibon}, {Ivezi{\'c}}, {Kent}, {Kron},
  {Lee}, {Lupton}, {Malanushenko}, {Malanushenko}, {Oravetz}, {Pan}, {Pier},
  {Price}, {Saxe}, {Schlegel}, {Simmons}, {Snedden}, {SubbaRao}, {Szalay}, \&
  {Weinberg}}]{Schneider2010}
{Schneider}, D.~P., {Richards}, G.~T., {Hall}, P.~B., {et~al.} 2010, \aj, 139,
  2360, \dodoi{10.1088/0004-6256/139/6/2360}

\bibitem[{{Selsing} {et~al.}(2016){Selsing}, {Fynbo}, {Christensen}, \&
  {Krogager}}]{Selsing2016}
{Selsing}, J., {Fynbo}, J.~P.~U., {Christensen}, L., \& {Krogager}, J.-K. 2016,
  \aap, 585, A87, \dodoi{10.1051/0004-6361/201527096}

\bibitem[{{Shen} {et~al.}(2020){Shen}, {Hopkins}, {Faucher-Gigu{\`e}re},
  {Alexander}, {Richards}, {Ross}, \& {Hickox}}]{ShenXuejian2020}
{Shen}, X., {Hopkins}, P.~F., {Faucher-Gigu{\`e}re}, C.-A., {et~al.} 2020,
  \mnras, 495, 3252, \dodoi{10.1093/mnras/staa1381}

\bibitem[{{Shen} \& {Kelly}(2012)}]{Shen2012}
{Shen}, Y., \& {Kelly}, B.~C. 2012, \apj, 746, 169,
  \dodoi{10.1088/0004-637X/746/2/169}

\bibitem[{{Shen} {et~al.}(2019){Shen}, {Wu}, {Jiang}, {Ba{\~n}ados}, {Fan},
  {Ho}, {Riechers}, {Strauss}, {Venemans}, {Vestergaard}, {Walter}, {Wang},
  {Willott}, {Wu}, \& {Yang}}]{Shen2019a}
{Shen}, Y., {Wu}, J., {Jiang}, L., {et~al.} 2019, \apj, 873, 35,
  \dodoi{10.3847/1538-4357/ab03d9}

\bibitem[{{Shin} {et~al.}(2020){Shin}, {Im}, {Kim}, {Hyun}, {Pak}, {Jeon},
  {Ji}, {Ahn}, {Byeon}, {Han}, {Hwang}, {Kim}, {Lim}, {Paek}, {Paek}, {Taak},
  {Choi}, {Hong}, {Jun}, {Kim}, {Kim}, {Kim}, {Kim}, {Kim}, {Lee}, {Lee},
  {Park}, {Park}, \& {Yoon}}]{ShinSuhyun2020}
{Shin}, S., {Im}, M., {Kim}, Y., {et~al.} 2020, \apj, 893, 45,
  \dodoi{10.3847/1538-4357/ab7bde}

\bibitem[{{Temple} {et~al.}(2021){Temple}, {Hewett}, \& {Banerji}}]{Temple2021}
{Temple}, M.~J., {Hewett}, P.~C., \& {Banerji}, M. 2021, \mnras, 508, 737,
  \dodoi{10.1093/mnras/stab2586}

\bibitem[{{Tsuzuki} {et~al.}(2006){Tsuzuki}, {Kawara}, {Yoshii}, {Oyabu},
  {Tanab{\'e}}, \& {Matsuoka}}]{Tsuzuki2006}
{Tsuzuki}, Y., {Kawara}, K., {Yoshii}, Y., {et~al.} 2006, \apj, 650, 57,
  \dodoi{10.1086/506376}

\bibitem[{{van der Walt} {et~al.}(2011){van der Walt}, {Colbert}, \&
  {Varoquaux}}]{numpy}
{van der Walt}, S., {Colbert}, S.~C., \& {Varoquaux}, G. 2011, Computing in
  Science Engineering, 13, 22

\bibitem[{{Venemans} {et~al.}(2007){Venemans}, {McMahon}, {Warren},
  {Gonzalez-Solares}, {Hewett}, {Mortlock}, {Dye}, \& {Sharp}}]{Venemans2007}
{Venemans}, B.~P., {McMahon}, R.~G., {Warren}, S.~J., {et~al.} 2007, \mnras,
  376, L76, \dodoi{10.1111/j.1745-3933.2007.00290.x}

\bibitem[{{Venemans} {et~al.}(2013){Venemans}, {Findlay}, {Sutherland}, {De
  Rosa}, {McMahon}, {Simcoe}, {Gonz{\'a}lez-Solares}, {Kuijken}, \&
  {Lewis}}]{Venemans2013}
{Venemans}, B.~P., {Findlay}, J.~R., {Sutherland}, W.~J., {et~al.} 2013, \apj,
  779, 24, \dodoi{10.1088/0004-637X/779/1/24}

\bibitem[{{Venemans} {et~al.}(2020){Venemans}, {Walter}, {Neeleman}, {Novak},
  {Otter}, {Decarli}, {Ba{\~n}ados}, {Drake}, {Farina}, {Kaasinen},
  {Mazzucchelli}, {Carilli}, {Fan}, {Rix}, \& {Wang}}]{Venemans2020}
{Venemans}, B.~P., {Walter}, F., {Neeleman}, M., {et~al.} 2020, \apj, 904, 130,
  \dodoi{10.3847/1538-4357/abc563}

\bibitem[{{Verner} {et~al.}(1996){Verner}, {Ferland}, {Korista}, \&
  {Yakovlev}}]{Verner1996}
{Verner}, D.~A., {Ferland}, G.~J., {Korista}, K.~T., \& {Yakovlev}, D.~G. 1996,
  \apj, 465, 487, \dodoi{10.1086/177435}

\bibitem[{{Vestergaard} \& {Wilkes}(2001)}]{Vestergaard2001}
{Vestergaard}, M., \& {Wilkes}, B.~J. 2001, \apjs, 134, 1,
  \dodoi{10.1086/320357}

\bibitem[{{Virtanen} {et~al.}(2020){Virtanen}, {Gommers}, {Oliphant},
  {Haberland}, {Reddy}, {Cournapeau}, {Burovski}, {Peterson}, {Weckesser},
  {Bright}, {van der Walt}, {Brett}, {Wilson}, {Jarrod Millman}, {Mayorov},
  {Nelson}, {Jones}, {Kern}, {Larson}, {Carey}, {Polat}, {Feng}, {Moore}, {Vand
  erPlas}, {Laxalde}, {Perktold}, {Cimrman}, {Henriksen}, {Quintero}, {Harris},
  {Archibald}, {Ribeiro}, {Pedregosa}, {van Mulbregt}, \&
  {Contributors}}]{scipy}
{Virtanen}, P., {Gommers}, R., {Oliphant}, T.~E., {et~al.} 2020, Nature
  Methods, 17, 261, \dodoi{https://doi.org/10.1038/s41592-019-0686-2}

\bibitem[{{Vito} {et~al.}(2016){Vito}, {Gilli}, {Vignali}, {Brandt},
  {Comastri}, {Yang}, {Lehmer}, {Luo}, {Basu-Zych}, {Bauer}, {Cappelluti},
  {Koekemoer}, {Mainieri}, {Paolillo}, {Ranalli}, {Shemmer}, {Trump}, {Wang},
  \& {Xue}}]{Vito2016}
{Vito}, F., {Gilli}, R., {Vignali}, C., {et~al.} 2016, \mnras, 463, 348,
  \dodoi{10.1093/mnras/stw1998}

\bibitem[{{Wang} {et~al.}(2019{\natexlab{a}}){Wang}, {Wang}, {Fan}, {Wu},
  {Yang}, {Neri}, \& {Yue}}]{WangFeige2019a}
{Wang}, F., {Wang}, R., {Fan}, X., {et~al.} 2019{\natexlab{a}}, \apj, 880, 2,
  \dodoi{10.3847/1538-4357/ab2717}

\bibitem[{{Wang} {et~al.}(2016{\natexlab{a}}){Wang}, {Wu}, {Fan}, {Yang}, {Yi},
  {Bian}, {McGreer}, {Yang}, {Ai}, {Dong}, {Zuo}, {Jiang}, {Green}, {Wang},
  {Cai}, {Wang}, \& {Yue}}]{WangFeige2016}
{Wang}, F., {Wu}, X.-B., {Fan}, X., {et~al.} 2016{\natexlab{a}}, \apj, 819, 24,
  \dodoi{10.3847/0004-637X/819/1/24}

\bibitem[{{Wang} {et~al.}(2017){Wang}, {Fan}, {Yang}, {Wu}, {Yang}, {Bian},
  {McGreer}, {Li}, {Li}, {Ding}, {Dey}, {Dye}, {Findlay}, {Green}, {James},
  {Jiang}, {Lang}, {Lawrence}, {Myers}, {Ross}, {Schlegel}, \&
  {Shanks}}]{WangFeige2017}
{Wang}, F., {Fan}, X., {Yang}, J., {et~al.} 2017, \apj, 839, 27,
  \dodoi{10.3847/1538-4357/aa689f}

\bibitem[{{Wang} {et~al.}(2018){Wang}, {Yang}, {Fan}, {Yue}, {Wu}, {Schindler},
  {Bian}, {Li}, {Farina}, {Ba{\~n}ados}, {Davies}, {Decarli}, {Green}, {Jiang},
  {Hennawi}, {Huang}, {Mazzucchelli}, {McGreer}, {Venemans}, {Walter}, \&
  {Beletsky}}]{WangFeige2018}
{Wang}, F., {Yang}, J., {Fan}, X., {et~al.} 2018, \apjl, 869, L9,
  \dodoi{10.3847/2041-8213/aaf1d2}

\bibitem[{{Wang} {et~al.}(2019{\natexlab{b}}){Wang}, {Yang}, {Fan}, {Wu},
  {Yue}, {Li}, {Bian}, {Jiang}, {Ba{\~n}ados}, {Schindler}, {Findlay},
  {Davies}, {Decarli}, {Farina}, {Green}, {Hennawi}, {Huang}, {Mazzuccheli},
  {McGreer}, {Venemans}, {Walter}, {Dye}, {Lyke}, {Myers}, \& {Haze
  Nunez}}]{WangFeige2019}
---. 2019{\natexlab{b}}, \apj, 884, 30, \dodoi{10.3847/1538-4357/ab2be5}

\bibitem[{{Wang} {et~al.}(2021){Wang}, {Yang}, {Fan}, {Hennawi}, {Barth},
  {Banados}, {Bian}, {Boutsia}, {Connor}, {Davies}, {Decarli}, {Eilers},
  {Farina}, {Green}, {Jiang}, {Li}, {Mazzucchelli}, {Nanni}, {Schindler},
  {Venemans}, {Walter}, {Wu}, \& {Yue}}]{WangFeige2021b}
---. 2021, \apjl, 907, L1, \dodoi{10.3847/2041-8213/abd8c6}

\bibitem[{{Wang} {et~al.}(2010){Wang}, {Carilli}, {Neri}, {Riechers}, {Wagg},
  {Walter}, {Bertoldi}, {Menten}, {Omont}, {Cox}, \& {Fan}}]{WangRan2010}
{Wang}, R., {Carilli}, C.~L., {Neri}, R., {et~al.} 2010, \apj, 714, 699,
  \dodoi{10.1088/0004-637X/714/1/699}

\bibitem[{{Wang} {et~al.}(2011){Wang}, {Wagg}, {Carilli}, {Neri}, {Walter},
  {Omont}, {Riechers}, {Bertoldi}, {Menten}, {Cox}, {Strauss}, {Fan}, \&
  {Jiang}}]{WangRan2011b}
{Wang}, R., {Wagg}, J., {Carilli}, C.~L., {et~al.} 2011, \aj, 142, 101,
  \dodoi{10.1088/0004-6256/142/4/101}

\bibitem[{{Wang} {et~al.}(2016{\natexlab{b}}){Wang}, {Wu}, {Neri}, {Fan},
  {Walter}, {Carilli}, {Momjian}, {Bertoldi}, {Strauss}, {Li}, {Wang},
  {Riechers}, {Jiang}, {Omont}, {Wagg}, \& {Cox}}]{WangRan2016}
{Wang}, R., {Wu}, X.-B., {Neri}, R., {et~al.} 2016{\natexlab{b}}, \apj, 830,
  53, \dodoi{10.3847/0004-637X/830/1/53}

\bibitem[{{Weigel} {et~al.}(2015){Weigel}, {Schawinski}, {Treister}, {Urry},
  {Koss}, \& {Trakhtenbrot}}]{Weigel2015}
{Weigel}, A.~K., {Schawinski}, K., {Treister}, E., {et~al.} 2015, \mnras, 448,
  3167, \dodoi{10.1093/mnras/stv184}

\bibitem[{{Wenzl} {et~al.}(2021){Wenzl}, {Schindler}, {Fan}, {Andika},
  {Ba{\~n}ados}, {Decarli}, {Jahnke}, {Mazzucchelli}, {Onoue}, {Venemans},
  {Walter}, \& {Yang}}]{Wenzl2021}
{Wenzl}, L., {Schindler}, J.-T., {Fan}, X., {et~al.} 2021, \aj, 162, 72,
  \dodoi{10.3847/1538-3881/ac0254}

\bibitem[{{W}es {M}c{K}inney(2010)}]{pandas_paper}
{W}es {M}c{K}inney. 2010, in {P}roceedings of the 9th {P}ython in {S}cience
  {C}onference, ed. {S}t\'efan van~der {W}alt \& {J}arrod {M}illman, 56 -- 61,
  \dodoi{10.25080/Majora-92bf1922-00a}

\bibitem[{{Willott} {et~al.}(2007){Willott}, {Delorme}, {Omont}, {Bergeron},
  {Delfosse}, {Forveille}, {Albert}, {Reyl{\'e}}, {Hill}, {Gully-Santiago},
  {Vinten}, {Crampton}, {Hutchings}, {Schade}, {Simard}, {Sawicki}, {Beelen},
  \& {Cox}}]{Willott2007}
{Willott}, C.~J., {Delorme}, P., {Omont}, A., {et~al.} 2007, \aj, 134, 2435,
  \dodoi{10.1086/522962}

\bibitem[{{Willott} {et~al.}(2010){Willott}, {Delorme}, {Reyl{\'e}}, {Albert},
  {Bergeron}, {Crampton}, {Delfosse}, {Forveille}, {Hutchings}, {McLure},
  {Omont}, \& {Schade}}]{Willott2010a}
{Willott}, C.~J., {Delorme}, P., {Reyl{\'e}}, C., {et~al.} 2010, \aj, 139, 906,
  \dodoi{10.1088/0004-6256/139/3/906}

\bibitem[{{Wu} {et~al.}(2015){Wu}, {Wang}, {Fan}, {Yi}, {Zuo}, {Bian}, {Jiang},
  {McGreer}, {Wang}, {Yang}, {Yang}, {Thompson}, \& {Beletsky}}]{Wu2015}
{Wu}, X.-B., {Wang}, F., {Fan}, X., {et~al.} 2015, \nat, 518, 512,
  \dodoi{10.1038/nature14241}

\bibitem[{{Wyithe} \& {Bolton}(2011)}]{Wyithe2011c}
{Wyithe}, J. S.~B., \& {Bolton}, J.~S. 2011, \mnras, 412, 1926,
  \dodoi{10.1111/j.1365-2966.2010.18030.x}

\bibitem[{{Yang} {et~al.}(2016){Yang}, {Wang}, {Wu}, {Fan}, {McGreer}, {Bian},
  {Yi}, {Yang}, {Ai}, {Dong}, {Zuo}, {Green}, {Jiang}, {Wang}, {Wang}, \&
  {Yue}}]{Yang2016}
{Yang}, J., {Wang}, F., {Wu}, X.-B., {et~al.} 2016, \apj, 829, 33,
  \dodoi{10.3847/0004-637X/829/1/33}

\bibitem[{{Yang} {et~al.}(2019){Yang}, {Wang}, {Fan}, {Wu}, {Bian},
  {Ba{\~n}ados}, {Yue}, {Schindler}, {Yang}, {Jiang}, {McGreer}, {Green}, \&
  {Dye}}]{YangJinyi2019b}
{Yang}, J., {Wang}, F., {Fan}, X., {et~al.} 2019, \apj, 871, 199,
  \dodoi{10.3847/1538-4357/aaf858}

\bibitem[{{Yang} {et~al.}(2020){Yang}, {Wang}, {Fan}, {Hennawi}, {Davies},
  {Yue}, {Banados}, {Wu}, {Venemans}, {Barth}, {Bian}, {Boutsia}, {Decarli},
  {Farina}, {Green}, {Jiang}, {Li}, {Mazzucchelli}, \&
  {Walter}}]{YangJinyi2020}
---. 2020, \apjl, 897, L14, \dodoi{10.3847/2041-8213/ab9c26}

\bibitem[{{Yang} {et~al.}(2021){Yang}, {Wang}, {Fan}, {Barth}, {Hennawi},
  {Nanni}, {Bian}, {Davies}, {Farina}, {Schindler}, {Ba{\~n}ados}, {Decarli},
  {Eilers}, {Green}, {Guo}, {Jiang}, {Li}, {Venemans}, {Walter}, {Wu}, \&
  {Yue}}]{YangJinyi2021}
---. 2021, \apj, 923, 262, \dodoi{10.3847/1538-4357/ac2b32}

\bibitem[{{Yue} {et~al.}(2022){Yue}, {Fan}, {Yang}, \& {Wang}}]{YueMinghao2022}
{Yue}, M., {Fan}, X., {Yang}, J., \& {Wang}, F. 2022, \apj, 925, 169,
  \dodoi{10.3847/1538-4357/ac409b}

\bibitem[{Zonca {et~al.}(2019)Zonca, Singer, Lenz, Reinecke, Rosset, Hivon, \&
  Gorski}]{Zonca2019}
Zonca, A., Singer, L., Lenz, D., {et~al.} 2019, Journal of Open Source
  Software, 4, 1298, \dodoi{10.21105/joss.01298}

\bibitem[{{Zuo} \& {Phinney}(1993)}]{ZuoLin1993}
{Zuo}, L., \& {Phinney}, E.~S. 1993, \apj, 418, 28, \dodoi{10.1086/173367}

\end{thebibliography}
\bibliographystyle{aasjournal}



\end{document}